\documentclass{article}
   \usepackage{graphicx}
   \usepackage{amsmath}
   \usepackage{amssymb}
   \usepackage{epstopdf}
   \usepackage{pdfsync}
   \usepackage{color}
   \usepackage{float}
   \usepackage{apjgalley}
\font\fiverm=cmr5             \font\sevenrm=cmr7

          \font\sixrm=cmr6

\def\reference{\par \noindent \hangafter=1 \hangindent=0.7 true cm}
\def\fsc{\alpha_{\hbox{\sevenrm f}}}                                
\def\machson{{\cal M}_{\hbox{\sixrm S}}}
\def\machalf{{\cal M}_{\hbox{\sixrm A}}}

\def\dover#1#2{\hbox{${{\displaystyle#1 \vphantom{(} }\over{
   \displaystyle #2 \vphantom{(} }}$}}
\def\teq#1{$\, #1\,$}                         
\def\ThetaBfone{\Theta_{\hbox{\sixrm Bf1}}}
\def\ThetaBftwo{\Theta_{\hbox{\sixrm Bf2}}}
\def\ThetaBsone{\Theta_{\hbox{\sixrm Bs1}}}
\def\ThetaBstwo{\Theta_{\hbox{\sixrm Bs2}}}
\def\ThetaBHTone{\Theta_{\hbox{\sixrm B\fiverm HT\sixrm 1}}}
\def\ThetaBHTtwo{\Theta_{\hbox{\sixrm B\fiverm HT\sixrm 2}}}

\def\thetautwo{\theta_{\hbox{\sixrm u2}}}
\def\thetautwoS{\theta_{\hbox{\sixrm u2S}}}
\def\betaonexS{\beta_{\hbox{\sixrm 1xS}}}
\def\betatwoxS{\beta_{\hbox{\sixrm 2xS}}}
\def\betatwozS{\beta_{\hbox{\sixrm 2zS}}}
\def\betaxS{\beta_{\hbox{\sixrm xS}}}
\def\betazS{\beta_{\hbox{\sixrm zS}}}
\def\betaxHT{\beta_{\hbox{\sixrm xHT}}}
\def\betazHT{\beta_{\hbox{\sixrm zHT}}}
\def\gammaoneS{\Gamma_{\hbox{\sixrm 1S}}}
\def\gammatwoS{\Gamma_{\hbox{\sixrm 2S}}}
\def\BxS{B_{\hbox{\sixrm xS}}}
\def\BzS{B_{\hbox{\sixrm zS}}}
\def\BxHT{B_{\hbox{\sixrm xHT}}}
\def\BzHT{B_{\hbox{\sixrm zHT}}}
\def\Bfone{B_{\hbox{\sixrm 1f}}}
\def\BHTone{B_{\hbox{\sixrm 1HT}}}
\def\BHTtwo{B_{\hbox{\sixrm 2HT}}}
\def\vect#1{\boldsymbol{\vec{#1}}}
\def\uoneHT{u_{\hbox{\sixrm 1HT}}}
\def\betaoneHT{\beta_{\hbox{\sixrm 1HT}}}
\def\thetascatt{\theta_{\hbox{\sevenrm scatt}}}

\def\pr{Phys. Rev.}                             
\def\aa{{Astron. Astrophys.}}

\def\apj{ApJ}

\def\app{Astroparticle Phys.}                   
\def\apss{Astr. Space Sci.}                     
\def\asr{Adv. Space Res.}                       
\def\grl{Geophys. Res. Lett.}                   
\def\jgr{J. Geophys. Res.}
\def\mnras{{M.N.R.A.S.}}
\def\prl{Phys. Rev. Lett.}                      
\def\prd{Phys. Rev. D}                          
\def\ssr{Space Sci. Rev.}                       

\begin{document}
\newcommand{\vol}[2]{$\,$\rm #1\rm , #2.}                 
\newcommand{\figureoutpdf}[5]{\centerline{}
   \centerline{\hspace{#3in} \includegraphics[width=#2truein]{#1}}
   \vspace{#4truein} \figcaption{#5} \centerline{} }
\newcommand{\twofigureoutpdf}[3]{\centerline{}
   \centerline{\includegraphics[width=3.8truein]{#1}
        \hspace{0.0truein} \includegraphics[width=3.8truein]{#2}}
        \vspace{-0.1truein}
    \figcaption{#3} }    
\newcommand{\twofigureoutpdfadj}[3]{\centerline{}
   \centerline{\includegraphics[height=3.0truein]{#1}
        \hspace{+0.2truein} \includegraphics[height=2.9truein]{#2}}
        \vspace{-0.1truein}
    \figcaption{#3} }    
\newcommand{\fourfigureoutpdf}[5]{\centerline{}
   \centerline{\includegraphics[width=3.7truein]{#1}
        \hspace{0.0truein} \includegraphics[width=3.7truein]{#2}}
        \vspace{-0.1truein}
   \centerline{\includegraphics[width=3.7truein]{#3}
        \hspace{0.0truein} \includegraphics[width=3.7truein]{#4}}
        \vspace{-0.1truein}
    \figcaption{#5} }    
%

\title{DIFFUSIVE ACCELERATION OF PARTICLES AT OBLIQUE,\\
     RELATIVISTIC, MAGNETOHYDRODYNAMIC SHOCKS}

    \author{Errol J. Summerlin}
    \affil{Heliospheric Physics Laboratory, Code 672,\\
        NASA's Goddard Space Flight Center, Greenbelt, MD 20770, USA \\  
        {\it errol.summerlin@nasa.gov}}

    \and

   \author{Matthew G. Baring}
   \affil{Department of Physics and Astronomy, MS 108,\\
      Rice University, Houston, TX 77251, U.S.A.\\
      {\it baring@rice.edu}}
\slugcomment{Accepted for publication in The Astrophysical Journal; 
to appear in the December 20, 2011 issue.}

\begin{abstract}
Diffusive shock acceleration (DSA) at relativistic shocks is expected to
be an important acceleration mechanism in a variety of astrophysical
objects including extragalactic jets in active galactic nuclei and gamma
ray bursts.   These sources remain good candidate sites for the
generation of ultra-high energy cosmic rays. In this paper, key
predictions of DSA at relativistic shocks that are germane to production
of relativistic electrons and ions are outlined.  The technique employed
to identify these characteristics is a Monte Carlo simulation of such
diffusive acceleration in test-particle, relativistic, oblique,
magnetohydrodynamic (MHD) shocks.  Using a compact prescription for
diffusion of charges in MHD turbulence, this approach generates particle
angular and momentum distributions at any position upstream or
downstream of the shock. Simulation output is presented for both small
angle and large angle scattering scenarios, and a variety of shock
obliquities including superluminal regimes when the de Hoffmann-Teller
frame does not exist. The distribution function power-law indices
compare favorably with results from other techniques.  They are found to
depend sensitively on the mean magnetic field orientation in the shock,
and the nature of MHD turbulence that propagates along fields in shock
environs. An interesting regime of flat spectrum generation is
addressed; we provide evidence for it being due to shock drift
acceleration, a phenomenon well-known in heliospheric shock studies. 
The impact of these theoretical results on blazar science is outlined.
Specifically, {\it Fermi}-LAT gamma-ray observations of these
relativistic jet sources are providing significant constraints on
important environmental quantities for relativistic shocks, namely the
field obliquity, the frequency of scattering and the level of field
turbulence.
\end{abstract}


\section{INTRODUCTION}
 \label{sec:intro}
Collisionless magneto-hydrodynamic (MHD) shocks are found in diverse
environments ranging from the inner heliosphere to the central regions
of distant galaxies and other astrophysical objects. Particle
acceleration at these collisionless shocks is believed to be a common
phenomenon in space plasmas. In the heliosphere, direct measurements of
accelerated non-thermal ions and electrons in various energy ranges at
the Earth's bow shock (e.g. Scholer et al., 1980, M\"{o}bius et al.,
1987 and Gosling et al., 1989) and interplanetary shocks (e.g., Sarris
\& Van Allen, 1974; Decker et al. 1981; Tan et al. 1988; Baring et al.
1997) indicate energization processes that are intimately connected to
shock environs. Outside the heliosphere, non-thermal particle
distributions are inferred from observed photon spectra of supernova
remnants, pulsar wind nebulae, blazars, and gamma-ray bursts (e.g.
Blandford \& Eichler, 1987, and references therein), all of which
possess supersonic outflows that are readily shocked.  Commonly, these
non-thermal distributions take the form of power-law tails that can
extend to thousands or millions of times the ambient thermal energies of
the particles.

First-order Fermi acceleration, often called diffusive shock
acceleration (DSA), is believed to be the primary acceleration mechanism
in most collisionless MHD shocks.  This phenomenon arises when charged
particles interact quasi-elastically with turbulent fields in the shock
layer, and are diffusively transported back and forth across the shock,
each time achieving a net gain in energy on average.  Monte Carlo
simulations of this process (see Jones and Ellison, 1991, and references
therein) have had great success in modeling shocks inside the
heliosphere and comparing them directly with in-situ measurements from
various spacecraft (e.g. Ellison et al., 1990b; Baring et al., 1997;
Summerlin \& Baring, 2006). It is quite likely that this same process is
responsible for the power-law tails inferred in astrophysical shocks,
including relativistic MHD discontinuites such as those believed to be
associated with blazars (e.g. see Stecker, Baring \& Summerlin 2007) and
gamma-ray bursts (e.g. see reviews by Piran 1999; M\'esz\'aros, 2001).

Early work on relativistic shocks was mostly analytic in the
test-particle approximation (e.g., Peacock 1981; Kirk \& Schneider 1987;
Heavens \& Drury 1988; Kirk \& Heavens 1989), where the accelerated
particles do not contribute significantly to the global MHD structure of
the shock.  Since such systems are inherently anisotropic, due to rapid
convection of particles through and away downstream of the shock, the
diffusion approximation cannot be applied. This renders analytic
approaches, such as solution of the diffusion-convection Fokker-Planck
equation, more difficult for ultra-relativistic upstream flows, though
advances can be made in special cases, such as the limit of extremely
small angle scattering (e.g. Kirk \& Schneider 1987; Kirk et al. 2000).
Accordingly, complementary Monte Carlo techniques, first developed for
non-relativistic shock applications by Ellison, Jones \& Eichler (1981),
have been employed for relativistic shocks by a number of authors,
including test-particle analyses for steady-state shocks of parallel and
oblique magnetic fields by Ellison et al. (1990a), Ostrowski (1991),
Bednarz \& Ostrowski (1998), Baring (1999), Niemiec \& Ostrowski (2004),
Ellison \& Double (2004) and Stecker, Baring \& Summerlin (2007). It is
such a simulational approach that is highlighted here; its accessibility
to broad dynamic ranges in momenta is extremely desirable, providing a
niche for Monte Carlo techniques in connecting with observations of
astronomical objects such as gamma-ray bursts (GRBs) and blazars.

It should be noted that the most comprehensive way to study dissipation,
acceleration and wave generation in collisionless shocks is with
particle-in-cell (PIC) simulations, where particle motion and field
fluctuations are obtained as solutions of the Newton-Lorentz and
Maxwell's equations. Relativistic PIC codes have blossomed to model
shocks in applications such as GRBs and pulsar wind termination shocks,
focusing largely, but not exclusively, on perpendicular shocks (e.g.
Gallant et al. 1992; Smolsky \& Usov 1996; Silva et al. 2003; Hededal et
al. 2004; Liang \& Nishimura 2004; Medvedev et al. 2005; Nishikawa et
al. 2005; and Spitkovsky 2008). These works have explored pair shocks,
ion-doped shocks, Poynting flux-dominated outflows, and low-field
systems with dissipation driven by the Weibel instability. PIC
simulations are dynamic in nature, and rarely achieve a time-asymptotic
state. None of these works has demonstrated the establishment of an
extended power-law that is required in modeling emission from gamma-ray
bursts and active galactic nuclei, though note the isolated recent
suggestion (Spitkovsky, 2008; Sironi \& Spitkovsky 2011) of a
non-thermal tail generated by diffusive transport.  The general
difficulty with explicitly seeing acceleration in PIC codes beyond true
thermalization is perhaps due to the severely restricted spatial and
temporal scales of the simulations, imposed by their intensive CPU and
memory requirements.  With the anticipated advances in computational
capability over the next decade, PIC simulations will become a much more
powerful tool for probing DSA. For a discussion of relativistic shock
acceleration, see Baring (2004).

To date, much simulational work on DSA at relativistic shocks has
focused on parallel systems (where the magnetic field direction is
parallel to the shock normal) in which particles experience frequent
small angle scatterings (SAS), as opposed to infrequent large angle
scatterings (LAS). In the limit of ultra-relativistic shock speeds, for
differential particle distributions \teq{dn/dp=p^{-\sigma}}, a power-law
index of \teq{\sigma\approx 2.23} is realized, as can be found
analytically (e.g. Kirk et al., 2000) and numerically (e.g. Bednarz \&
Ostrowski, 1998; Baring, 1999; Ellison \& Double 2004).  However, it is
not necessary to assume that SAS is the dominant scattering mechanism,
nor is it warranted in some situations: the phase space for the
character of small angle scattering to be realized shrinks with
increasing shock Lorentz factor. Moreover, many astrophysical shocks,
such as those in blazar jets, are either not parallel, or not
ultra-relativistic. Clearly, a more robust examination of the parameter
space is desirable if one is to characterize the emission coming from
these objects, and use it to probe their shocked plasma environments.

To effect such a goal, here we have extended our Monte Carlo DSA code
(Summerlin \& Baring, 2006) to include shocks of arbitrary speed and
obliquity, including the trans-relativistic regime. Additionally, we
generally presume an electron-positron plasma shock, following current
thinking on the nature of GRB outflows (e.g. Piran 1999; M\'esz\'aros
2001) and blazar jets, though the results apply equally well to
ion-dominated relativistic shocks.  The global structure of the shocks
is defined via the Rankine-Hugoniot relations, solved along the lines of
previous expositions (e.g. Double et al., 2004).  Principal output
includes complete momentum and angular distributions, at different
distances upstream and downstream of the shock.  To demonstrate the
validity of the simulation, and to distinguish its particular character,
comparisons are made with both theoretical and simulation results of
other papers (principally Kirk \& Heavens, 1989; Kirk et al., 2000;
Ellison \& Double, 2004; Niemiec \& Ostrowski 2004). More importantly,
we expand on these previous works by exploring the parameter space for
oblique relativistic shocks comprehensively, focusing on the shock
obliquity, turbulence levels, and parameters encapsulating the
microphysics of the turbulent interactions, as key variables determining
the high energy power-law index of the particle distribution.

We find that, in relativistic shocks, unlike in non-relativistic shocks,
the microphysics of the turbulence becomes an important factor in
determining both the value of the power-law index, and how many decades
in energy particles are accelerated before a power law is achieved.
Particles undergoing infrequent large angle scatterings consistently
produce harder power laws than their SAS counterparts and take many more
decades in energy to realize a smooth power-law.  It is also apparent
that the power-law index is critically dependent upon the subluminality,
versus superluminality, of the shock, as discussed in
Sec.~\ref{sec:results}.  We find that, as do Ellison and Double (2004)
and Baring (2004), in superluminal shocks, the power-law rapidly becomes
softer with decreasing levels of turbulence and increasing obliquity,
due to the difficulty particles have returning to the shock once they
have crossed to the downstream side.

In distinct contrast, in the case of subluminal shocks, a decreased
amount of turbulence and increased obliquity can actually render the
acceleration process far more efficient as particles undergo the
coherent process of shock drift acceleration (SDA), where some particles
persistently gyrate in the shock layer, preferentially gaining energy
due to the kinking of the magnetic field. In the limit of no cross-field
diffusion and a de-Hoffmann Teller frame velocity of nearly $c$,
explored theoretically by Kirk and Heavens (1989) using semi-analytic
solutions to the diffusion-convection equation, an extremely low value
of the power-law index around $\sigma=1$ becomes possible.  However,
with our simulation, we are able to more readily isolate how such flat
distributions arise. In marginally subluminal shocks with SAS operating,
a small fraction of high energy particles are reflected off the shock by
the kink in the magnetic field.  For those that do reflect, the angular
distribution for subsequent shock encounters is such that the
transmission region is almost entirely depleted, resulting in virtually
100\% reflection at each shock encounter. These particles essentially
become trapped and are accelerated to very high energies very quickly,
before they are eventually lost downstream. The extremely low levels of
turbulence necessary to permit SDA to act unabated almost certainly do
not occur in Nature, but the effects of SDA can be seen to a lesser
degree in shocks with more realistic parameters. In general, it can be
concluded that the power-law indices in relativistic shocks can sample a
broad range, depending on the three basic system parameters explored
here.  After outlining our simulation technique in
Sec.~\ref{sec:MCtechnique} and summarizing our method for determining
the shock jump conditions in Sec.~\ref{sec:RHrel}, our results are
presented in Sec.~\ref{sec:results}, and then interpreted in the context
of blazars in Sec.~\ref{sec:blazar}.

\section{The Monte Carlo Simulation Technique}
  \label{sec:MCtechnique}
The Monte Carlo Simulation technique employed in this paper closely
follows the pioneering work on this method by Ellison, Jones \& Eichler
(1981) and Ellison \& Eichler (1984); see Jones \& Ellison (1991) for a
review. It is a test-particle simulation that models convection and
diffusion of charges in a turbulent, shocked flow, complementing the
analytic approach of Bell (1978) that was extended to the relativistic
regime by Peacock (1981). It has been successfully applied in a variety
of environments including, the Earth's bow shock (Ellison, M\"obius \&
Paschmann 1990b), interplanetary shocks (e.g Baring et al. 1997;
Summerlin \& Baring 2006), the solar wind termination shock (see Ellison
et al. 1999), supernova remnants (see Baring et al. 1999; Baring \&
Summerlin 2007), and in the regime of highly relativistic shocks that is
generally encountered in extragalactic contexts (e.g. Ellison et al.
1990a; Ellison \& Double 2004; Stecker, Baring \& Summerlin 2007).  The
code models particle gyration about bulk magnetic fields in convecting
fluid flows, while having their trajectories perturbed by embedded
hydromagnetic turbulence. The perturbations mediate spatial diffusion
that permits some small fraction of particles to transit the shock front
multiple times, kinematically sampling the difference in flow speeds on
either side of the shock, and thereby being accelerated via first-order
Fermi (or diffusive) shock acceleration (see Bell 1978; Jones \& Ellison
1991). The code is fully relativistic and transitions seamlessly from
non-relativistic to relativistic flow regimes; it also treats arbitrary
orientations of the mean magnetic field.

The simulation space is divided into a distinct number of grid zones
distributed along the x-axis, which is here defined to be the direction
normal to the planar shock surface. The boundaries of these grid zones
are locations where the bulk properties of the fluid (flow speeds,
magnetic fields, etc.) can change. The values of these fluid properties
are specified {\it a priori}, and for the test-particle implementation
of the simulation in this paper, have fixed values throughout the
simulation runs. In the simulations presented in this paper a simple
step function shock is used with only 2 gridzones: one upstream, and one
downstream.  The field and fluid quantities in these two zones are
related by the fully-relativistic, Rankine-Hugoniot jump conditions, as
discussed in Sec.~\ref{sec:RHrel} below.  This construction facilitates
the generalization to non-linear acceleration regimes (e.g. Ellison \&
Eichler 1984; Ellison, Baring \& Jones 1996; see also Ellison \& Double,
2002, for the first treatment of non-linear modification of relativistic
shocks), where the energetic particles contribute to the grid-by-grid
specification of MHD quantities constrained by energy/momentum flux
conservation.

Particles are injected isotropically into the system anywhere along the
x-axis, though usually an upstream injection is adopted.  The energy
distribution of injected particles can be either mono-energetic, a
thermal Maxwell-Boltzmann form at any temperature (relativistic or
non-relativistic), or a power-law distribution in momentum of arbitrary
index. For non-relativistic shocks with thermal particle injection, the
code automatically calculates the Rankine-Hugoniot shock jump conditions
to ascertain the downstream fluid and field vector values.  For
relativistic shocks, the jump condition solution technique is
necessarily more complicated, as described in Sec. 3.  This solution is
accomplished outside the simulation program, and the jump conditions are
then input manually as initial conditions for the simulation runs.  The
code can also include multiple species of charged particles (e.g.
treating hydrogenic and pair plasmas, and even contributions from
helium) besides the test-particles in the determination of the jump
conditions. After particles are injected into the upstream fluid, they
are allowed to gyrate in the local bulk magnetic field, convecting with
the fluid, until it is determined that a phenomenological scattering
occurs.

The effects of magnetic turbulence are simulated by specifying a local
{\it fluid frame} mean free path for particle diffusion, given by
\begin{equation}
   \lambda \; =\; \lambda_{0} \left( \frac{r_{g}}{r_{g1}} \right)^{\alpha} \; \propto\; p^{\alpha}
   \quad ,\quad \lambda_{0}=\eta r_{g1}\quad ,
 \label{eq:mfp}
\end{equation}
where \teq{r_{g}=pc/(q B)} is the gyroradius of an ion or electron of
momentum \teq{p=mv}, mass \teq{m}, and charge \teq{q} in a magnetic
field \teq{B=\vert\hbox{\bf B}\vert}. Also \teq{r_{g1}=mu_{1x}c/(q B)}
is the gyroradius of an ion with a speed \teq{v} equal to the velocity
component, \teq{u_{1x}}, of the far upstream flow normal to the shock
plane; here \teq{x} denotes the direction normal to the shock.  Without
loss of generality, the mean free path scale \teq{\lambda_{0}} is set
proportional to \teq{r_{g1}} with constant of proportionality \teq{\eta}
defined via Eq.~(\ref{eq:mfp}). This phenomenological prescription for
scattering was adopted in numerous papers outlining results from the
Monte Carlo technique, including Ellison, Jones \& Eichler (1981),
Ellison, Jones \& Reynolds (1990a), Ellison, Baring \& Jones (1995,
1996), and Stecker, Baring \& Summerlin (2007). Following this and other
previous Monte Carlo work, for simplicity, we set \teq{\alpha=1}, a
specialization that is appropriate for traveling interplanetary shocks:
see Ellison et al. (1990a,b), Mason et al. (1983) and Giacalone et al.
(1992) for discussions about the micro-physical expectations  for
\teq{\alpha}.   The simulation can easily accommodate other values of
\teq{\alpha}, however, the spectral results are somewhat insensitive to
the choice of this parameter --- its dominant effect is to modify the
relative scale lengths for diffusion at different particle momenta. 
Since \teq{\lambda \geq r_{g}} is required for physically meaningful
diffusion resulting from gyro-resonant wave-particle interactions, the
\teq{\alpha=1} case is also motivated on fundamental grounds. The mean
free path represents the spatial scale in the local fluid frame on which
the momentum vector is deflected by $\pi/2$, on average.  Note that for
diffusion that is driven by non-gyroresonant interactions with field
turbulence, perhaps grown via filamentation or Weibel instabilities, it
is quite possible to sample \teq{\eta < 1} regimes, especially when the
ambient magnetic field is quite low.  Diffusion in this domain resembles
the Bohm limit of \teq{\eta =1} for gyroresonant diffusion, and
accordingly the distributions for shock-accelerated charges are only
mildly dependent on \teq{\eta} when it is less than unity. For high
Alfv\'enic Mach number shocks, the scattering is approximately elastic
in the fluid frame, i.e., \teq{\vert \hbox{\bf p}\vert} is conserved in
this frame for interactions with field turbulence that perturb a
particle's pitch angle \teq{\theta}, gyrophase and orbital gyrocenter.

When the Alfv\'enic Mach number \teq{\machalf} is low, the Alfv\'en
waves move with appreciable speed in the fluid frame, so that partial
inelasticity in scatterings arises. This yields second order, stochastic
diffusion contributions.  While these can be routinely modeled in the
simulation, inspection of Eqs.~(8) and~(10) of Pryadkho \& Petrosian's
(1997) quasi-linear stochastic acceleration formalism clearly indicates
that the stochastic contribution to the spatial diffusion coefficients
is smaller than the first order Fermi one by the order of
\teq{1/\machalf^2}. For the efficient generation of the high energy
power-law tails that are the primary focus of this paper, the
astrophysical shocks of interest generally have large enough Alfv\'enic
Mach numbers to neglect the effects of second-order acceleration.
However, note that for near-luminal shocks at slightly suprathermal
energies, particles are generally unable to convect upstream against the
downstream flow and are inexorably swept downstream. In this energy
regime, other mechanisms acting in the shock environs such as
second-order Fermi acceleration or electrostatic cross-shock potentials
may noticeably broaden/heat the downstream distribution function.  This
can then enhance injection into the first-order acceleration process,
and thereby affect the normalization of the power-law tail that results,
particularly in cases of strongly-inhibited injection.  Notwithstanding,
treatment of stochastic acceleration effects will be deferred to future work.

The simplest invocation of scattering is to isotropize the fluid frame
momentum over the surface of the sphere in momentum space (Ellison et
al., 1990a). This is large angle scattering (LAS), and physically
corresponds to large magnetic disturbances that completely disrupt
trajectories of particles. To model moderate or even smaller
disturbances, each scattering event can be restricted to a much smaller
solid angle, i.e. can be isotropic on a conical sector of a momentum
sphere.  The angular extent of this spherical sector
\teq{\delta\theta_{max}} becomes an additional parameter for the
diffusion.  Then multiple scattering events are required to realize a
full mean free path.  This is the scattering construct that is employed
in this paper. The relationship between \teq{\delta\theta_{max}} and
\teq{\lambda} was originally developed in Ellison et al. (1990a), but is
more succinctly presented in Ellison and Double (2004) via
\begin{equation}
   \delta\theta_{max} \;=\; \sqrt{\dover{12 \pi r_g}{\lambda N}}\quad ,
 \label{eq:delthetamax}
\end{equation}
where \teq{r_g} is the gyro-radius, and \teq{N} is the number of times
per gyroradius the particle is scattered.   The limit of small angle
scattering (SAS) corresponds to \teq{N\gg 1}, for which the increment
\teq{\delta\hbox{\bf p}} in momentum in a scattering satisfies
\teq{\vert \delta\hbox{\bf p}\vert /\vert\hbox{\bf p}\vert\sim\delta\theta_{max}}.
In practice, as will become evident below, for relativistic shocks the
SAS domain is realized when the scattering angle satisfies
\teq{\delta\theta_{max}\ll 1/\Gamma_1}, where
\teq{\Gamma_1=1/\sqrt{1-u_{1x}^2/c^2}} is the bulk Lorentz factor of the
upstream fluid in the shock rest frame.

Cross-field diffusion emerges naturally from this scattering mechanism
since, at every scattering, the particle's momentum vector is shifted in
the local fluid frame, with the resulting effect that the gyrocenter of
the particle is shifted randomly by a distance of order
\teq{r_g\sin\theta} in the plane orthogonal to the local field.
Transport perpendicular to the field is then governed by a kinetic
theory description, so that the ratio of the spatial diffusion
coefficients parallel (\teq{\kappa_{||}=\lambda v/3}) and perpendicular
(\teq{\kappa_{\perp}}) to the magnetic field is given by
\teq{\kappa_{\perp}/\kappa_{||}=1/(1+\eta^2)} (see Forman, Jokipii \&
Owens 1974; Ellison, Baring \& Jones 1995, for detailed expositions).
Hence, \teq{\eta} couples directly to the amount of cross-field
diffusion and is a measure of the level of turbulence in the system,
i.e., is an indicator of \teq{\langle \delta B/B \rangle}. The
quasi-isotropic diffusion case of \teq{\eta=1} constitutes the Bohm
diffusion limit, presumably corresponding to \teq{\langle \delta
B/B\rangle \approx 1}.

As will become clear in Secs.~\ref{sec:obq_sublumin} 
and~\ref{sec:obq_superluminal}, in oblique relativistic
shocks, the resulting energy spectrum is critically dependent upon both
\teq{\eta}, due to the necessity of cross-field diffusion, and the
scattering angle \teq{\delta\theta_{max}}, due to beaming effects,
producing a broad range of power law indices.  For small angle
scattering (SAS) regimes, \teq{\delta\theta_{max}<1/\Gamma_1}, there is
little variation in the power law tails when other parameters are held
constant, since the scatter angle is now less than the relativistic
beaming angle, and the diffusion process becomes insensitive to the
scattering kernel. Except for Sec.~\ref{sec_LAS}, SAS is deployed 
throughout this
paper. Examples of the differences between small and large angle
scattering in relativistic shocks can be seen Fig. 2 of Stecker, Baring
\& Summerlin (2007), and also in Fig.~\ref{fig:SAS_LAS_plot}.

In between each of the \teq{N} scatterings per mean free path, the code
calculates shock frame gyro-orbit trajectories using a semi-analytic
solver rather than the more popular Bulirsch-Stoer method (Stoer \& 
Bulirsch, 1980). Using the properties of the magnetized fluid, the shock
frame position as a function of time is easily derived analytically. The
particle is then moved along this analytic trajectory until one of two
conditions is met: (a) the particle scatters or (b) the particle reaches
the edge of a grid zone. The solution for the time it takes a particle
to reach the edge of a grid zone must be performed numerically, since it
involves roots of a transcendental equation of motion in the shock frame
-- the simulation employs a standard bisection technique for this
purpose. When a particle crosses a grid zone boundary, the local fluid
properties change, and the trajectory is recalculated and the
propagation continues. When distances between scatterings are many
gyro-radii, the semi-analytic method can go from one scattering to the
next in one step covering many gyro-orbits in a single computational
step. The Bulirsch-Stoer method will always require at least several
steps per gyro-orbit due to the curvature of the trajectory. However, if
particles scatter many times in one gyro-radius, the increased overhead
of the semi-analytic method makes it slower than the Bulirsch-Stoer
method, but not unreasonably so.

Particles that do not immediately return to the shock may isotropize in
the downstream reference frame once they have traveled, on average, one
mean free path. At this juncture, an analytical formula developed
originally by Bell (1978) and later shown to be applicable to
relativistic shocks by Peacock (1981; see also Jones \& Ellison 1991)
can be used to calculate the probability \teq{P_r} that a particle
heading downstream through a \teq{y}-\teq{z} plane at a particular
distance \teq{x} downstream will return upstream of this plane:
\begin{equation}
   P_r \; =\; \left(\dover{v_f-u}{v_f+u}\right)^{2}\quad .
 \label{eq:prret}
\end{equation}
In the above equation, \teq{u} is the local downstream flow speed, and
\teq{v_f} is the speed of the particle in this fluid frame. Particles
that are deemed to fail to return are removed from the system.  For
those ascertained to be returning, their vector velocity components are
also determined probabilistically. The particles are isotropic in the
local fluid frame and have constant energy in the downstream frame of
reference thanks to the elastic scattering off magnetic turbulence. So,
the probability of a particle of a given fluid frame momentum returning
with a particular angle cosine with respect to the shock normal,
\teq{\mu_{s}} , can be found for arbitrary values of the particle speed
and downstream flow speed. The details of this calculation and final
result can be found in the Appendix, specifically
Eq.~(\ref{eq:flux_ang_dist}).  Employing this result, a simple
accept-reject method (Garcia, 2000, Ch. 11) can be used to select a
value for \teq{\mu_{s}} for particles determined to have returned. This
statistical decision algorithm circumvents excessive computations of
extensive downstream diffusion that are irrelevant to the acceleration
process; accordingly, it speeds up the simulation dramatically.  Using
the correct angular distribution of returning particles, i.e.
Eq.~(\ref{eq:flux_ang_dist}), is essential, guaranteeing that the
complete distribution function of particles anywhere upstream of the
probability of return plane is independent of the choice of \teq{x},
provided \teq{x>\lambda} and isotropy in the fluid frame is satisfied at
\teq{x}.

\newpage

For simulation output, accounting of particles in distinct momentum bins
is documented.  As a result of statistical losses in the downstream
region, when less than half of the particles originally recorded in a
given momentum bin are retained, the remaining particles are ``split"
into two particles each with half the ``counting" value of the original.
This technique of particle splitting allows the simulation to maintain
good statistics over a large energy range. This extensive energy range
is one of the primary advantages that the Monte-Carlo technique has over
other types of simulations. Compared to hybrid plasma simulations and
particle-in-cell (PIC) simulations, Monte-Carlo simulations are
computationally inexpensive, allowing the simulation to be run long
enough for particles to be accelerated to very high energies, well above
that of the incoming upstream ballistic flow, in a reasonable amount of
time.

In the test particle implementation employed here, the characteristics
of the shock and the functional form of the turbulence are specified
{\it a priori}. The test particle approximation is entirely appropriate
unless there is a significant fraction of the total energy present in
the accelerated particles. Since the distribution of these particles is
only known after the simulation accelerates them, the shape of the shock
can not be adjusted to account for their existence until after the
simulation is run. For non-relativistic shocks, Ellison and Eichler
(1984) developed a feedback loop technique where subsequent runs
calculate the modified hydrodynamic shock structure, based upon the
accelerated particle distributions of the previous iterations; this
non-linear acceleration method is not employed here. Also, since the
choice of the scattering mechanism can affect both injection and
acceleration of particles, it can strongly impact non-linear
modifications for relativistic shocks.  The influence of different
scattering scenarios in such non-linear acceleration systems will be the
subject of future work.

This implementation also does not retain accounting of the amount of
time the particle would have spent downstream of this ``return'' plane. 
In the event that acceleration time information is needed, a
retrodictive approach described first in Jones (1978) and later applied
directly to a Monte-Carlo simulation in Ellison et al. (1990a) can be
used.  One important finding is that the interplay between energy
boosting and time dilation effects leads only to modest changes (Baring
2002) in the acceleration time at plane-parallel relativistic shocks
compared with standard non-relativistic shock formalism (Forman, Jokipii
\& Owens 1974).  The consideration of particle acceleration times is
beyond the scope of the present work, and will be deferred to a future
investigation.

\section{Magnetohydrodynamic Jump Conditions For Oblique 
   Relativistic Shocks}
 \label{sec:RHrel}
In the case of relativistic shocks, the shock jump conditions are
considerably more difficult to solve than the non-relativistic solutions
presented in Decker (1988), due to the impact of length contraction and
time dilation effects on the structure of the six conservation
equations. There are different approaches to solving the
Rankine-Hugoniot jump conditions in relativistic MHD discontinuities,
surveyed in Double et al. (2004;  see also Gerbig \& Schlickeiser 2011,
for a recent exposition). Our approach here builds upon previous work by
Ballard \& Heavens (1991) that formulates the Rankine-Hugoniot
conditions in the de Hoffmann-Teller frame (de Hoffmann and Teller,
1950; hereafter HT) in a manageable form.  The HT frame is a shock rest
frame in which there are no {\bf u}\teq{\times}{\bf B} drift electric
fields. This can be obtained from the local fluid frame by boosting
along {\bf B}, but can also be obtained as a combination of two boosts
along the axes of the coordinate system to avoid a rotation of the
coordinate system. The system of equations is then transformed from the
HT frame into the normal incidence frame (NIF, in which the upstream
plasma flow is parallel to the shock normal or $\hat x$ direction),
arriving at a system of three comparatively simple simultaneous
equations in which the terms that become imaginary in super-luminal
shocks are no longer present.  These three equations are solved
numerically after the J\"uttner-Synge equation of state is invoked to
connect key thermodynamic quantities such as pressure and enthalpy, to
the temperatures of the upstream and downstream relativistic
Maxwell-Boltzmann distributions.  This method encompasses a broad range
of shock conditions, specifically ranges of sonic and Alfv\'enic Mach
numbers, and transitions seamlessly from subluminal to superluminal
regimes.  Our results are compared directly with that of the work by
Double et al. (2004), highlighting similarities, and also differences
that result from a specific approximation to the downstream equation of
state employed in that work.

Before embarking upon the construction and reduction of the jump
conditions, a brief summary of the subscript conventions adopted here
for the different frames of reference is given.  The `f' subscript will
denote a quantity defined in the rest frame of the upstream (subscript
1) or downstream (subscript 2) plasma. HT frame variables will be
subscripted with an `HT.' To distinguish NIF frame quantities from those
measured in the fluid or HT frames, they will denoted by an `S'
subscript for the shock frame. Additionally, \teq{\Theta_{\hbox{\sixrm
B}}} will always refer to an angle the magnetic field \teq{\vect{B}}
makes with the shock normal, and \teq{\theta_{\hbox{\sixrm u}}} will
refer to the angle a plasma flow makes with the shock normal.  When the
HT frame is found via a single boost along the direction of the magnetic
field, the field components are identical in the local fluid and HT
frames, often the `f' and `HT' subscripts will be explicitly omitted for
compactness of notation, i.e., \teq{B_1\equiv B_{1\hbox{\sixrm f}}\equiv
\BHTone}, etc.

\begin{figure*}[t]
\twofigureoutpdf{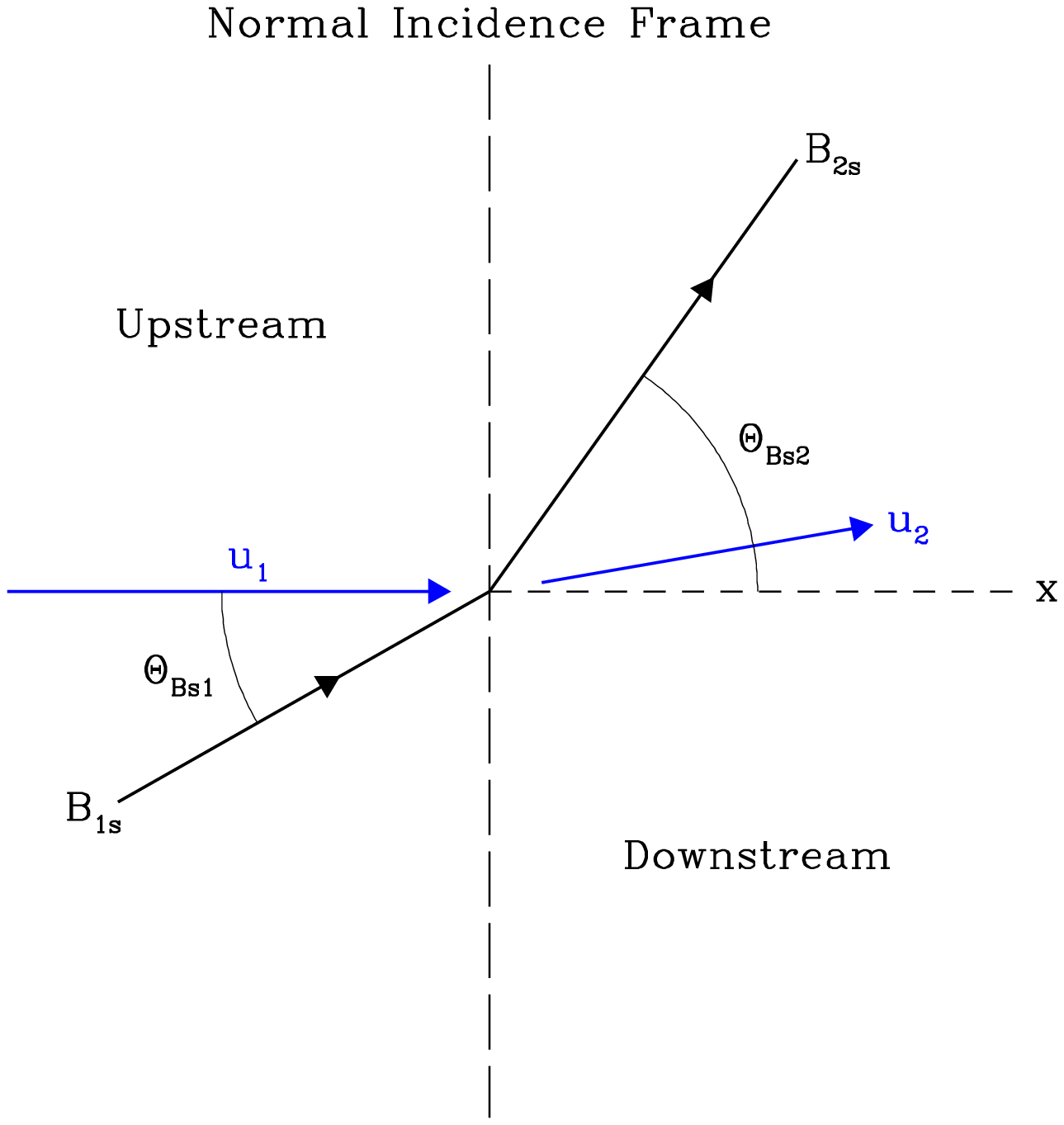}{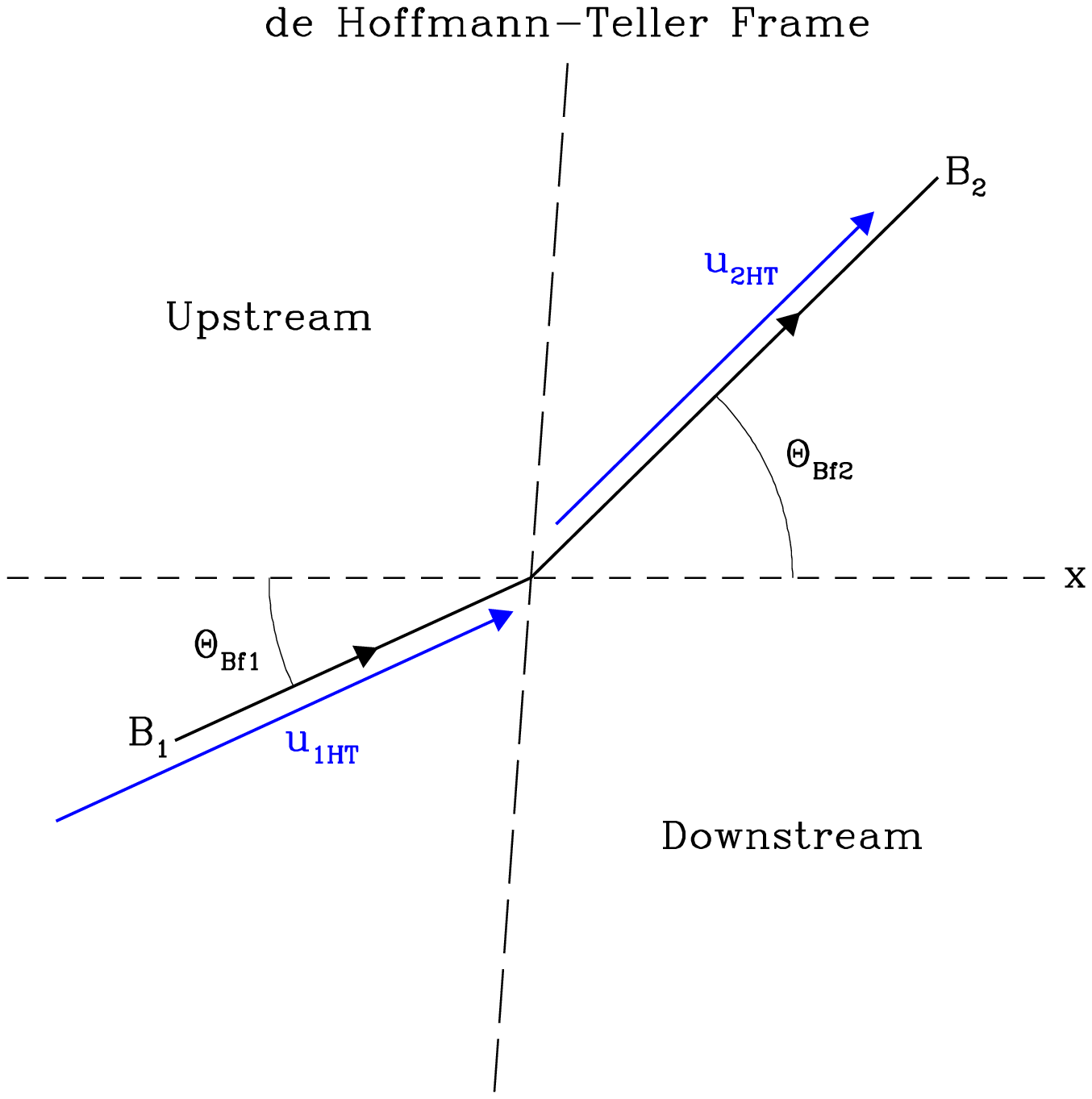}{
The geometry in the normal incidence (NIF; left panel) and 
de-Hoffmann Teller (HT; right panel) shock rest frames.  Upstream 
flow speeds in the two reference frames are related by
\teq{\uoneHT\equiv \betaoneHT c= u_1/\cos\ThetaBfone}.
Upstream and downstream quantities are denoted by subscripts 
\teq{1} and \teq{2}, respectively.  In general, the NIF field angle 
\teq{\ThetaBsone} differs from the fluid frame/HT frame value 
\teq{\ThetaBfone\equiv\ThetaBHTone}, and likewise for the downstream angles.  
Also, in cases where the HT frame is obtained by boosting along the field 
fluid frame direction, the shock plane in the HT frame is rotated from that in the 
NIF due to relativistic aberration effects. For the two-step fluid-to-HT 
frame transformation protocol adopted here, the shock planes in the 
NIF and HT frames are coincident, i.e. no rotation is involved.
 \label{fig:shock_geometry} }
\end{figure*}

The character of the solutions to this system of equations is controlled
by two key parameters, basically the relative scaling of the upstream
thermal energy or pressure \teq{P_1}, and the fluid frame magnetic field 
energy density \teq{B_{1\hbox{\sixrm f}}^2/(8\pi )} to the upstream ram 
pressure \teq{\rho_1u_{1x}^2}.  Here, \teq{u_{1x}=\betaonexS c} is the 
velocity component of the upstream fluid normal to the shock, in the NIF.  
Accordingly, we define these via the sonic ($\machson$) and 
Alfv\'enic ($\machalf$) Mach numbers:
\begin{equation}
   \machson^2 \; =\; \dover{\rho_1 u_{1x}^2}{\gamma_{g1}P_1}\quad ,\quad
   \machalf^2 \; =\; \dover{4\pi\rho_1 u_{1x}^2 }{B_1^2}\quad .
 \label{eq:son_alf_mach}
\end{equation}
These are conventional definitions for non-relativistic shocks, and
their extension to oblique discontinuities and relativistic systems does
not lead to unique choices.  For example, subjectivity is involved in
deciding between \teq{u_{1x}} and \teq{u_1}, and similarly for
\teq{B_{1x\hbox{\sixrm f}}} versus \teq{B_{1\hbox{\sixrm f}}}.  Hence, we
adopt the above definitions (as did Double et al. 2004), for which
\teq{\gamma_{g1}} is the upstream adiabatic gas index, discussed further in
Sec.~\ref{sec:EOS}, so that \teq{\gamma_{g1} P_1/\rho_1} is the square of
the upstream sound speed.

\subsection{The de Hoffmann-Teller Frame Solution}
 \label{sec:HTsoln}
For subluminal flows, where \teq{\beta_{1x}/\cos \ThetaBfone <1}, the HT
frame is an obvious choice in which the shock jump conditions can be
written, since therein the jump conditions reduce to a simple form
because the fluid flows along the magnetic field lines and there is no
\teq{\vect{u} \times \vect{B}} electric field. For the time being, we
will restrict considerations to these types of shocks and later
trivially generalize the results to include superluminal shocks. Four of
the shock jump equations are defined by the conservation of the mass,
momentum (2 components), and energy fluxes across the shock interface
are conserved. The remaining two derive from the electromagnetic field
constraints \teq{{\bf \nabla \cdot} \vect{B} =0} and \teq{{\bf \nabla
\times} \vect{E} = \vect{0}}, the latter of which is trivial in the HT
frame, because \teq{\vect{E}=\vect{0}} identically everywhere.

The form of these jump conditions in the HT frame has been derived
previously by Ballard and Heavens (1991). Those equations are reproduced
below with the notable exception that the subscript `y' used in their
paper has been replaced with the subscript `z' to avoid confusion when
comparing to other works. Here, the \teq{x}-direction defines the normal
to the shock plane in the HT frame, the magnetic field everywhere lies
in the \teq{x}-\teq{z} plane, and the \teq{y}-axis defines the direction
of \teq{\vect{u} \times \vect{B}} drift velocities.  All quantities save
\teq{\Gamma}, \teq{\vect{\beta}}, and \teq{\vect{B}} are defined in the
frame where the plasma is stationary hereafter referred to as the
``local fluid frame" or ``upstream/downstream rest frame".  For the
present, those 3 quantities are defined in the HT frame.  Setting
\teq{c=1}, as is done throughout this paper, conservation of mass or
particle number flux along the shock normal gives
\begin{equation}
   \Gamma_1\beta_{1x} \,\rho_1 \; =\; \Gamma_2 \beta_{2x} \,\rho_2
 \label{eq:mass_cons_HT}
\end{equation}
where \teq{\rho_i} denotes mass density, and subscripts 1 and 2 
denote upstream and downstream quantities (labelled by \teq{i} 
in general). Throughout this subsection, HT subscripts will be omitted, 
but implied. Also, \teq{\beta} is the flow speed written as a fraction 
of the speed of light, and \teq{\Gamma =1/\sqrt{1-\beta_x^2-\beta_z^2}}
is the Lorentz factor associated with the flow speed \teq{\beta}. 
Conservation of the \teq{x} and \teq{z} components of momentum flux 
gives, respectively,
\begin{eqnarray}
   \Gamma_1^2 \beta_{1x}^{2}\, w_1 + P_1 + \dover{B_{1z}^2}{8\pi} 
      &=&  \Gamma_2^2 \beta_{2x}^{2}\, w_2 + P_2 + \dover{B_{2z}^2}{8\pi} \nonumber\\[-5.5pt]
 \label{eq:mom_cons_HT}\\[-5.5pt]
   \Gamma_1^2 \beta_{1x} \beta_{1z}\, w_1 - \dover{B_{1x} B_{1z}}{4\pi} 
      &=& \Gamma_2^2 \beta_{2x} \beta_{2z}\, w_2 - \dover{B_{2x} B_{2z}}{4\pi} \nonumber
\end{eqnarray}
This corrects an obvious typographical error in Eq.~(26) of Ballard and
Heavens (1991) in their terms involving the enthalpies
\teq{w_i=e_i+P_i}.  The internal energy \teq{e_i}, which includes the
rest mass energy density, can be related to \teq{P_i} and \teq{\rho_i}
through an equation of state, as is addressed in Sec.~\ref{sec:EOS}
below. In the HT frame, conservation of energy flux is simply
\begin{equation}
   \Gamma_1^2 \beta_{1x}\, w_1 \; =\; \Gamma_2^2 \beta_{2x}\, w_2 \quad .
 \label{eq:erg_cons_HT}
\end{equation} 
Here, the magnetic field contributions to the stress-energy tensor 
(see, for example, Eq. (21) of Double et al. 2004) cancel to zero
precisely because of the pair of equations
\begin{equation}
   \dover{\beta_{1z}}{\beta_{1x}}\; =\; 
                   \dover{B_{1z}}{B_{1x}}\;\equiv\;\tan\ThetaBHTone  \quad ,\quad
   \dover{\beta_{2z}}{\beta_{2x}}\; =\; 
                   \dover{B_{2z}}{B_{2x}}\;\equiv\;\tan\ThetaBHTtwo\quad ,
 \label{eq:HT_def}
\end{equation}
that define the specific choice of the HT frame.  The absence of such
magnetic terms in the energy flux, combined with the compact nature of
the momentum flux conditions, underline the attractive simplicity of
adopting the HT frame (compare, for example, with the electromagnetic
stress tensor contributions to the momentum fluxes in Eqs.~(25) and~(26) 
of Double et al. 2004). The trivial \teq{{\bf \nabla \times} \vect{E} = \vect{0}} 
can be eliminated, effectively being replaced by the HT frame definitions 
in Equation~(\ref{eq:HT_def}). Finally, the Maxwell equation 
\teq{{\bf \nabla \cdot} \vect{B} = \vect{0}} defining the absence 
of magnetic monopoles gives
\begin{equation}
   B_{1x}\; =\; B_{2x}\quad ,
 \label{eq:Bx_cons}
\end{equation}
unaltered by relativistic generalization because it is intrinsically
relativistic.

Following Ballard and Heavens, Eqs.~(\ref{eq:HT_def})
and~(\ref{eq:Bx_cons}) can be used to eliminate the z-components
\teq{B_{1z}} and \teq{B_{2z}}, and the downstream x-component
\teq{B_{2x}}.  Their solutions were defined in terms of two upstream
parameters \teq{A = \Gamma_1w_1/\rho_1} and \teq{C = B_{1x}^2/[
4\pi\rho_{1}\gamma_{g1} \beta_{1x} ]}.  Here, as an alternative listing, we
observe that the ratio \teq{C/A} appears repeatedly in the resulting
subset of processed equations, so we define 
this ratio via
\begin{equation}
   \psi \; =\; \dover{B_{1x}^2}{4\pi \Gamma_1^2\beta_{1x} w_1}
   \;\equiv\; \dover{\cos^2\ThetaBHTone}{\machalf^2}
         \, \dover{\beta_{1x}\rho_1}{\Gamma_1^2w_1}\quad ,
 \label{eq:psi_def}
\end{equation}
which, as a relativistically-modified ratio of magnetic to thermal (plus
rest mass) energy density, is essentially an adaptation of the inverse
of the upstream plasma {\it beta parameter} \teq{\beta_{\hbox{\sixrm P}}
= 8\pi P_1/B_1^2} to oblique, relativistic MHD flows. The second
prescription for \teq{\psi} uses the Alfv\'enic Mach number definition
in Eq.~(\ref{eq:son_alf_mach}), together with identity of  {\it total}
magnetic fields in the fluid and HT frames, i.e. \teq{\BHTone =\Bfone}.
The energy flux equation is most easily manipulated, dividing
Eq.~(\ref{eq:erg_cons_HT}) by the mass conservation in
Eq.~(\ref{eq:mass_cons_HT}):
\begin{equation}
   \dover{\Gamma_1w_1}{\rho_1} \; =\; \dover{\Gamma_2w_2}{\rho_2}
   \;\equiv\; \dover{w_2}{\rho_2\sqrt{1-\beta_{2x}^2-\beta_{2z}^2}}\quad .
 \label{eq:erg_scaled_HT}
\end{equation}
This is just the constant \teq{A} employed by Ballard \& Heavens (1991).
Next, dividing the z-component of momentum conservation in 
Eq.~(\ref{eq:mom_cons_HT}) by Eq.~(\ref{eq:erg_cons_HT}) solves 
for \teq{\beta_{2z}}:
\begin{equation}
   \beta_{2z}\; =\; \beta_{2x}\, \dover{\beta_{1z}}{\beta_{1x}}
      \left( \dover{\beta_{1x} - \psi }{\beta_{2x} - \psi}\right)\quad .
 \label{eq:u2z_HT}
\end{equation}
This can be inserted into Eq.~(\ref{eq:erg_scaled_HT}), eliminating 
\teq{\beta_{2z}}.  Observe that viable jump conditions are only realizable 
when \teq{\psi < \beta_{2x}}.  This is equivalent to requiring that the 
total Mach number be greater than unity. Finally, the  x-component of 
momentum conservation in Eq.~(\ref{eq:mom_cons_HT}) can be 
divided by Eq.~(\ref{eq:erg_cons_HT}), producing
\begin{eqnarray}
   &&\beta_{1x} + \dover{P_{1}}{\Gamma_1^2 \beta_{1x}w_1} 
       + \dover{\psi}{2}  \left( \dover{\beta_{1z}}{\beta_{1x}}\right)^2\nonumber\\[-5.5pt]
 \label{eq:px_HT}\\[-5.5pt]
   & =&\beta_{2x} + \dover{P_{2}}{\Gamma_1^2 \beta_{1x}w_1}
      + \dover{\psi}{2}  \left( \dover{\beta_{1z}}{\beta_{1x}}\right)^2
         \left( \dover{\beta_{1x} - \psi }{\beta_{2x} - \psi}\right)^2\;\; .\nonumber
\end{eqnarray}
Here, expressing the ratio \teq{\beta_{1z}/\beta_{1x} =
\tan\ThetaBHTone} in terms of the de Hoffmann-Teller field angle
\teq{\ThetaBHTone}, a constant for the shock structure, yields an
alternative algebraic form.  Observe also that the second term on the
right hand side is proportional to \teq{P_2/w_2} multiplied by
\teq{w_2/w_1}; the second factor can be expressed using
Eq.~(\ref{eq:erg_scaled_HT}), and the first is a function of the
downstream temperature \teq{T_2} through the equation of state, to be
defined in Sec.~\ref{sec:EOS}.

Eq.~(\ref{eq:erg_scaled_HT}), with Eq.~(\ref{eq:u2z_HT}) inserted,
and~(\ref{eq:px_HT}) constitute a system of two simultaneous equations
with unknowns \teq{P_2}, \teq{w_2} and \teq{\beta_{2x}}. However,
\teq{w_2} will be related in section \ref{sec:EOS} to \teq{P_2} via an
equation of state, rendering the system numerically solvable. This set
of equations is simple and elegant, being virtually as compact as the
system for jump conditions at relativistic, plane-parallel, hydrodynamic
shocks (e.g. Blandford \& McKee 1976).  However, their validity is
technically restricted to subluminal regimes where the HT frame formally
exists. Therefore, to realize broader applicability, it is necessary to
transform them to the normal incidence shock rest frame, and thereafter
explore their numerical solution.

\vspace{20pt}

\subsection{Transforming to the Normal Incidence Shock Frame}
 \label{sec:trans->NIF}
In subluminal cases where the HT frame exists, one can define a boost
velocity \teq{\beta_t} in the \teq{\hat{z}} direction that transforms
from the normal incidence frame into the HT frame. The two key input
quantities in this regard are \teq{\betaonexS}, the shock speed in the
upstream fluid frame, and \teq{\ThetaBfone}, the angle between the shock
velocity and the magnetic field vector in the upstream fluid frame.  A
third parameter that is derivative of these two is the HT frame field
angle \teq{\ThetaBHTone}.  As discussed by Kirk \& Heavens (1989), there
is a lack of uniqueness in defining field and flow angles in the de
Hoffmann-Teller frame, up to rotations.  Here, we will adopt the
following sequence of boosts to effect Lorentz transformation to the HT
frame from the local fluid frame: this will be performed by first
boosting by \teq{\betaxS \hat{x}} along the shock normal to the NIF, and
then boosting by \teq{\beta_t \hat{z}} in the shock plane to arrive at
the HT frame.  The planes of the shock in both the NIF and HT frames are
thereby coincident.  This yields a convenient definition of
\teq{\ThetaBHTone} (and \teq{\ThetaBHTtwo}), and is the preferred
protocol for our simulation due to the enhanced simplicity it permits
for modeling particle convection and gyration in the shock layer.
However, it should be emphasized that a single boost along the field
vector \teq{\vect{B}} from the fluid to HT frames yields an aberration
of the shock plane: it is rotated relative the NIF shock plane, as
described in Ballard \& Heavens (1991), and is illustrated in
Fig.~\ref{fig:shock_geometry}.  Such a rotation leads to a need to
account for it when defining field and flow angles with respect to the
shock plane, an unnecessary complication.  The two-step fluid-to-HT
frame transformation approach adopted here was also the preference of
Kirk \& Heavens (1989).

The flow velocities in the NIF and HT frames of reference are related via 
standard Lorentz transformations
\begin{eqnarray}
   \betaxHT &=& \dover{\betaxS}{\Gamma_t (1+\beta_t \betazS)}\quad , \nonumber\\[-5.5pt]
 \label{eqn:uxuzHT}\\[-5.5pt]
   \betazHT &=& \dover{\betazS +\beta_t}{(1+ \beta_t \betazS )}\quad .\nonumber
\end{eqnarray}
These relations can be applied to both the 
upstream and downstream sides of the shock; their subscripts 1,2
have been suppressed here for the sake of compactness.
In the upstream region, where the definition of the NIF requires 
that \teq{\betazS=0}, these equations distill down to
\teq{\betaxHT = \betaxS/\Gamma_t} and \teq{\betazHT = \beta_t},
respectively.  Subsequently, taking the ratio of these two upstream equations,
one can express the boost speed 
\teq{\beta_t} and Lorentz factor \teq{\Gamma_t =(1-\beta_t^2)^{-1/2}} in 
terms of \teq{\betaonexS} and \teq{\ThetaBHTone}:
\begin{eqnarray}
   \Gamma_t\beta_t & =& \betaonexS \tan\ThetaBHTone\quad .\nonumber\\[-5.5pt]
 \label{eq:HT->NIFboost2}\\[-5.5pt]
   \beta_t & =& \dover{\betaonexS \tan\ThetaBHTone}{
      \sqrt{1 + \betaonexS^2 \tan^2\ThetaBHTone}}\quad , \nonumber
\end{eqnarray}
Since, for flux conserving jump conditions in MHD discontinuities, the 
HT frame is identical for both upstream and downstream locations,
it can be inferred that the identities in Eq.~(\ref{eq:HT->NIFboost2}) can be
written also in terms of downstream quantities, merely via the substitutions 
\teq{\betaonexS\to\betatwoxS} and \teq{\ThetaBHTone\to\ThetaBHTtwo}.

The relationship between the magnetic field components in the two 
frames of reference is similarly routinely derived via standard transformation 
equations for electromagnetic fields:
\begin{equation}
   \BxHT \; =\; \dover{\BxS}{\Gamma_t}\quad ,\quad \BzHT \; =\; \BzS\quad ,
 \label{eq:BxBzHT}
\end{equation}
noting that the equation for \teq{\BxHT} is only one part of the full 
transformation equations, which also transform the NIF drift electric field 
in the \teq{{\hat y}} direction exactly to zero in the HT frame.  The ratios 
of the equations in Eq.~(\ref{eq:BxBzHT}) then simply yield 
\begin{equation}
   \tan\ThetaBHTone \; =\; \Gamma_t \tan\ThetaBsone\quad ,\quad
   \tan\ThetaBHTtwo \; =\; \Gamma_t \tan\ThetaBstwo
 \label{eq:ThetaBHT}
\end{equation}
for the upstream and downstream HT frame field angles to 
the shock normal.  These are recognizable as aberration formulae for the 
electromagnetic/photon field, with the NIF frame field obliquity always 
being less than that in the de Hoffmann-Teller frame.  Combining this 
result with the Eq.~(\ref{eq:HT->NIFboost2}) yields the relationship
\begin{equation}
   \beta_t \; =\; \betaonexS \tan\ThetaBsone\quad
 \label{eq:HT->NIFboost}
\end{equation}
removing a reference to the HT frame from Eq.~(\ref{eqn:uxuzHT}).
The subluminal condition for the existence of the HT frame, written in terms 
of NIF quantities, is then \teq{\betaonexS \tan\ThetaBsone < 1}.

To close this sequence of boost algebra, one needs the relation between 
field angles in the fluid frames and the NIF.  These are derived in the same 
manner as Eq.~(\ref{eq:ThetaBHT}), yielding
\begin{equation}
   \tan\ThetaBsone \; =\; \gammaoneS \tan\ThetaBfone\;\; ,\quad 
   \tan\ThetaBstwo \; =\; \gammatwoS \tan\ThetaBftwo \quad ,
 \label{eq:ThetaBNIF}
\end{equation}
where \teq{\gammaoneS = 1/\sqrt{1-\betaonexS^2}}, and 
\teq{\gammatwoS = 1/\sqrt{1-\betatwoxS^2-\betatwozS^2}\,}.
As a result, Eq.~(\ref{eq:HT->NIFboost}) can be rewritten using 
\teq{\tan\ThetaBHTone = \gammaoneS\Gamma_t\tan\ThetaBfone} to
yield a boost speed expressed entirely in terms of input quantities:
\begin{equation}
   \beta_t \; =\; \gammaoneS\betaonexS\,\tan\ThetaBfone\quad .
 \label{eq:beta_t}
\end{equation}
This then routinely rearranges so that the subluminal \teq{\beta_t < 1}
condition becomes the familiar \teq{\betaonexS/\cos\ThetaBfone < 1}.

By replacing the HT frame quantities in Eqs.~(\ref{eq:erg_scaled_HT}),
(\ref{eq:u2z_HT}) and (\ref{eq:px_HT}) with their shock frame
equivalents via Eqs. (\ref{eqn:uxuzHT}), (\ref{eq:ThetaBHT}), and
(\ref{eq:HT->NIFboost}), the Rankine-Hugoniot relations then become a
system of 3 equations with unknowns \teq{P_2}, \teq{w_2},
\teq{\betatwoxS} and \teq{\betatwozS} that possesses a non-singular
mathematical character in the NIF frame at the luminal interface
\teq{\betaonexS/\cos\ThetaBfone = 1}. The system now retains only
information about upstream and downstream fluid frame field angles and
thermodynamic quantities, and the transformation velocities required to
get to the NIF frame from the fluid frames. It must be emphasized that
an attractive characteristic of this methodology is that significant
cancellations remove any terms that become imaginary or unphysical in a
superluminal shock, revealing a smooth mathematical transition of
solutions from subluminal to superluminal regimes. The specification of
an equation of state that relates \teq{P_2} to \teq{w_2} closes this
system, rendering it amenable to numerical solution.

\subsection{The Equation of State}
 \label{sec:EOS}
Assuming there are no shear stresses and axial symmetry about the
magnetic field, the pressure tensor is diagonal. One can then form an
isotropic pressure \teq{P= (P_{\parallel}+2P_{\perp})/3}, where
\teq{P_{\parallel}} and \teq{P_{\perp}} denote the pressure components
respectively parallel to and perpendicular to the mean magnetic field. 
Then, the ``adiabatic'' gas index \teq{\gamma_g}, the approximate ratio
of specific heats, can parametrize the equation of state via the
adiabatic expansion law for an ideal gas:
\begin{equation}
   PV^{\gamma_g} \; =\; \hbox{constant}.  
 \label{eq:adiabat}
\end{equation}
Here \teq{\gamma_g} ranges between \teq{5/3} for a non-relativistic, 
compressible gas, to \teq{4/3} for an ultra-relativistic gas. With the 
specification of this index, the internal (thermal) energy density 
\teq{P/(\gamma_g -1)} is related to the pressure, so that the total internal 
energy density plus the rest mass energy density can be written
\begin{equation}
   e \; =\; \dover{P}{\gamma_g-1} +\rho \quad ,
 \label{eq:e_adiabatic}
\end{equation}
where, again, c=1 has been assumed, as will be done throughout the rest
of this work.  Reintroducing the subscripts \teq{i=1,2} to label upstream and 
downstream fluid frames, this leads to the forms for the enthalpies 
that are deployed in the Rankine-Hugoniot relations:
\begin{equation}
   w_i \; =\; e_i+P_i \; =\; \dover{\gamma_{gi} P_i}{\gamma_{gi} - 1} + \rho_i\quad .
 \label{eq:enthalpy_adiabatic}
\end{equation}
The particular values of the \teq{\gamma_{gi}} can be expressed as a 
moment of the fluid frame particle momentum distributions upstream 
and downstream, and so can apply to both thermal and non-thermal 
populations.  While they are simply prescribed in the non-relativistic 
and ultra-relativistic asymptotic cases, a more precise formulation is 
required to treat the mildly-relativistic domain.

Here it is assumed that the background plasma possesses a relativistic thermal
Maxwell-Boltzmann distribution that defines the J\"uttner-Synge equation
of state (e.g. Synge 1957).  Then the temperature \teq{T} can be the
sole thermodynamic parameter, and all other thermodynamic quantities can
be prescribed in terms of it.  The equation of state depends on the
number of species, their masses, and the state of thermal equilibrium
between the various species, i.e the temperature equilibration or
otherwise.  For simplicity, here a single component plasma is adopted,
appropriate for an electron-positron pair plasma as might be encountered
in relativistic jets in extragalactic sources such as gamma-ray bursts
or blazars.  Equations of states for electron-ion and other mixed
species gases are addressed in Ballard \& Heavens (1991).  For a pair
plasma, the enthalpy can be written in terms of modified Bessel
functions:
\begin{equation}
   \dover{w_i}{\rho_i} \; =\; R(\tau_i) + \tau_i\quad ,\quad
   R(\tau) \; =\; 3\tau + \dover{K_1(1/\tau)}{K_2(1/\tau)}\quad ,
 \label{eq:wi_synge}
\end{equation}
where the \teq{K_{i}}s are modified Bessel functions of the second 
kind (e.g., see pp.~708-715 of Arfken and Weber, 2001), and
\begin{equation}
   \tau_i\; =\; \dover{kT_i}{m_e} \; =\; \dover{P_i}{\rho_i}\;\; ,\quad i=1,2\;\; ,
\end{equation}
is the dimensionless pair temperature (in units of \teq{c=1}).  This
equation of state can treat arbitrary sonic Mach numbers, in distinct
contrast to the approximation employed by Double et al. (2004),
discussed below, that uses kinematics in high \teq{\machson} shocks to
specify the downstream pressure.

\begin{figure*}[t]
\centerline{\includegraphics[width=6.5in]{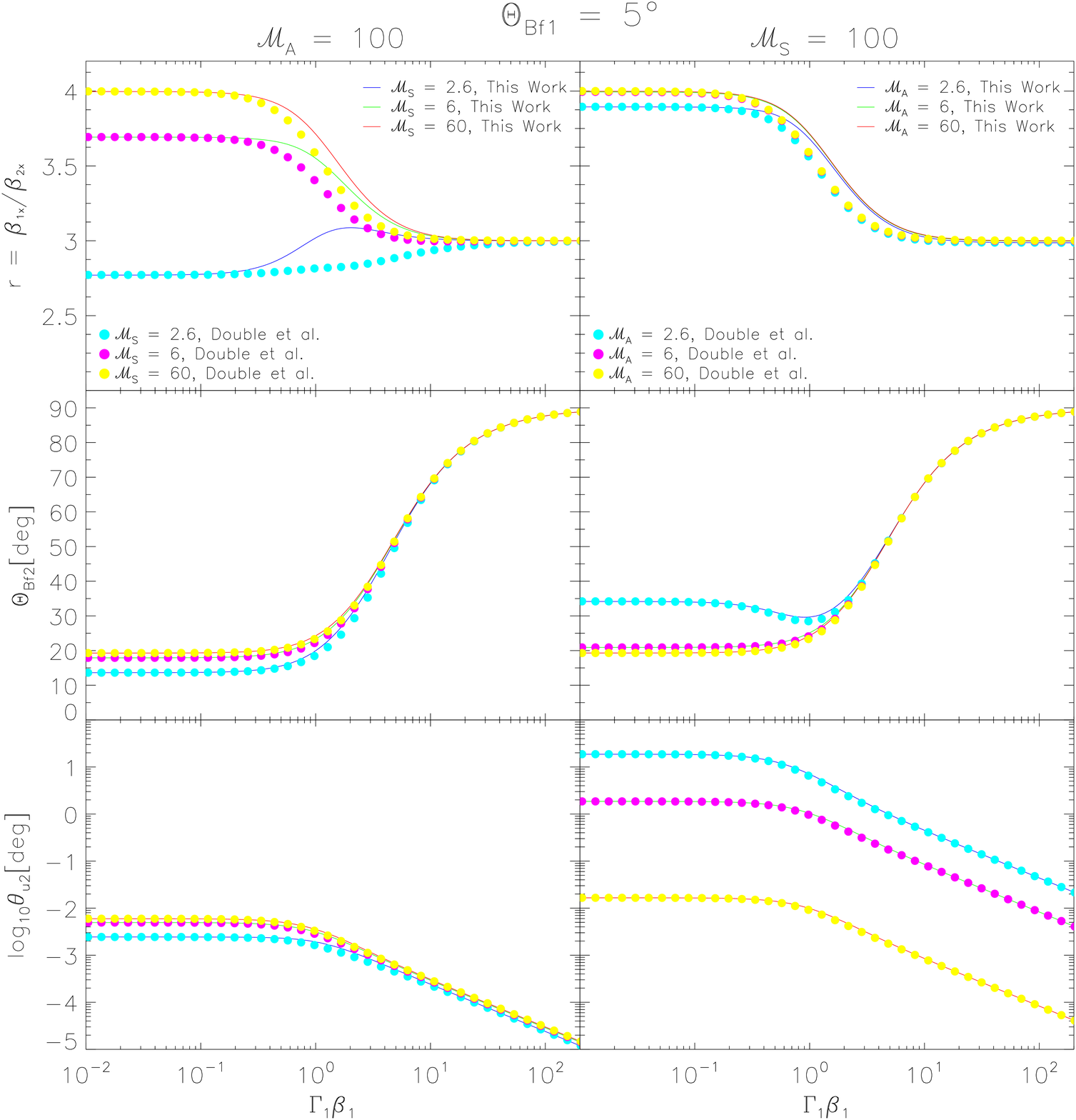}}
\caption{\small Rankine-Hugoniot relations for \teq{r}, the
compression ratio, \teq{\ThetaBftwo}, the angle the downstream fluid
frame magnetic field makes with the shock normal, and \teq{\thetautwo},
the angle the downstream flow makes with the shock normal, as functions
of the upstream NIF rapidity of the shock, \teq{\Gamma_1\beta_1\equiv\gammaoneS\betaonexS}.
Solutions are displayed for various values of the Alfv\'{e}nic
($\machalf$) and sonic ($\machson$) Mach numbers, with the angle the
upstream magnetic field makes with the shock normal, \teq{\ThetaBfone},
set to \teq{5^{\circ}}. Solid lines are new results from this work using
the J\"uttner-Synge (J-S) equation of state and Pade' approximation
described by Eq.~(\ref{eq:R_approx_Pade}). Dotted curves represent 
results from the previous work of Double et al. (2004).
 \label{fig:RH_ThetaB=5deg} }
\end{figure*}

A modest disadvantage of the J\"uttner-Synge equation of state lies in
the complexity of the Bessel function; it is not conducive to either
analytical or numerical solutions of Eqs.~(\ref{eq:erg_scaled_HT}),
(\ref{eq:u2z_HT}) and (\ref{eq:px_HT}). However, noting the asymptotic
behavior of Eq.~(\ref{eq:wi_synge}), namely \teq{R(\tau) \rightarrow 
3\tau} as \teq{\tau\rightarrow \infty} , and \teq{R(\tau) \rightarrow 
1+3\tau/2} as \teq{\tau \rightarrow 0}, a remarkably good approximation
for the function \teq{R(\tau)} is given by a Pad\'e approximation of
3rd-order:
\begin{equation}
   R(\tau) \;\approx\; \dover{2+7\tau+12\tau^{2}+6\tau^{3}}{2+4\tau+2\tau^{2}}\quad .
 \label{eq:R_approx_Pade}
\end{equation}
This approximation is accurate to 0.25\% over the entire domain and 
is slightly less algebraically complicated than the approximation in 
Eq.~(38) of Double et al. (2004).  By inverting
Eq.~(\ref{eq:enthalpy_adiabatic}) to obtain \teq{\gamma_{gi}}, and using 
Eqs.~(\ref{eq:wi_synge}) and~(\ref{eq:R_approx_Pade}), one can find 
\teq{\gamma_{gi}} as a function of \teq{\tau_i}:
\begin{equation}
\label{eqn:gammaofT}
   \gamma_{gi} \; =\; 1 + \dover{\tau_i}{R(\tau_i)-1}
      \;\approx\; \dover{5+14\tau_i+8\tau_i^2}{3+10\tau_i+6\tau_i^2}\quad .
\end{equation}
Inserting the approximation from Eq.~(\ref{eq:R_approx_Pade}) in 
Eq.~(\ref{eq:wi_synge}) provides \teq{w_i/\rho_i} as a function of 
only \teq{\tau_i}, eliminating the fourth unknown quantity in 
Eqs.~(\ref{eq:erg_scaled_HT}), (\ref{eq:u2z_HT}) and (\ref{eq:px_HT});
this is the protocol adopted for the numerical solution of the 
Rankine-Hugoniot relations.

\subsection{Numerical Solutions for the Jump Conditions}
 \label{sec:jumpcond}

\begin{figure*}[t]
\centerline{\includegraphics[width=6.5in]{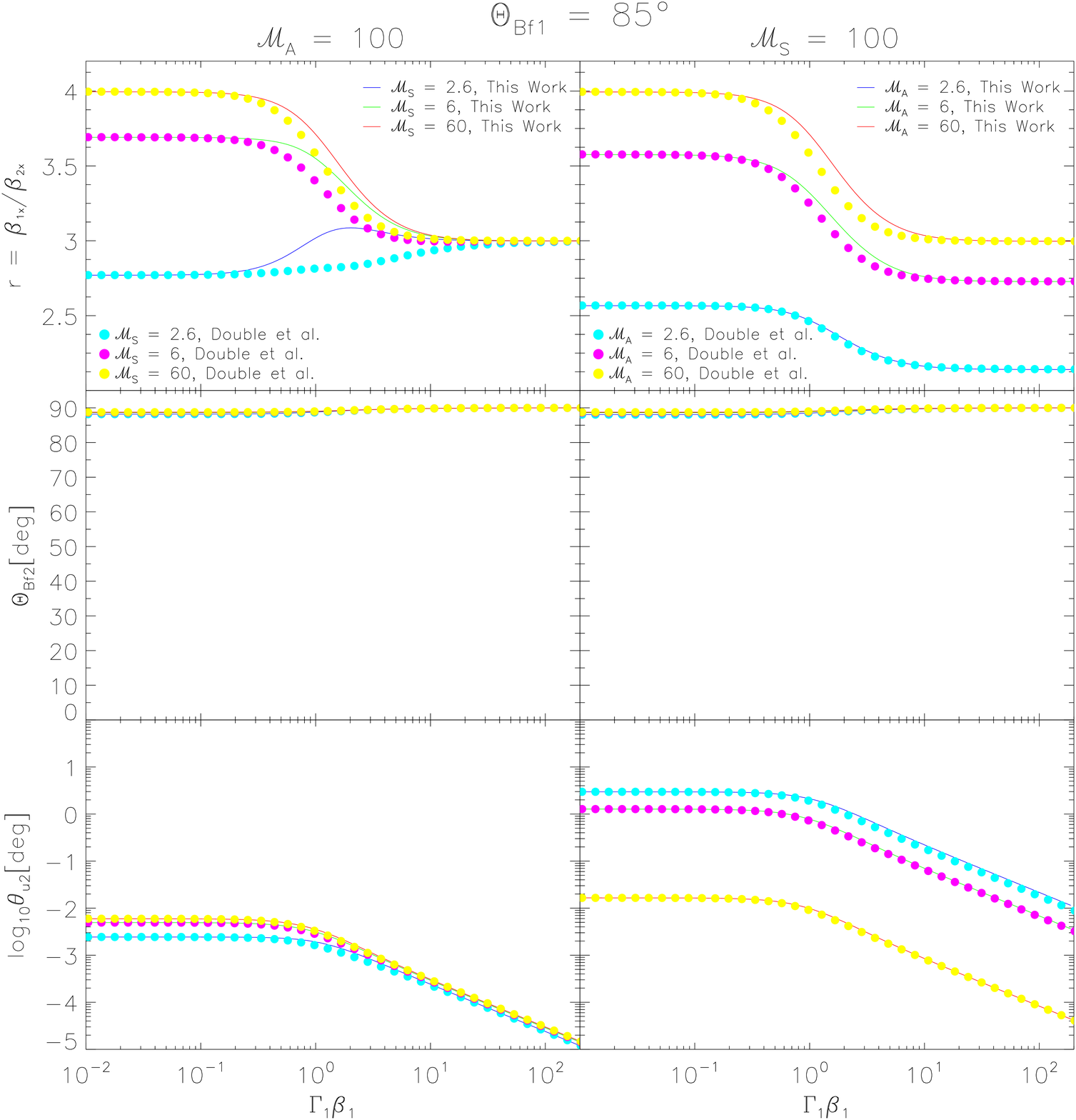}}
\caption{\small Solutions to the Rankine-Hugoniot relations for \teq{r},
\teq{\ThetaBftwo}, and \teq{\thetautwo}, as in
Figure~\ref{fig:RH_ThetaB=5deg}.  However, here, the upstream field
obliquity is \teq{\ThetaBfone = 85^{\circ}}, constituting
quasi-perpendicular shocks.  Values of the Alfv\'{e}nic ($\machalf$) and
sonic ($\machson$) Mach numbers are as labelled.  Again, solid lines are
new results from this work using the J\"uttner-Synge equation of state,
while the dotted curves represent the solutions of Double et al. (2004).
 \label{fig:RH_ThetaB=85deg} }
\end{figure*}

The resulting three equations were solved numerically using Mathematica.
For the case of a plane-parallel shock, we compared directly with
solutions displayed in Fig. 1b of Heavens \& Drury (1988) where
downstream flow speeds are found as a function of upstream flow speeds
for parallel (\teq{\ThetaBfone = 0^{\circ}}) electron-positron shocks at
various temperatures using the J\"uttner-Synge equation of state just as
we do. We find no observable differences between our results and theirs
for plane-parallel shocks. For the case of oblique shocks,
representative solutions, as functions of the upstream fluid rapidity
\teq{\Gamma_1\beta_1\equiv\gammaoneS\betaonexS}, are displayed in
Figs.~\ref{fig:RH_ThetaB=5deg} (quasi-parallel case)
and~\ref{fig:RH_ThetaB=85deg} (quasi-perpendicular shock).  The plots
exhibit the velocity compression ratio \teq{r=\betaonexS/\betatwoxS},
and the downstream fluid frame field and fluid NIF velocity angles to
the shock normal.  Observe that hereafter, the subscript ``S'' will be
omitted when referring to NIF values for \teq{\beta_i} and
\teq{\Gamma_i}.  For the sake of comparison, these Figures also display
equivalent plots for the same sonic and Alfv\'enic Mach numbers [as
defined in Eq.~(\ref{eq:son_alf_mach})] and shock obliquities, taken
from Double et al. (2004).  It is evident that the solutions here
closely match those of Double et al. in both the non-relativistic and
ultra-relativistic regimes. The jump conditions reveal several
characteristics, such as the declining compression ratio with declining
Mach numbers of either variety, and small fluid deflections at the shock
in the ultra-relativistic regime.  There are noticeable differences
between our solutions and those of Double et al. (2004) in the
trans-relativistic domain, but mainly for just the compression ratio.

These differences in the two works, especially apparent for low sonic
Mach numbers, are the result of two different assumptions regarding the
equation of state both upstream and downstream of the shock.  In this
work, the effective \teq{\gamma_{g1}} for a given flow speed and Mach
number can be determined using equations (\ref{eqn:gammaofT}) and
(\ref{eq:son_alf_mach}).  For \teq{\machson=100, 60}, and 6, the
upstream values of \teq{\gamma_{g1}} stay within 1\% of the nominal
non-relativistic value of 5/3. However, in the \teq{\machson=2.6} case,
we find \teq{\gamma_{g1}\approx 1.6} for large values of \teq{\Gamma_1
\beta_1}. In their work, Double et al. (2004) make an approximation
assuming a cold upstream flow (i.e a large sonic mach number with
upstream \teq{\gamma_{g1}=5/3}), resulting in the following equation
relating the downstream \teq{\gamma_{g2}} to the downstream flow speed
rather than to the downstream pressure:
\begin{equation}
   \gamma_{g2}\; =\; 
   \dover{\Gamma_{rel}\beta_{rel}^{2}}{3(\Gamma_{rel}-1)}+1
   \;\equiv\; \dover{1+4\Gamma_{rel}}{3\Gamma_{rel}} \quad ,
 \label{eqn:39ofdouble}
\end{equation}
where
\begin{equation}
   \beta_{rel} \; =\;\dover{\beta_1-\beta_2}{1-\beta_1\beta_2}\quad ,\quad
   \Gamma_{rel} \;\simeq\; (1-\beta_{rel}^{2})^{-1/2}\quad .
\end{equation}
Note that the slight angle between the upstream and downstream NIF flow
velocity vectors spawns the approximation for \teq{\Gamma_{rel}}; the
details and justification of this approximation can be found in section
3.1 of their paper. Assuming that \teq{\gamma_{g1}=5/3} then results in a
small \teq{\sim 4}\% discrepancy in \teq{\gamma_{g2}} relative to results from 
our J\"uttner-Synge equation of state, in the lowest Mach number cases. 
From the plots, clearly the numerical evaluations of the compression
ratio are sensitive to the choice of the form of the downstream equation
of state, i.e., \teq{\gamma_{g2}}.

In the limit that \teq{\Gamma_1\beta_1} approaches infinity, the details
of the equation of state become irrelevant, and a simple analytic
solution can be found to approximate the results of both approaches.
Defining
\begin{equation}
 \label{eqn:q_def}
   q \; =\; \dover{\machalf^2}{\sin^2\ThetaBfone} \, \dover{w_1}{\rho_1} 
      \; =\; \left(\frac{\machson^2(\gamma_{g1}-1)+1}{\machson^2(\gamma_{g1}-1)}\right)  
            \, \dover{\machalf^2}{\sin^2\ThetaBfone} \quad ,
\end{equation}
one can write the asymptotic limit for the compression ratio as
\begin{equation}
   r \;\approx\;  \sqrt{4+4q+\dover{q^2}{4}} - 1 - \dover{q}{2} 
      \quad, \quad \Gamma_1 \gg 1 \quad , 
 \label{eq:asymp_r}
\end{equation}
which corresponds to Eq.~(47) of Double et al. (2004).  The 
downstream fluid deflection and magnetic fields angles possess 
corresponding asymptotic forms for \teq{\Gamma_1 \gg 1} of
\begin{eqnarray}
   \tan\ThetaBftwo &\approx&  \Gamma_1\tan\ThetaBfone\,\sqrt{r^{2}-1}
      \quad ,\nonumber\\[-5.5pt]
\label{eq:asymp_TB2Tu2}\\[-5.5pt]
   \tan\thetautwoS &\approx& \dover{3-r}{2\Gamma_1\tan\ThetaBfone}
      \quad . \nonumber
\end{eqnarray}
When combined with Eq.~(\ref{eq:Bx_cons}), the asymptotic equation 
for \teq{\tan\ThetaBftwo} becomes Eq.~(40) of Double et al. (2004).  
Clearly, the fluid deflection is very small for ultra-relativistic flows, 
the hallmark of the extreme inertia of the upstream fluid.  One can see that, 
for the range of sonic Mach numbers explored here, the critical parameter 
\teq{q} is a comparatively weak function of the sonic Mach number inducing 
less than a 50\% change in \teq{q} as the sonic mach number varies from 
2.6 to 60.  In these cases, the Alfv\'{e}nic Mach number and the upstream 
magnetic field angle \teq{\ThetaBfone} are more important for determining 
the asymptotic behavior of the jump conditions.  It is also evident that
since \teq{r\approx 3} when \teq{\Gamma_1\gg 1} and \teq{q\gg 1}, the
downstream fluid deflection angle \teq{\thetautwoS} in the shock frame
is extremely small.

\section{Results}
 \label{sec:results}
The simulation we have developed is capable of handling both subluminal
and superluminal shocks of arbitrary obliquity. It can also simulate the
effects of SAS or LAS with varying levels of cross-field diffusion
controlled through the \teq{\eta} parameter. These broad capabilities
encapsulate a range of properties that are relevant to astrophysical
shock environs such as those in extragalactic jets in gamma-ray bursts
and blazars.  Moreover, they allow us to examine and expand upon a
variety of previous investigations, including the semi-analytic studies
of Kirk et al. (2000) and Kirk and Heavens (1989) as well as other
simulations such as Ellison and Double (2004), and Niemiec and Ostrowski
(2004). The following subsections highlight our key results in distinct
parameter regimes: parallel shocks, oblique subluminal shocks, oblique
superluminal shocks, and finally, LAS scenarios. Each of these sections
provide physical scenarios where the power-law index is substantially
different from the ``canonical'' \teq{\sigma=-2.23} (where
\teq{dn/dp=p^{-\sigma}}) result, which we demonstrate only holds at the
shock location in parallel, ultra-relativistic (\teq{\Gamma_1\gg 1})
shocks with a SAS scenario, concurring with previous work. Altering this
specific scenario yields power-law indices that depend on the
microphysics (LAS vs. SAS), shock obliquity (\teq{\ThetaBfone}), and
turbulence parameter (\teq{\eta}) as well as the location relative to
the shock front. A brief interpretation of these results in the context
of blazars is offered in Section~\ref{sec:blazar}.

\subsection{Parallel Shocks}
 \label{sec:parallel}
Parallel shocks possess the important simplification that cross-field
diffusion is immaterial.  Accordingly, the model parameter \teq{\eta}
does not impact the spectra at the shock, and serves only to define the
diffusive scale along the shock normal.  For the case of relativistic
parallel shocks, the canonical result a \teq{\sigma= 2.23} power law
spectrum was highlighted in the semi-analytic study of Kirk et al.
(2000), but had been found previously by Monte Carlo simulations
(Bednarz \& Ostrowski 1998; Baring 1999) and confirmed also by Ellison
\& Double (2004).  However, exhibited results from these studies were
spatially restricted to the immediate vicinity of the shock. The
simulation presented here provides the opportunity to expand upon these
studies and explore the spatial evolution of the particle distribution
as well. The semi-analytical work of Kirk et al. (2000) provides the
best basis for benchmarking simulated distributions at the shock
discontinuity. Accordingly, in this study, shock parameters are chosen
in order to facilitate this comparison. The eigenfunction method of Kirk
et al. (2000) built upon the earlier seminal work of Kirk \& Schneider
(1987) as an approach to solving the diffusion-convection equation in
the neighborhood of a relativistic shock.  Kirk et al. used this
technique to generate power-law indexes and angular distributions for
accelerated particles at a strong, weakly-magnetized plane-parallel
shock in the SAS limit.  In this case, accelerated particles are defined
as particles whose Lorentz factor far exceeds that of the shock, so that
distribution characteristics in the injection domain are not traced. 
However, the injection process is modeled in Monte Carlo simulation
approaches, and we show results probing this domain in later sections.

\begin{figure*}[t]
\twofigureoutpdf{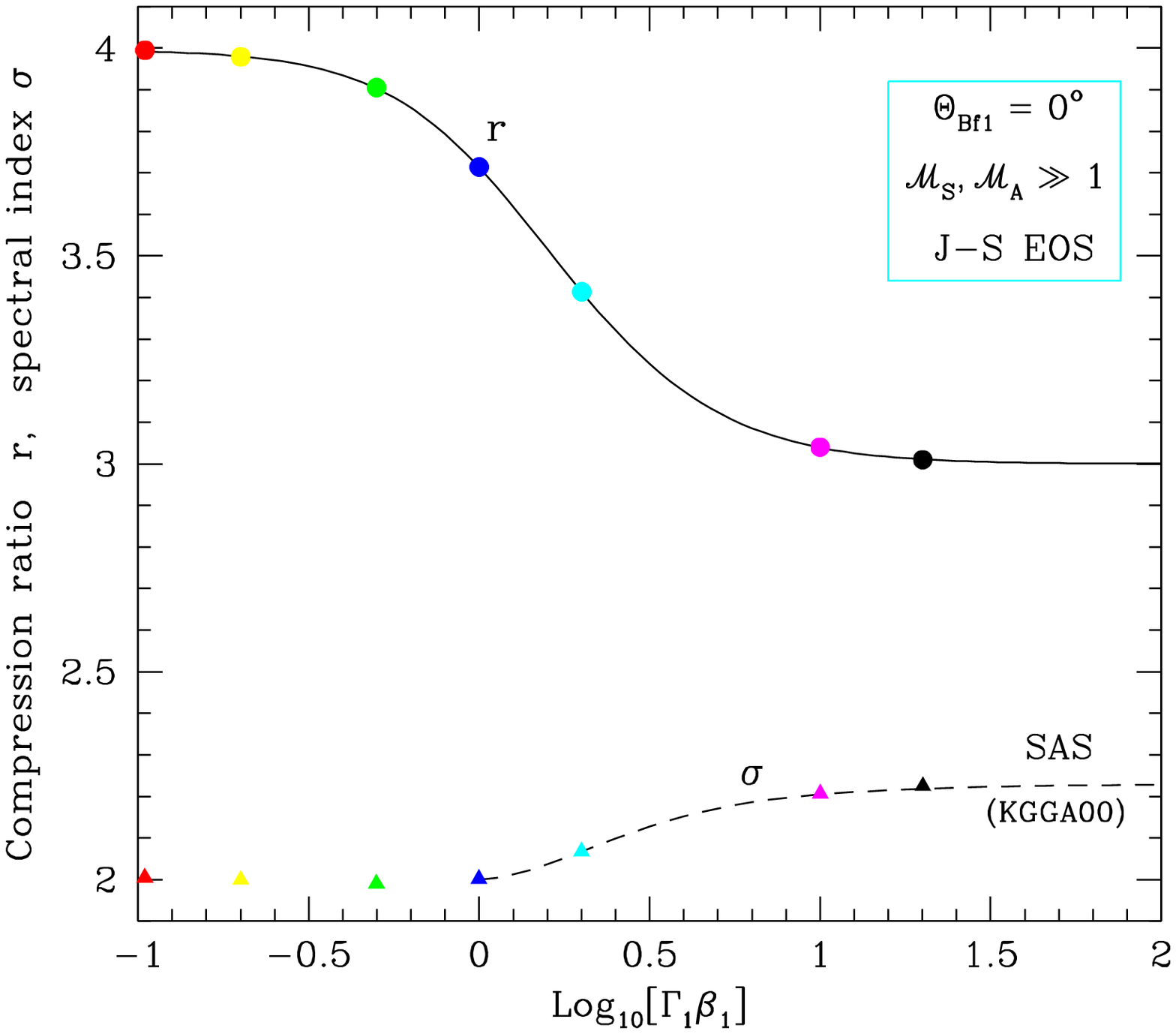}{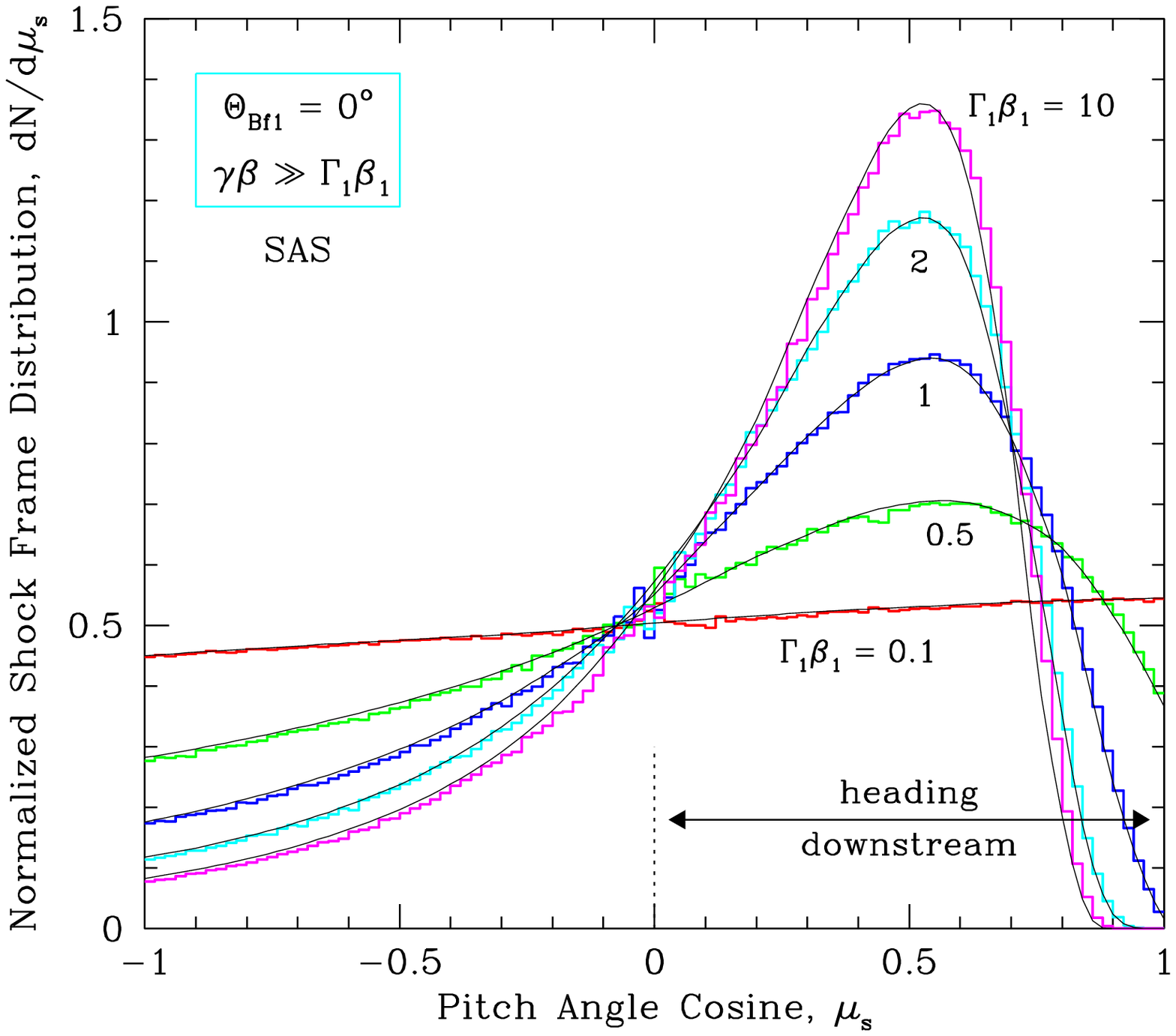}{
{\it Left Panel}: The Rankine-Hugoniot MHD compression ratio \teq{r} and 
spectral indices \teq{\sigma} for non-thermal particle 
distributions, as functions of the shock rapidity \teq{\Gamma_1\beta_1}, 
for plane-parallel, \teq{\ThetaBfone = 0^{\circ}}, shocks.  The compression ratio
is computed according to the protocols of Section~\ref{sec:RHrel}.  The 
points for \teq{r} and \teq{\sigma} correspond to select shock speeds 
with simulation data; most of these possess angular distributions illustrated 
in the right panel.  The simulation runs were restricted to the small angle scattering
(SAS) regime, for which the dashed curve corresponds to the low-magnetization
semi-analytic determinations of \teq{\sigma} in  Figure~4 of Kirk et al. (2000;
labelled KGGA00).
{\it Right Panel}: NIF frame angular distribution functions, normalized to unity, 
for parallel (\teq{\ThetaBfone =0^{\circ}}), relativistic shocks with
bulk rapidities \teq{\Gamma_1\beta_1} as labelled, and compression 
ratios \teq{r=3.995, 3.905, 3.714, 3.414} and \teq{3.04}, from lowest to highest 
speed (see points in left panel).  Distributions were measured at the shock (\teq{x=0}) 
and sampled only high energy particles with 
rapidity \teq{\gamma\beta\gg\Gamma_1 \beta_1} in each case.  
The simulation results are the histograms, acquired for the small angle
scattering (SAS) case, and the smooth curves 
are the semi-analytic solutions that Kirk et al. (2000) obtained (see their Fig.~3)
to the diffusion-convection equation in the SAS limit.
 \label{fig:anis_Kirkcomp} }
\end{figure*}

Fig.~\ref{fig:anis_Kirkcomp} displays shock compression ratios, and our
Monte Carlo results for spectral indices and angular distributions at
the shock in the NIF, for parallel shocks spanning a range of rapidities
\teq{\Gamma_1\beta_1}. Moreover, it exhibits corresponding results from
Kirk et al. (2000), and clearly illustrates that we find excellent
agreement between Monte Carlo simulation results and Fig. 3 of their
work.  To facilitate comparison, we adopted the J\"uttner-Synge EOS
here, though we note that details of the shock parameters for Fig.~3 of
Kirk et al. (2000) were not explicitly stated in their paper.  This is,
effectively, a case approximating that of the red curve in Fig.
\ref{fig:RH_ThetaB=5deg} here, save that \teq{\ThetaBfone =0^{\circ}}.
This minor change actually simplifies the Rankine-Hugoniot solution and
is shown as the solid black curve in left panel of
Fig.~\ref{fig:anis_Kirkcomp}.  The spectral index \teq{\sigma} is a
monotonically-increasing function of \teq{\Gamma_1\beta_1}, as in Kirk
et al. (2000) and Baring (2004), reflecting the increased energization
per shock crossing cycle that competes against the influence of a
declining compression ratio. The angular distributions in the right
panel of the figure closely match those from Fig. 3 of Kirk et al.
(2000), all measured at the shock discontinuity.  In this panel,
\teq{0<\mu_s<1} cases are for particles heading downstream, and \teq{-1
<\mu_s < 0} are charges moving upstream. The distributions exhibit an
increase in convective beaming downstream as the upstream flow speed
increases. Such distributions were obtained as angle-dependent fluxes,
and then divided through by the flux weighting factor \teq{\beta\mu_s}
before normalization.  This introduces the apparent statistical noise in
the neighborhood of \teq{\mu_s=0}.

\begin{figure*}[t]
\twofigureoutpdf{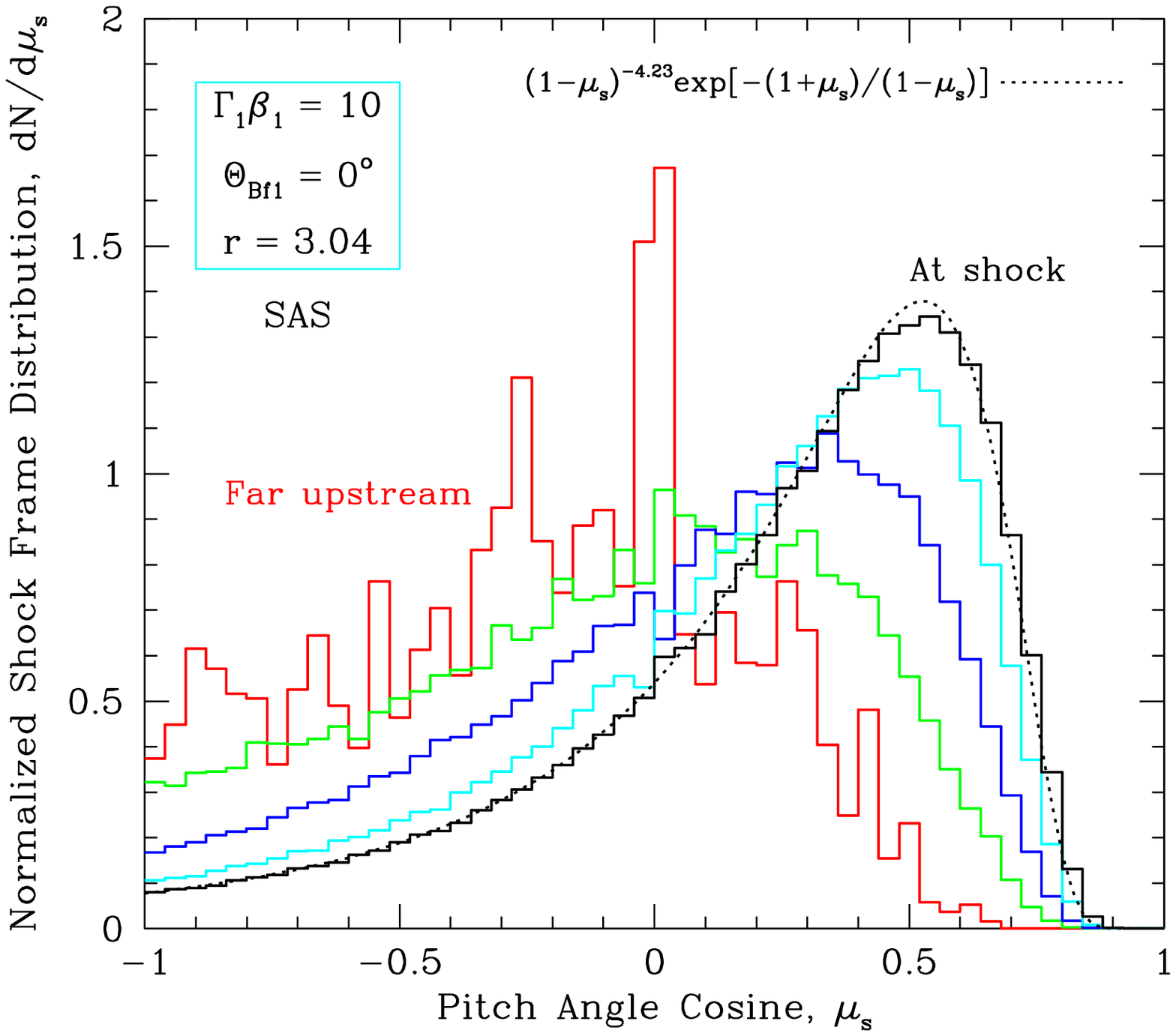}{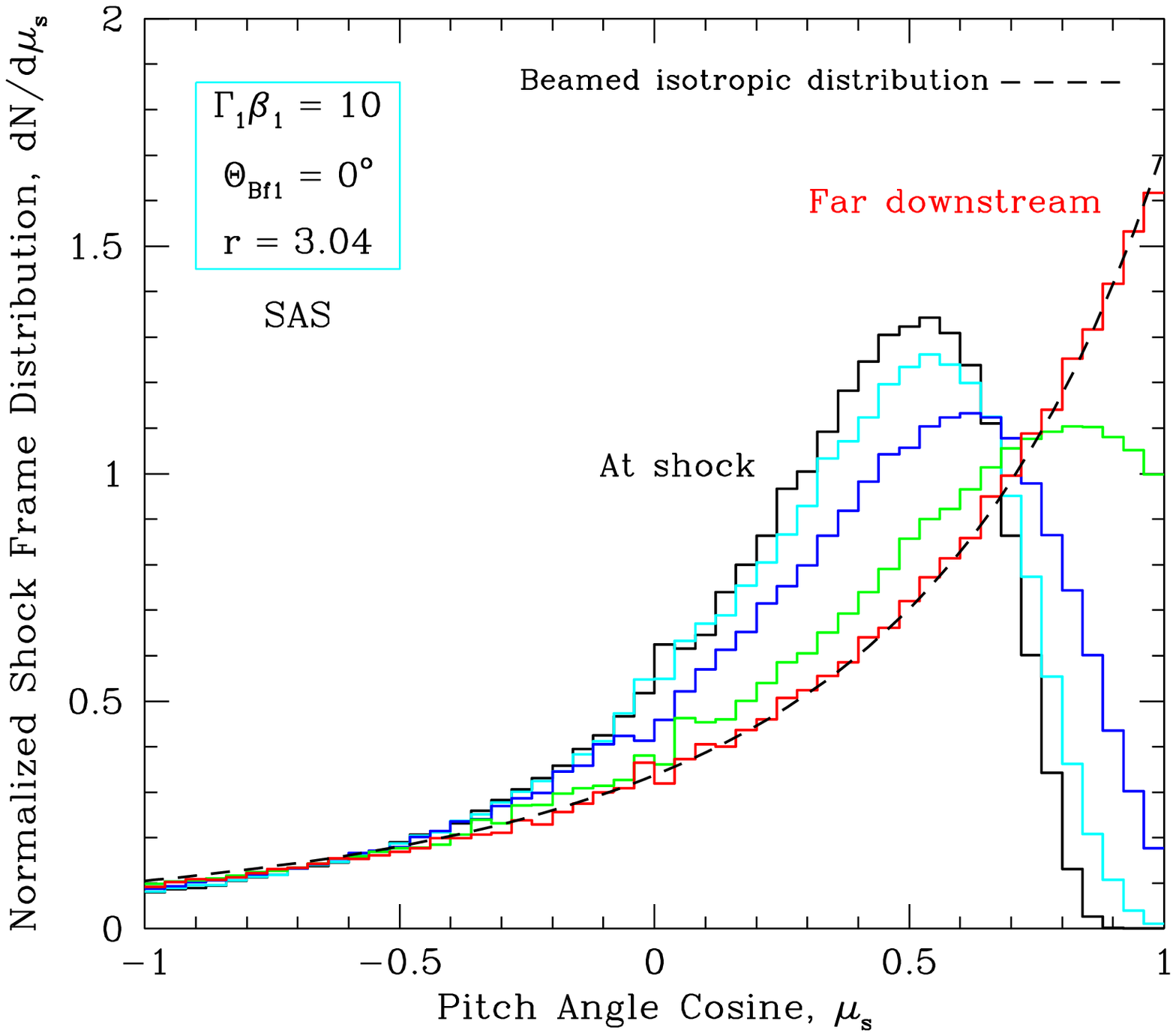}{
Normalized NIF frame angular distribution functions for high energy particles with 
rapidity, \teq{\gamma\beta\gg\Gamma_1 \beta_1}, upstream (left panel) 
and downstream (right panel) of the shock at various distances.  The simulation 
run was for a parallel (\teq{\ThetaBfone =0^{\circ}}), ultra-relativistic shock with
\teq{\Gamma_1\beta_1 =10} and compression ratio \teq{r=3.04}, and diffusion 
in the SAS limit. 
{\it Left panel}:
The black histogram is the distribution function at the shock and can be
compared directly to the dotted line, which is an analytic result from Eq.~(23) of
Kirk et al. (2000). The other 4 distribution functions are taken at increasingly large 
distances upstream of the shock.  In units of \teq{\Gamma_1\beta_1m_{p}/qB}, the 
cyan curve is at \teq{x=-20}, the blue curve at \teq{x=-80}, the green curve 
at \teq{x=-320}, and the red curve at \teq{x=-1280}. Distribution functions 
determined at larger distances upstream suffer from poor statistics, since
few particles are able to diffuse so far upstream against the relativistic flow.
{\it Right panel}:
As in the left panel, the black histogram is the distribution function at the shock,
with the other histograms now corresponding to \teq{x=400} (cyan), 
\teq{x=1600} (blue), \teq{x=6400} (green), and \teq{x=25600} (red).  The 
dashed line represents an isotropic distribution in the downstream fluid frame, 
as viewed by an observer in the NIF frame where the shock is stationary.
 \label{fig:ang_dist_gambet10} }
\end{figure*}

To delve deeper into the anisotropies incurred in relativistic shocks,
in Fig.~\ref{fig:ang_dist_gambet10} we examine the
\teq{\Gamma_1\beta_1=10} case in more detail, extending the angular
distribution illustrations to locations upstream (left panel) and
downstream (right panel) of the shock front. Again the distributions
correspond to high energy particles with rapidity,
\teq{\gamma\beta\gg\Gamma_1\beta_1}. In each panel, the black histogram
is the distribution function at the shock, exhibited in
Fig.~\ref{fig:anis_Kirkcomp}. The origin of the shape of the angular
distributions can be understood qualitatively. In non-relativistic shock
scenarios, high energy particles of speeds far in excess of that of the
shock realize isotropy in all relevant frames of reference. However, in
relativistic shocks, even particles traveling very close to the speed of
light can no longer be considered isotropic in all relevant frames and
at all positions. Consider a relativistic particle returning to the
upstream side of the shock from the downstream side. The upstream fluid
frame velocity vector of this particle is initially pointed upstream. As
the trajectory is perturbed via SAS seeded by field turbulence, the
velocity vector in the upstream fluid frame performs a random walk.
Because of relativistic beaming effects, once the particle's path is
perturbed by the small angle \teq{\theta>1/\Gamma_1} in the upstream
fluid frame, the shock frame x-component of the velocity (or angle
cosine \teq{\mu_s}) becomes positive, sweeping the particle back to the
shock before it has the chance to isotropize in the upstream fluid
frame. Accordingly, the parameter space around \teq{\mu_s=1} is
under-populated (actually exponentially suppressed) because the
upstreaming particles have not had enough time to diffuse from
$\mu_{s}<0$ to $\mu_{s}\gtrsim 0.9$ before they are convected through
the shock and downstream. This feature is critical to the
hyper-efficient reflection in oblique relativistic shocks discussed in
Sec. \ref{sec:obq_sublumin}. Note that at various non-zero obliquities,
similar NIF frame angular distributions are elicited in the simulation
at the shock, but are not shown.

It is interesting to note that, as the particle detection plane is moved
upstream, the domain of population suppression near \teq{\mu_s=1}
expands to lower \teq{\mu_s}. This is because particles that have angle
cosines closer to \teq{\mu_s\sim -1} are more likely to penetrate
further upstream, so that when diffusing outside the Lorentz cone, they
are less likely to populate near \teq{\mu_s\sim 1}. This skews the
distribution progressively towards more negative \teq{\mu_s}.
Additionally, the probability of particles reaching a position \teq{x}
upstream exponentially declines with \teq{\vert x\vert} on diffusive
lengthscales (e.g. see Lee 1983, and Summerlin \& Baring 2006, for
illustrations of this in non-relativistic, heliospheric shock contexts).
Accordingly, distribution functions taken at larger distances upstream
suffer from poor statistics. Thus, the upstream distribution functions
exhibited in the left panel of Fig.~\ref{fig:ang_dist_gambet10} are
normalized to have the same area for display purposes.

The evolution of the distribution function downstream of the shock is
shown in the right hand panel of Fig. \ref{fig:ang_dist_gambet10},
ranging from the distribution found at the shock (black histogram) to an
isotropic distribution in the downstream plasma frame (red histogram).
As the particles move downstream, the relativistic beaming that biases
the distribution to higher average values of \teq{\mu_s} is
progressively enhanced, and they eventually isotropize in the downstream
fluid frame (the red histogram in the right panel).  The dashed line in
that panel is the angular (density) distribution function in the shock
frame, \teq{dN_s/d\mu_s}, for particles that are isotropic in the fluid
frame with a power-law distribution \teq{dN_f/dp_f\equiv 4\pi
p_f^2f({\vec r},\, p_f) \propto p_f^{-\sigma}}. Here \teq{f({\vec r},\,
p_f)} is the fluid frame phase space distribution function, which is a
Lorentz invariant.  Hence in the shock rest frame, the angular
distribution satisfies \teq{dN_s/d\mu_s \propto p_s^2\, p_f^{-(\sigma
+2)}} for a fixed choice of \teq{p_s}, which is imposed in this example
by truncating the NIF distribution at a lower limit of \teq{p_s=p_0}. 
For ultrarelativistic particles, the relationship between \teq{p_s} and
\teq{p_f} is given simply by the photon aberration formula
\teq{p_f=p_s\Gamma (1-\beta\mu_s)}.  Accordingly, the angular
distribution in the NIF for isotropy in the fluid frame scales as
\teq{dN_s/d\mu_s \propto (1-\beta\mu_s)^{-(\sigma+2)}}.  This is what is
illustrated in Fig. \ref{fig:ang_dist_gambet10}, which for the
\teq{\Gamma_1\gg 1} case reduces to that for values \teq{\beta\approx
1/3} downstream and \teq{\sigma = 2.21}.

This evolution of the anisotropy has consequences for the observed
power-law index downstream of the shock. Because the average value of
\teq{\mu_s} for the returning particles is lower than it would be for
particles that are isotropized in the downstream fluid frame, the
average bulk flow speed of the accelerated particles is lower than
\teq{u_{2x}}.  As the angular distribution function diffusively evolves
toward isotropy in the downstream fluid frame at larger \teq{x}, the
average velocity of the particles also increases, asymptotically
approaching the higher bulk velocity of the downstream thermal
particles. This necessarily reduces the density of the high energy
particles by conservation of particle number flux. The scale length for
the evolution of the distribution function is approximately the particle
mean free path. Thus, higher energy particles with typically longer mean
free paths isotropize farther downstream than do particles of lower
energy. Accordingly, the spectral shape of the distribution downstream
of the shock evolves, illustrated in Fig. \ref{spat_dep_spec}, in a
manner that correlates with the spatial changes in the angular
distribution.

The black curve in Fig. \ref{spat_dep_spec} displays the distribution
function at the shock. However, just downstream (cyan curve), the low
energy particles have isotropized, thereby increasing their bulk flow
speed and lowering their density, to conserve particle number flux. 
Higher energy particles in this curve have not yet fully isotropized and
possess slightly slower bulk speeds and thus, higher densities. So, if
one is measuring distributions somewhat downstream of a relativistic
shock, the observed power-law will be harder than the canonical
\teq{\sigma=-2.23} result realized exactly at \teq{x=0}. At large
distances downstream of the shock, accelerated particles of all
rapidities far in excess of \teq{\Gamma_2\beta_{2x}} acquire the same
beamed, isotropic distribution shown in the right panel of
Fig.~\ref{fig:ang_dist_gambet10}, so that their cumulative density
adjustments during downstream diffusion are identical, and the power-law
index returns to that realized at the shock.  This variation of the
power-law index with the position of a downstream observer relative to
the relativistic shock represents a fundamental shift from
non-relativistic shocks, where the distribution function is isotropic in
all relevant reference frames, and the spectral index is virtually
independent of position when downstream of the shock.  To our knowledge,
this is the first time this effect in relativistic shocks has been
highlighted in the literature.

\figureoutpdf{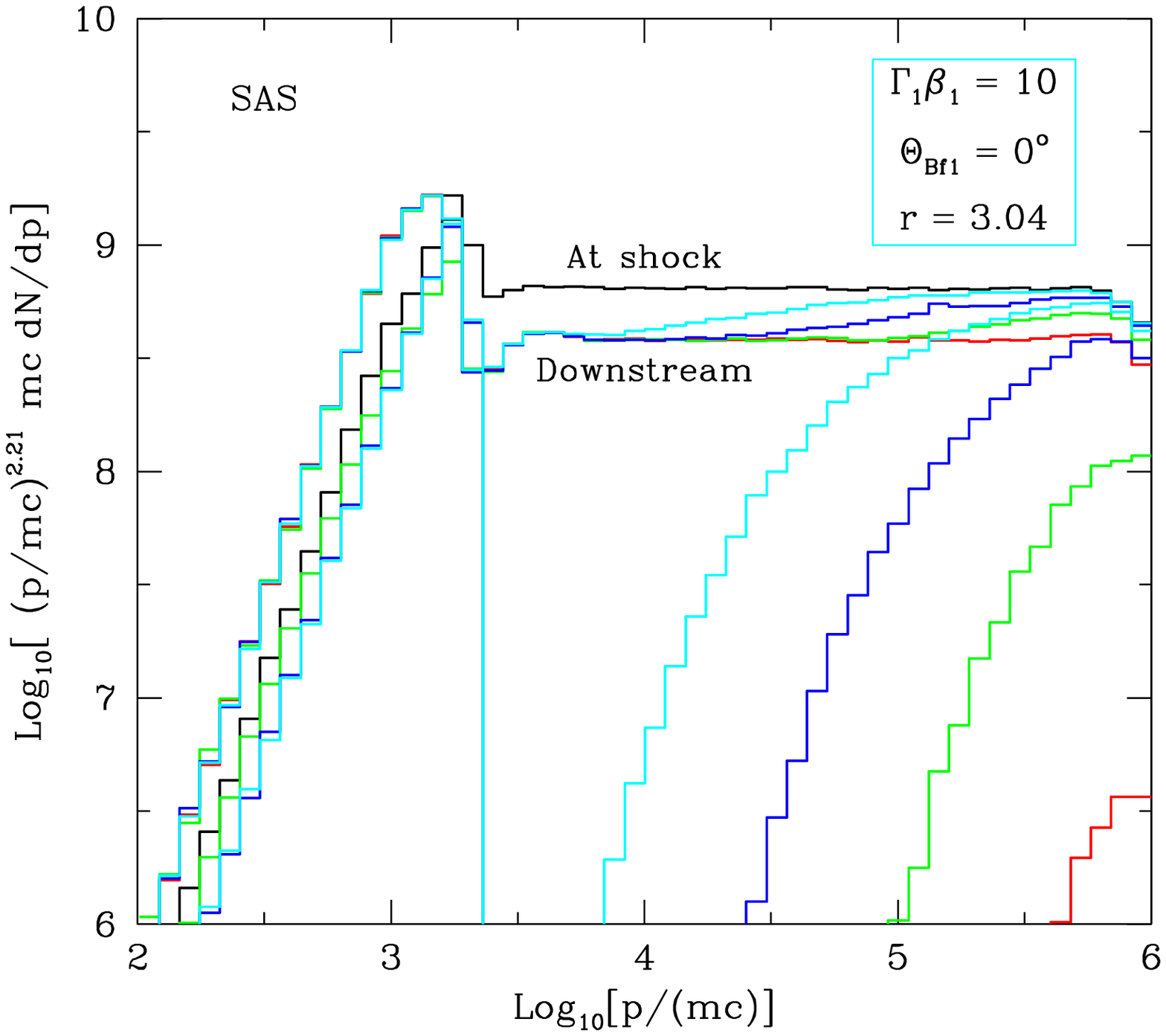}{3.8}{0.0}{-0.2}{
Accelerated particle distribution functions at positions upstream and
downstream of the shock. The positions and color coding corresponding to
Fig.~\ref{fig:ang_dist_gambet10}, with the histograms that fall sharply
with decreasing momentum corresponding to the upstream positions, and
the histograms that make only a small adjustment from the distribution
function at the shock corresponding to downstream positions.  To clearly
illustrate the differences between these distribution functions, the
differential density distribution \teq{dN/dp} has been multiplied by
\teq{p^{2.21}} so as to generate zero power-law slope at the shock. The
paucity of low energy particles at the upstream positions is due to
their limited contraflow mobility, and is seen in non-relativistic
shocks as the ``convective peel-off" effect described in Summerlin \& Baring
(2006), see text for a discussion. 
The spectral variations downstream of the shock are a result of 
energy-dependent density compression, and are addressed in the text.
 \label{spat_dep_spec} }

\subsection{Oblique, Subluminal Shocks}
 \label{sec:obq_sublumin}
While small shifts in the power-law index can occur based on observation
location and energy range in parallel shocks, the introduction of
non-zero magnetic field obliquity creates more substantial
ranges of power-law indices for the non-thermal particle component.  In
non-relativistic shocks, the power-law index is independent of magnetic
field obliquity, and depends only on the compression ratio:
\teq{\sigma = (r+2)/(r-1)} (Bell, 1978; Drury, 1983; Jones \& Ellison
1991). In relativistic shocks, the spectral index varies dramatically
with obliquity, as will be exemplified in due course. In particular, the
character of the spectral index with respect to field obliquity hinges
critically on whether the shock is subluminal or superluminal. Thus, our
study of oblique relativistic shocks is divided into two sections to
treat these parameter regimes separately.

\begin{figure*}[t]
\twofigureoutpdf{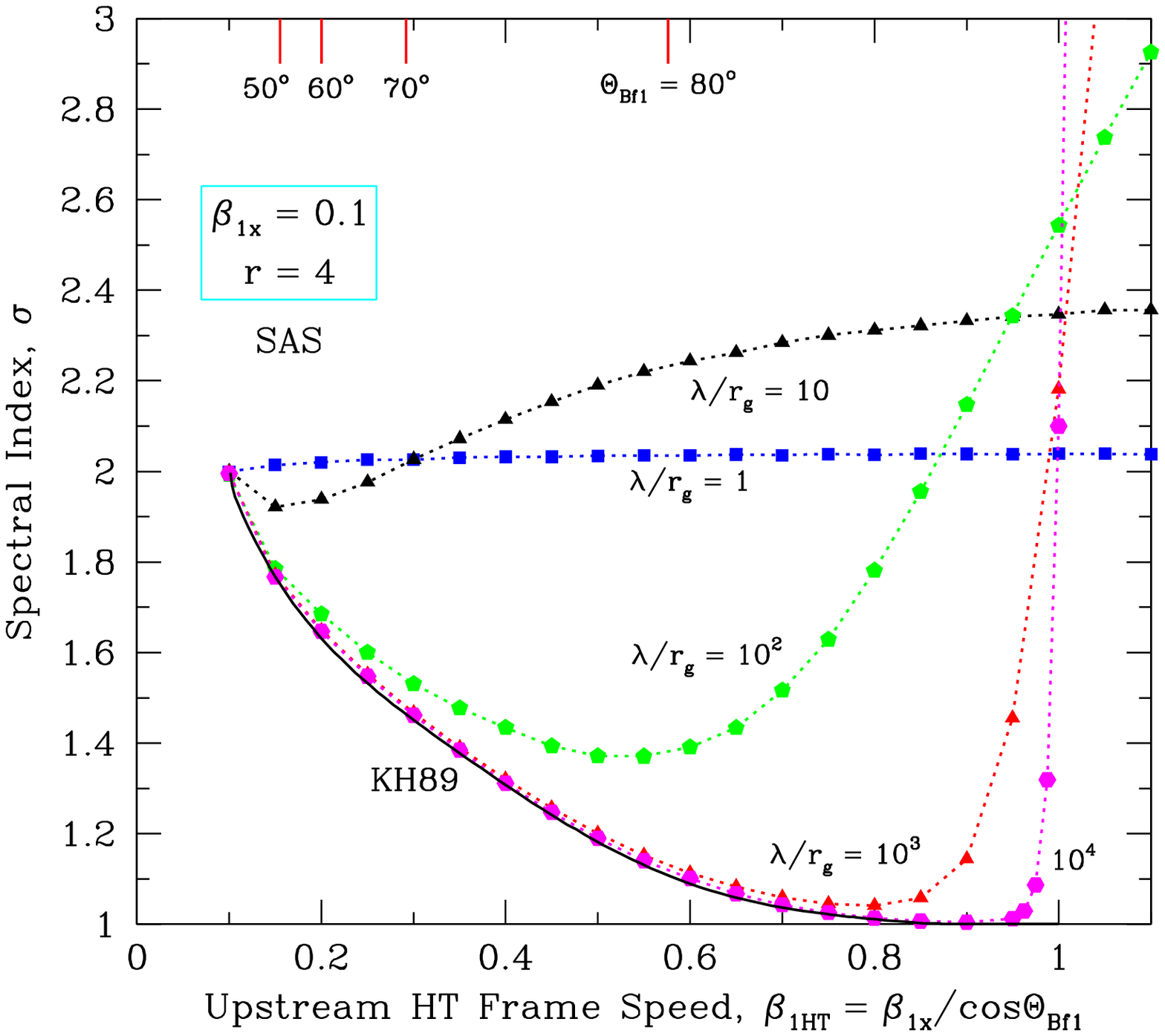}{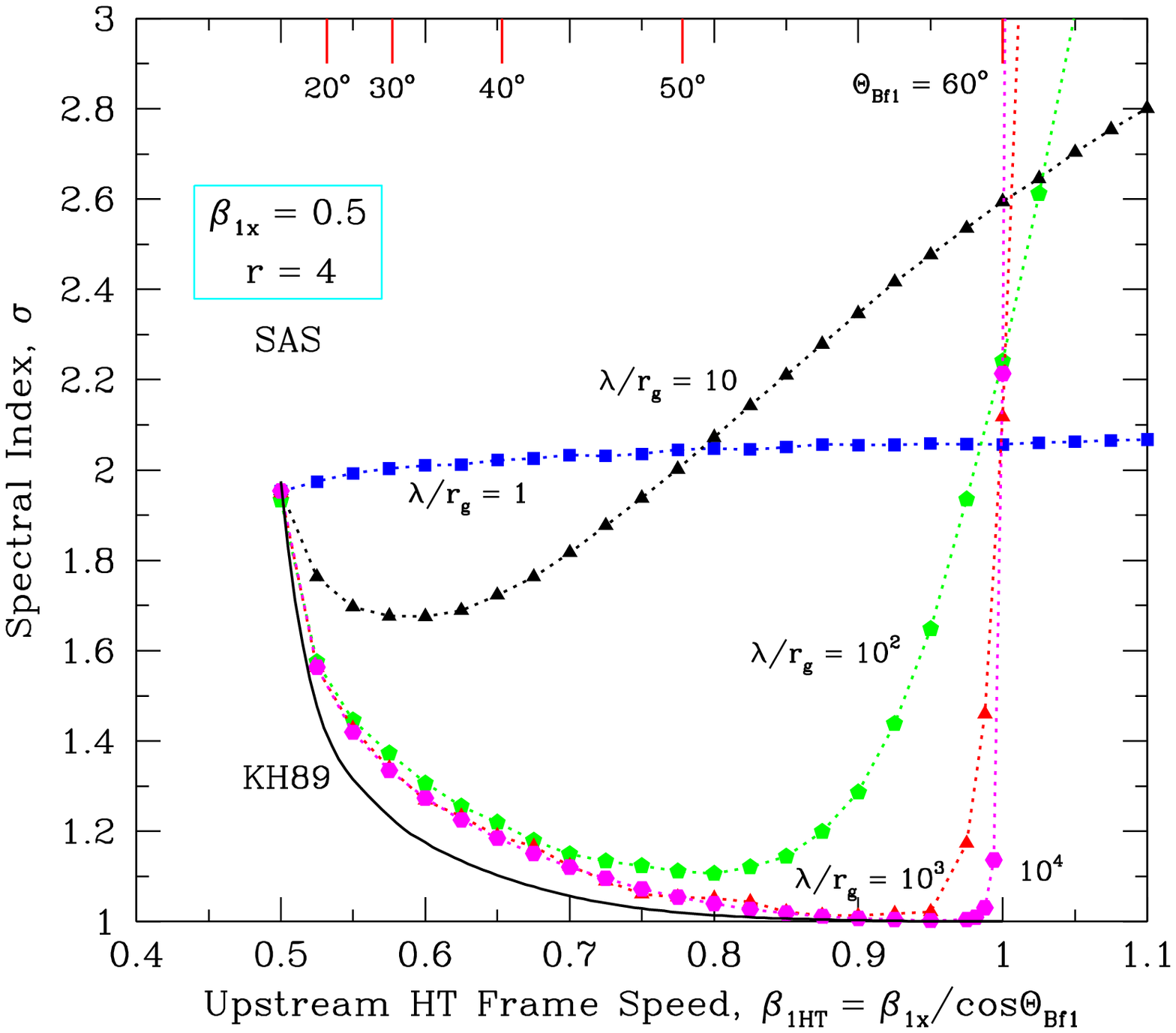}{
Power-law indices for simulation runs in the limit of small angle
scattering (SAS), for an almost non-relativistic
shock of upstream flow speed \teq{\beta_{1x}\equiv u_{1x}/c =0.1}
({\it left panel}), and a mildly-relativistic case with 
\teq{\beta_{1x}\equiv u_{1x}/c =0.5} ({\it right panel}), for an
MHD velocity compression ratio \teq{r=u_{1x}/u_{2x}=4}. The indices
are displayed as functions of the effective de Hoffmann-Teller frame
upstream flow speed \teq{\beta_{\hbox{\sevenrm 1HT}} =
\beta_{1x}/\cos\ThetaBfone}, with select values of the fluid frame field
obliquity \teq{\ThetaBfone} marked at the top of the panel.  Obliquities
for which \teq{\betaoneHT >1} constitute superluminal
shocks.  The displayed simulation index results were obtained for
different diffusive mean free paths \teq{\lambda} parallel to the mean
field direction (see text), namely \teq{\lambda/r_g=1} (blue squares),
\teq{\lambda/r_g=10} (black triangles), \teq{\lambda/r_g=10^2} (green
pentagons), \teq{\lambda/r_g=10^3} (red triangles) and
\teq{\lambda/r_g=10^4} (magenta hexagons), as labelled.  The lightweight
black curve at the bottom labelled KH89 defines the semi-analytic result
from Kirk \& Heavens' (1989) solution to the diffusion-convection
equation, corresponding to \teq{\lambda/r_g\to\infty}.
 \label{U1_U5_kirk_comp} }
\end{figure*}

\begin{figure*}[t]
\twofigureoutpdf{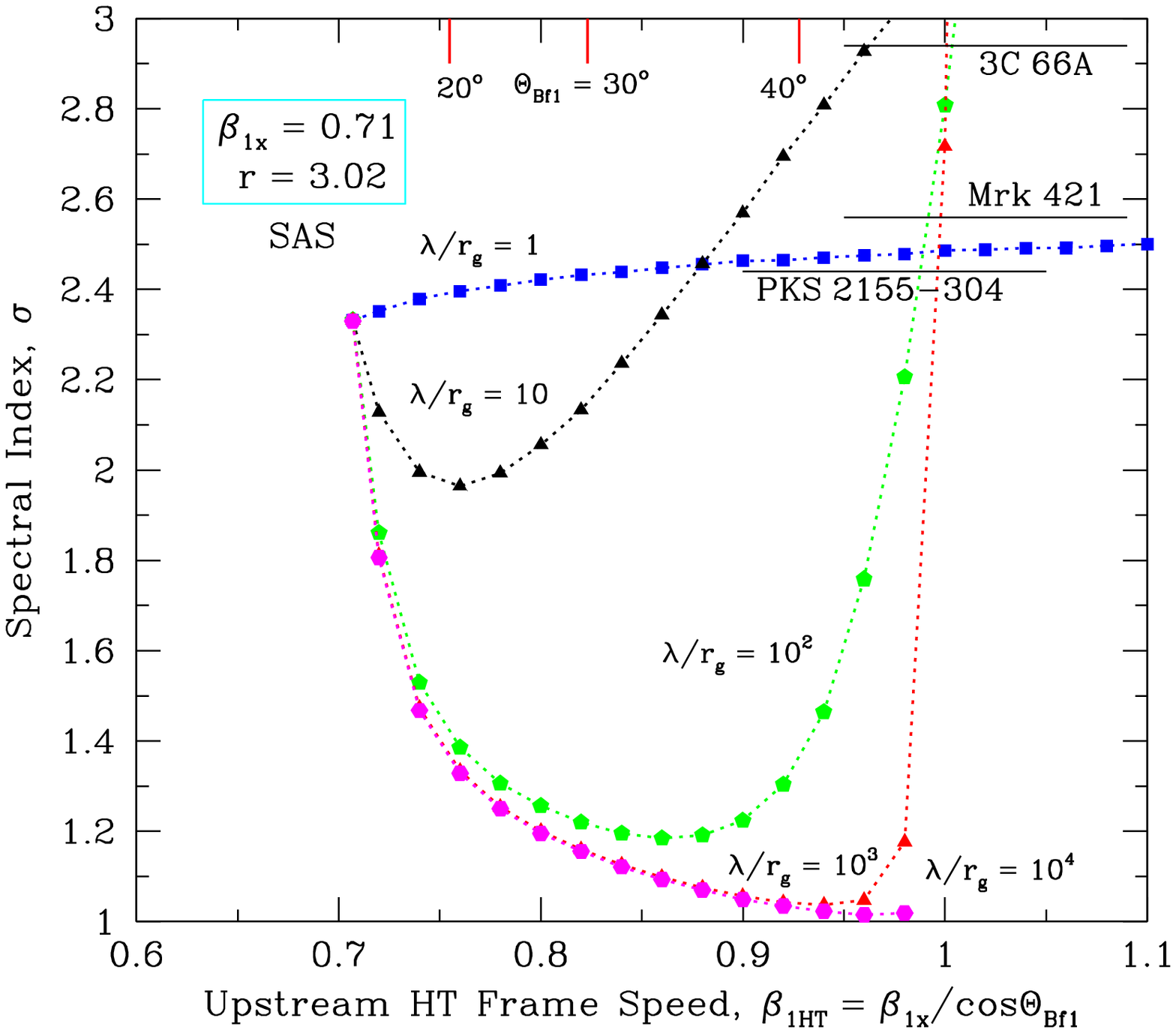}{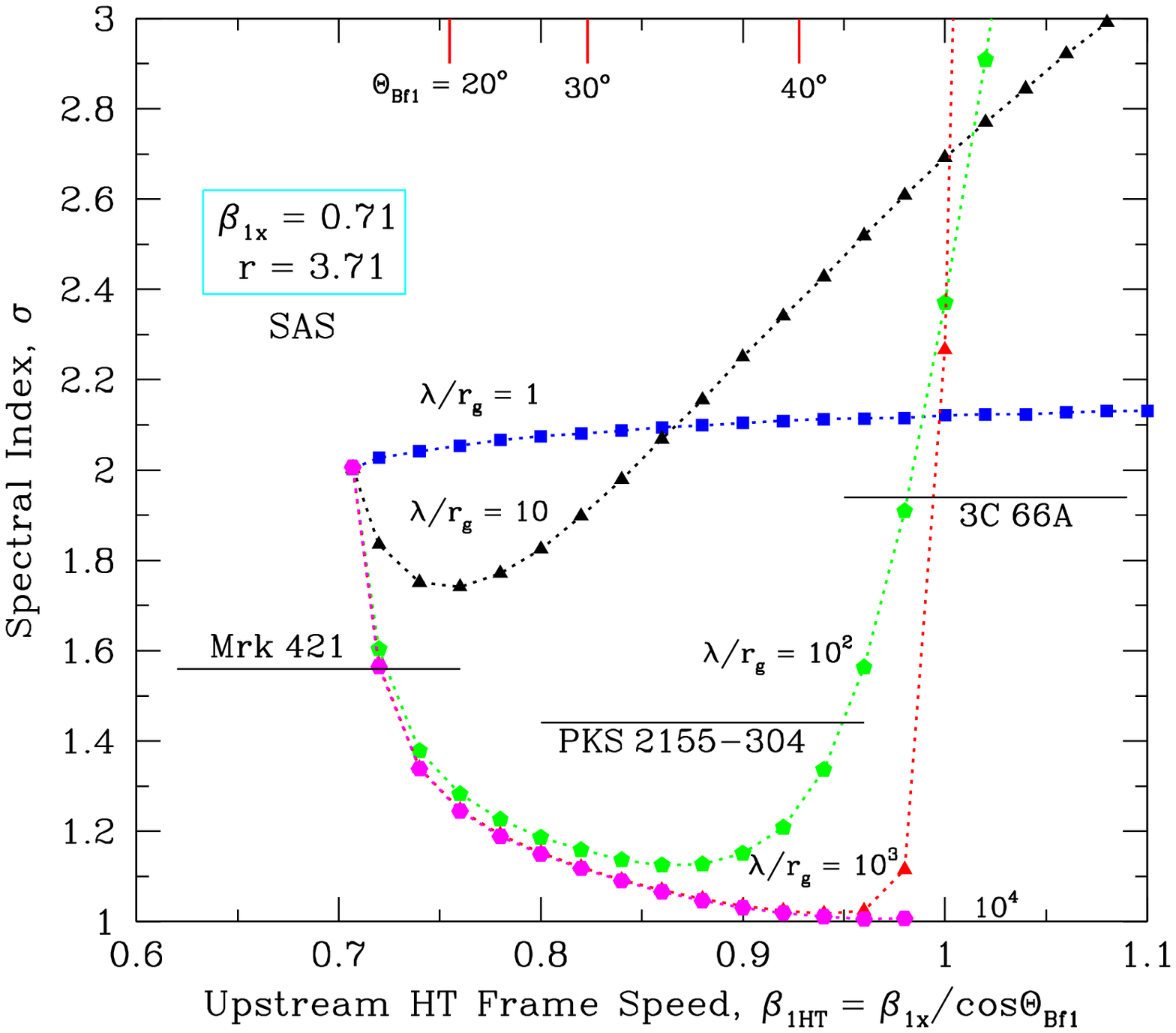}{
Power-law indices for simulation runs in the SAS limit, as in
Fig.~\ref{U1_U5_kirk_comp}, but now for shock parameters more
appropriate to the internal shocks associated with the relativistic jets
that are believed to be the source of GRBs. Here, \teq{\beta_{1x}\equiv
u_{1x}/c =0.71}, and the compression ratio and sonic Mach number are now
\teq{r=u_{1x}/u_{2x}=3.02} and \teq{\machson =2.6} ({\it left panel}),
and \teq{r=u_{1x}/u_{2x}=3.71} and \teq{\machson =60} ({\it right
panel}), calculated via the Rankine-Hugoniot relations derived in
Sec.~\ref{sec:RHrel}. The indices are again plotted versus the effective
de Hoffmann-Teller frame upstream flow speed \teq{\beta_{\hbox{\sevenrm
1HT}} = \beta_{1x}/\cos\ThetaBfone}, and selected fluid frame field
obliquities \teq{\ThetaBfone} are as marked at the top. The different
diffusive mean free path cases were again \teq{\lambda/r_g=1} (blue
squares), \teq{\lambda/r_g=10} (black triangles), \teq{\lambda/r_g=10^2}
(green pentagons), \teq{\lambda/r_g=10^3} (red triangles) and
\teq{\lambda/r_g=10^4} (magenta hexagons), as labelled. Also depicted
are marker indices for three {\it Fermi} blazars, Mrk 421, 3C 66A and
PKS 2155-304; they apply for arbitrary \teq{\betaoneHT}, and are
truncated in the horizontal direction to aid clarity of the figure.
These mark the approximate expectation for \teq{\sigma}, uncertain to
roughly \teq{\pm 0.2}, for an interpretation of the {\it Fermi}-LAT
gamma-ray spectral indices as uncooled inverse Compton scattering ({\it
left panel}) and strongly-cooled upscattering ({\it right panel}); see
Sec.~\ref{sec:blazar} for a discussion.
 \label{U1_071_r302_r371} }
\end{figure*}

Consider first subluminal, oblique shocks in the small angle
scattering limit. The first emphasis will be on the power-law behavior
of the accelerated portion of the population; later on the injection
efficiency will be addressed.   In the simulation, for each run, the
power-law regime is determined on an individual basis by inspection and
can begin anywhere from 5 to 100 times the mean injection (i.e.
approximately downstream thermal) energy. A least squares fit in log-log
space is used to determine the slope \teq{\sigma}. Results are depicted
in Fig.~\ref{U1_U5_kirk_comp} for  \teq{\beta_{1x}\equiv u_{1x}/c =0.1,
0.5}, and in Fig.~\ref{U1_071_r302_r371} for \teq{\beta_{1x} =0.71}, for
different values of the turbulence or cross-field diffusion parameter
\teq{\eta=\lambda/r_{g}}. The power-law index \teq{\sigma} is plotted as
a function of the HT frame dimensionless speed \teq{\betaoneHT
=\beta_{1x}/\cos\ThetaBfone}.  It is clear that there is a considerable
range of indices \teq{\sigma} for non-thermal particles accelerated in
mildly relativistic, oblique shocks. The essence of this array of indices 
and the global trends with \teq{\ThetaBfone} and \teq{\eta}
were outlined briefly in Baring \& Summerlin (2009) and Baring (2011), 
though a fuller interpretation ensues below.

A feature of this plot is that the dependence of \teq{\sigma}
on field obliquity is non-monotonic.  When \teq{\lambda /r_g\gg 1}, the
value of \teq{\sigma} at first declines as \teq{\ThetaBfone} increases
above zero, leading to very flat spectra.  As \teq{\betaoneHT}
approaches and eventually exceeds unity, this trend reverses, and
\teq{\sigma} then rapidly increases with increasing shock obliquity as
the shocks become superluminal.  This dramatic steepening is a
consequence of inexorable convection of particles away downstream of the
shock. The only way to ameliorate this rapid increase in index is to
reduce \teq{\eta =\lambda /r_g} to values below around \teq{10}.  Physically,
this corresponds to increasing the hydromagnetic turbulence to high
levels that force the particle diffusion to approach isotropy:
\teq{\kappa_{\perp}/\kappa_{\parallel} = 1/(1+\eta^{2})} in 
a kinetic theory description (e.g., Forman, Jokipii, and Owens (1974).  This
renders the field direction immaterial, and the shock behaves much like
a parallel, subluminal shock in terms of its diffusive character. 
Note that this general character is also 
evinced in the very recent diffusion-convection equation analysis 
of Bell, Schure \& Reville (2011) at shocks of lower speeds.  Figure~1 
in their paper clearly illustrates that the distribution index 
hardens with increasing obliquity when the shock is well within the 
subluminal regime and softens when the luminal boundary 
\teq{\cos\ThetaBfone = \beta_{1x}} is approached or crossed in 
quasi-perpendicular (and sometimes 
non-relativistic) shocks, unless the frequency of scattering is raised 
to the Bohm limit, for which the index then depends only weakly on 
the field obliquity.

In studying this case, we again choose to use previous
semi-analytic analyses as a benchmark for comparison: the work of Kirk
and Heavens (1989, KH89 hereafter) is ideally suited for this purpose.
KH89 calculated the spectral index of non-thermal particles at oblique,
trans-relativistic shocks using the eigenfunction technique of Kirk \&
Schneider (1987) to solve the diffusion-convection equation.  Their
analysis was restricted to situations where particles do not diffuse
across field lines; i.e., their collision operator contains only a pitch
angle scattering term. They also assumed that particles conserve their
magnetic moment on crossing the shock, a standard analytic
simplification. Results from Fig. 2 of their work are exhibited in
Fig.~\ref{U1_U5_kirk_comp}. Note that their exploration was done
exclusively in the HT frame and was thus limited to subluminal shocks.
An interesting product of their work was the appearance of power-laws
harder than both the non-relativistic and ultra-relativistic parallel
shock results in Bednarz \& Ostrowski, (1998) and Kirk et al. (2000),
achieved as \teq{\uoneHT} approaches the speed of light, but
\teq{u_{1x}} remains mildly-relativistic. This is an idealized result,
because the limit of zero cross-field diffusion does not occur in
Nature, since field turbulence abounds in astrophysical shocks,
and is needed to drive acceleration.  The
Monte Carlo technique is ideally suited to examining how close to zero
cross-field diffusion one must get to approach the particular analytical
case explored by KH89.  Fig.~\ref{U1_U5_kirk_comp} clearly indicates
that when \teq{\lambda/r_g\gtrsim 10^3} the KH89 zero-cross field
diffusion indices are closely reproduced for \teq{\beta_{1x}=0.1} and
well approximated for \teq{\beta_{1x}=0.5}.  The physical origin for
these extremely hard power-laws will be discussed in
Sec.~\ref{sec:shock_drift}. 

To allow a direct comparison with the results of KH89, we adopted 
the same compression ratio of \teq{r=4}, and the same formulation for the 
relationship between the upstream and downstream magnetic fields. This 
formulation is found in equations (2), (3) and (4) of their work and is summarized 
below. It assumes a weak magnetic field that does not influence the plasma 
motion (i.e. \teq{\machalf\gg1}); all the simulation runs used to generate 
Fig.~\ref{U1_U5_kirk_comp} satisfied this high Alfv\'enic Mach number criterion. 
\begin{eqnarray}
   r &=& \dover{\beta_{1x}}{\beta_{2x}} \nonumber\\[-5.5pt]
 \label{eq:KH89_MHD_assump}\\[-5.5pt]
   \dover{\BHTone}{\BHTtwo} & = &
       \sqrt{r^{2}-\Gamma_1^{2}(r^{2} - 1)\, \beta_{1x}^2\, 
       \left[ \dover{1}{\betaoneHT^2} - 1 \right] }\quad .\nonumber
\end{eqnarray}
Given the upstream quantities above, and the compression ratio, $r$, we 
can solve for \teq{\BHTtwo} and use our knowledge that $B_{x}$ is constant 
across the shock along with an appropriate Lorentz transformation 
(see Sec. \ref{sec:trans->NIF}) to find the appropriate downstream 
value for \teq{B_{z}} in any reference frame.

In Fig.~\ref{U1_U5_kirk_comp}, while impressive agreement with the
solutions of KH89 arises for \teq{\beta_{1x}=0.1} when \teq{\eta\gtrsim
10^{3}}, for \teq{\beta_{1x}=0.5}, we find that our simulation indices
match the results of KH89 at \teq{\betaoneHT \gtrsim 0.5} and just below
\teq{\betaoneHT=1}, but are noticeably softer (higher \teq{\sigma}) in
the central part of the curve. Increasing \teq{\eta} as high as
\teq{10^6} creates no appreciable change in the resulting power-law
index from that of \teq{\eta=10^{4}}: we believe we have reached the
asymptotic limit of our simulation. Monte Carlo results for
\teq{\beta_{1x}=0.3} are not depicted, but are similar to those for the
\teq{\beta_{1x}=0.5} case, and also exhibit a modest difference from the
KH89 determinations of \teq{\sigma} at intermediate values of
\teq{\betaoneHT}, while matching at the \teq{\betaoneHT} endpoints. We
contend that the reason for the modest discrepancy between the two
approaches probably lies in the assumption of conservation of magnetic
moment employed by KH89. This assumption facilitates an analytic result
but does not precisely describe orbiting particle reflection and
transmission properties at an oblique shock discontinuity (see Terasawa
1979; Drury 1983; Pesses \& Decker 1986; for non-relativistic shock
expositions). For parallel or quasi-perpendicular (in this case
nearly-luminal) shocks, the magnetic moment is conserved, and the two
approaches converge. For obliquities in between, there is slight
non-conservation of magnetic moment, and the precise tracing of
gyro-orbits in the shock layer, as is enacted in the Monte Carlo
technique, introduces modest but appreciable increases in \teq{\sigma}.

It is imperative to go beyond the artificial \teq{r=4} exploration,
since relativistic shocks are somewhat weaker in their compression.  To
this end we produced similar index plots for parameters more appropriate
to internal shocks in GRBs and blazars using the Rankine-Hugoniot
relations derived earlier.  Specifically, Fig.~\ref{U1_071_r302_r371}
displays Monte Carlo results for compression ratios that satisfy the
J\"uttner-Synge equation of state, \teq{{\cal M}_s=2.6} (\teq{r=3.02})
and \teq{{\cal M}_s=60} (\teq{r=3.71}), with the Alfv\'{e}nic Mach
number assumed large. The results mirror those in
Fig.~\ref{U1_U5_kirk_comp} in terms of overall character, with a large
range of indices, \teq{\sigma\sim 1} in near-luminal cases when
\teq{\lambda /r_g\gtrsim 10^3}, and a rapid steepening of the
non-thermal distribution in superluminal cases unless \teq{\lambda
/r_g\lesssim 10}. A particular index inferred from the radiation
spectrum of a single astronomical source can be accommodated by a range
of choices for shock speed, Mach numbers, field obliquity, and
turbulence parameter \teq{\eta}.  This interpretative aspect is the
subject of Section~\ref{sec:blazar}, with a focus on blazars.

Finally, note that Figs.~\ref{U1_U5_kirk_comp}
and~\ref{U1_071_r302_r371} were prepared specifically with diffusion
seeded by gyro-resonant interactions between charges and MHD turbulence
in mind.  In such cases, scattering descriptions are only physical if
\teq{\eta \geq 1} in Eq.~(\ref{eq:mfp}), i.e. above the Bohm limit. Yet
\teq{\eta < 1} regimes for diffusion can be realized for
non-gyroresonant interactions with field turbulence that is perhaps
grown via filamentation or Weibel instabilities.  Trial simulation runs
were performed in this \teq{\eta <1} domain, and it was found that the
distribution was not very sensitive to the choice of \teq{\eta}; for
example, reducing \teq{\eta} to 0.1 flattened the spectrum for the
\teq{\beta_{1x}=0.71}, \teq{\betaoneHT = 0.96}, \teq{r=3.71} case by an
index of around 0.1 relative to that displayed in
Fig.~\ref{U1_071_r302_r371}.  This behavior is a consequence of
diffusion in this ``sub-Bohm'' domain resembling that for the Bohm limit
of \teq{\eta =1} for gyroresonant diffusion. A more complete exploration
of this domain is deferred to future work.

\subsubsection{The Action of Shock Drift Acceleration}
 \label{sec:shock_drift}
\begin{figure*}[t]
\twofigureoutpdf{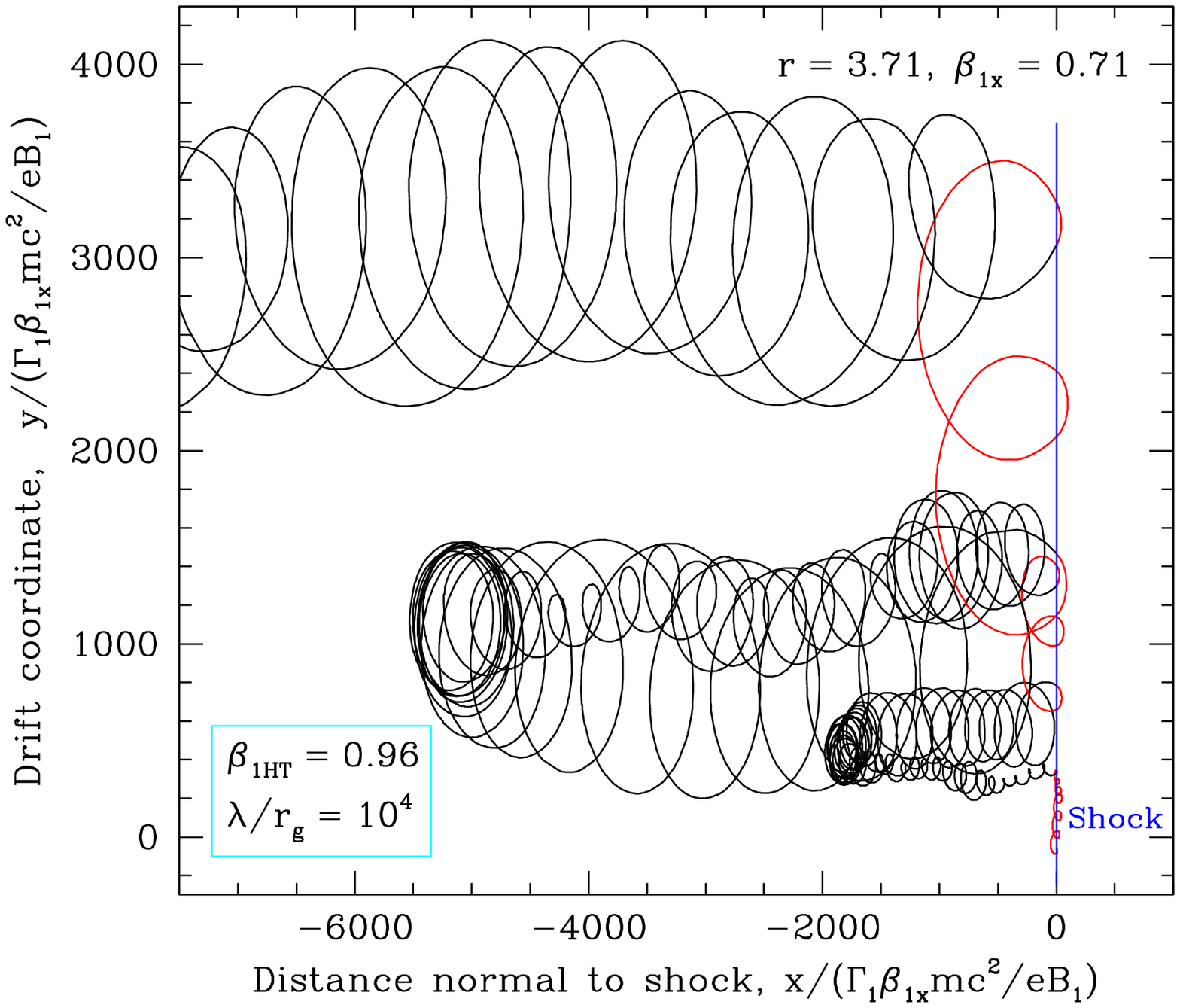}{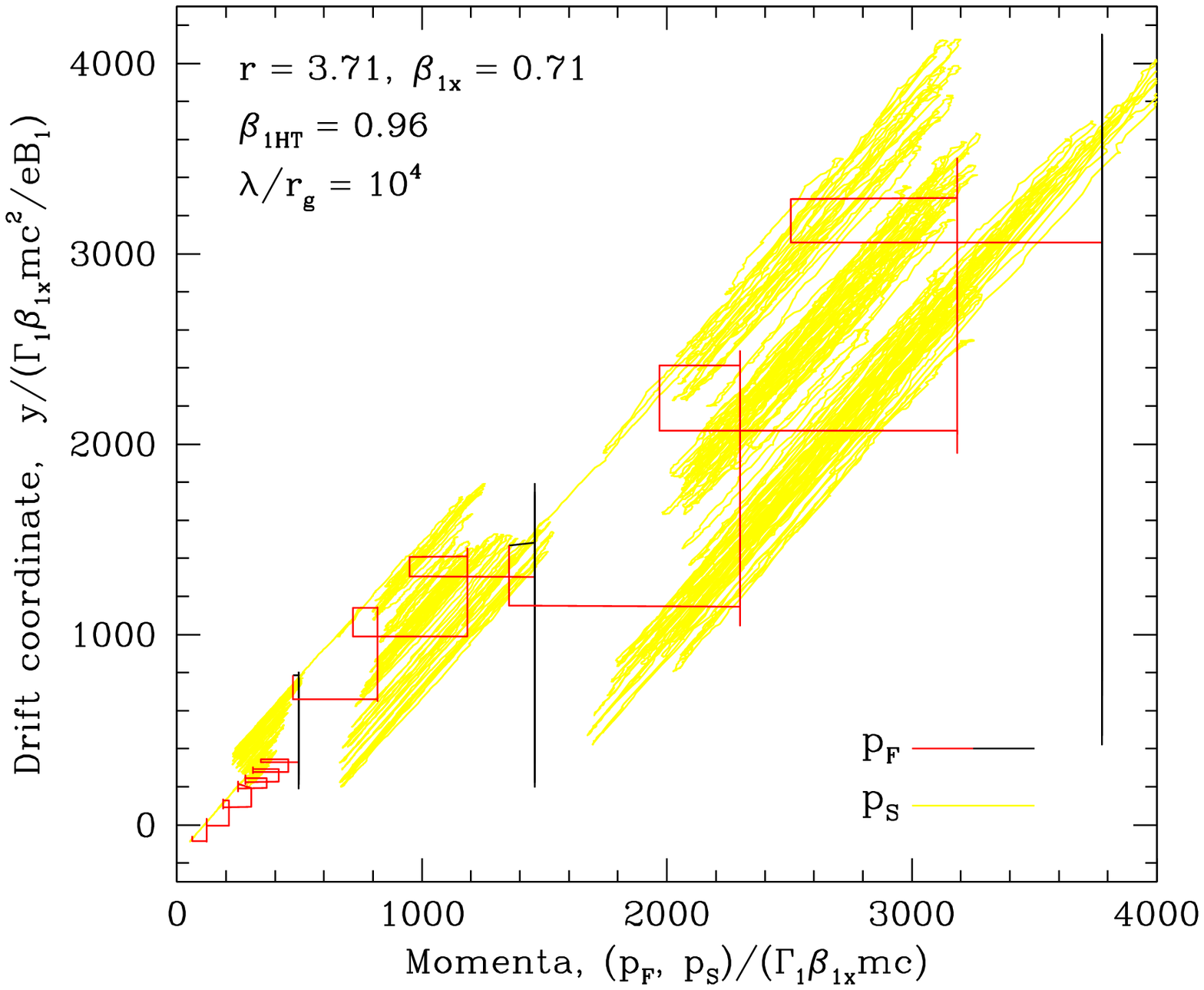}{
{\it Left panel}: A sample particle trajectory depicting strong SDA in
an oblique mildly relativistic shock with de Hoffmann-Teller frame
speed, \teq{\betaoneHT =0.96}, NIF shock speed, \teq{\beta_{1x}=0.71}, and compression
ratio, \teq{r=\beta_{1x}/\beta_{2x}=3.71} in a low turbulence environment
with \teq{\lambda /r_g=10^4}. 
The projection is onto the \teq{x}-\teq{y} plane,
where the {\bf u}\teq{\times}{\bf B} drifts lie in the
\teq{y}-direction.  This portion illustrates two key features of
shock drift acceleration: coherent trapping in the shock layer (colored red), interspersed
with upstream excursions (black) after reflection via approximate conservation of 
magnetic moment. This particular particle gained orders of 
magnitude in energy before finding a pitch angle small enough to allow 
transmission and subsequent escape downstream.
{\it Right panel}: The particle's position in the drift coordinate ($y$)
direction, as a function of the magnitude of the momentum in the fluid frame,
\teq{p_f} and the shock frame, \teq{p_s}.   This illustrates a core
property of shock drift acceleration: that over large times both \teq{p_f} and 
\teq{p_s} display approximately linear trends with the drift \teq{y}. 
Prolonged energy gains occur only during shock layer gyrations, 
when the fluid frame momentum exhibits a ``rectangular hysteresis'' (red). 
The shock frame momentum possesses perturbed
oscillatory temporal behavior (yellow) during the intervening upstream excursions.
 \label{fig_SDA_traj_mom} }
\end{figure*}

The bottom line of the preceding exposition is that the power-law index
achieved in subluminal oblique shocks can be considerably less than even
the \teq{\sigma=(r+2)/(r-1)} result for non-relativistic shocks.  For
moderate obliquities and \teq{\eta=\lambda /r_g\gtrsim 10^3} it can
become as hard as \teq{\sigma=1}. A power-law this hard can only be
achieved in the case that particle escape from the acceleration region
is miniscule. We illustrate here that the high \teq{\eta} cases that
have low \teq{\sigma} are subject to strong shock drift acceleration
(SDA), offering a close examination of the trajectories of energized
particles that reveals prolonged retention in the acceleration process. 
A small fraction of particles incident from upstream can be reflected at
the shock because they have suitable pitch angles, and these seed the
retention in the acceleration process. A sample trajectory and
associated momenta for a select particle undergoing such acceleration is
displayed in Fig.~\ref{fig_SDA_traj_mom}.  The particle was injected at
super-thermal energies to circumvent improbable injection from the
thermal population, a property that is discussed later in this
subsection. The trajectory highlights two hallmarks of shock drift
acceleration: coherent trapping in the shock layer, interspersed with
upstream excursions after reflection at the shock (see Decker \& Vlahos
1986, for illustrations in non-relativistic shock contexts).  The
reflection condition can be estimated by assuming conservation of
magnetic moment in the HT frame, i.e., requiring that
\teq{(1-\mu_{p1}^{2})/\BHTone=(1-\mu_{p2}^{2})/\BHTtwo}, where
\teq{\mu_{pi}} is the particle's pitch angle cosine in the upstream
(i=1) or downstream (i=2) HT frame. This assumption is technically valid
only when particles gyrate a large number of times in the shock layer
(e.g., see Drury 1983), which arises when the gyroperiod is far inferior
to the time it takes to convect one gyroradius downstream, i.e. when
\teq{p_s\gg\Gamma_1 \betaoneHT}.  For magnetic moment conservation,
given \teq{\BHTtwo > \BHTone}, it is clear that there are some values of
\teq{\mu_{p1}} for which \teq{\mu_{p1} \geq 0} cannot be satisfied. In
these cases, particles are reflected rather than transmitted, and the
shock is acting as a magnetic bottleneck.

As particles gyrate in the shock layer, the work done \teq{dW}
on a charge can be computed using the Lorentz force, resulting in an equation 
\vspace{-5pt}
\begin{equation}
   (mc)^2\, \gamma\, \dover{d\gamma}{dt}
   \; \equiv\; \hbox{\bf p}.\dover{d\hbox{\bf p}}{dt} \; =\; q\, \hbox{\bf p}.\left\{ \hbox{\bf E} 
                  + \dover{\hbox{\bf v}}{c} \times \hbox{\bf B} \right\}
                  \;\equiv\; q\, \hbox{\bf p}.\hbox{\bf E} \quad ,
 \label{eq:shock_drift}
\end{equation}
where {\bf E} is the {\bf u}$\times${\bf B} drift field.  In the uniform
{\bf B} fields either \newpage\noindent upstream or downstream, the
energy gains and losses acquired during a gyroperiod exactly cancel, so
that no net work is done, \teq{dW=mc^2d\gamma =0}. In contrast, when a
charge's gyromotion straddles the shock discontinuity, it samples the
different electromagnetic field on either side of the shock for
different times, with the net acceleration on the upstream side of the
shock being greater than the deceleration on the downstream side of the
shock. Such an asymmetry in energy increments is explicitly evident in
Eq.~(6) of Jokipii's (1982) exposition on SDA in non-relativistic shocks
(see also Webb, Axford \& Terasawa 1983, for relativistic cases). This
energization can be seen in the counter-clockwise rotations of the
\teq{p_f} curve in Fig. \ref{fig_SDA_traj_mom}.  In this curve, vertical
motion indicates travel upstream or downstream of the shock where
\teq{p_f} is constant. The horizontal lines are shock crossings where
the local \teq{p_f} changes in transits of the shock discontinuity. By
relating elapsed time during SDA to increments in the drift coordinate
\teq{y}, it is simply shown (Jokipii 1982) that 
\begin{equation}
   dW \;\equiv\; mc^2d\gamma\; =\; qE_y\, dy\quad ,
 \label{eq:SDA_workdone}
\end{equation}
or \teq{d(p/mc) \propto (eB_1/mc^2)\, dy} in the relativistic limit of
\teq{\gamma\gg 1}. This energy gain scales linearly with displacement
along \teq{y}, with the proportionality constant depending on the shock
obliquity. Such a linear scaling (in \teq{p_s}) is the punchline of the
right panel of Fig.~\ref{fig_SDA_traj_mom}, where \teq{y} effectively
represents a time coordinate during shock drift episodes. This
energization proportionality is clearly evident during shock
interactions for the selected particle, and is an established hallmark
of SDA.  But notably, on larger scales, for both \teq{p_f} and
\teq{p_s}, it is also an approximate description of the cumulative
energization spanning multiple shock encounters. This follows as a
consequence of the constancy of \teq{p_f} and only moderate changes to
\teq{y} during upstream excursions; i.e. material energy changes arise
only during shock interactions. The \teq{p_s} curve's behavior shows
smooth and rapid energization during gyrations in the shock layer
followed by a quasi-oscillatory epoch for \teq{p_s} coupled with random
spatial diffusion during the upstream excursions. The gradual increase
of \teq{p_s} during the upstream excursions results from the angular
diffusion of the particle's fluid frame momentum vector in the upstream
region. This increases \teq{p_s} as the fluid frame velocity vector
becomes more aligned with the flow vector. The widening of the gyrations
in \teq{y} is also due to angular diffusion gradually increasing the
pitch angle upstream of the shock.

In typical SDA, a reflected particle gains energy and is sent back
upstream to encounter the shock again. However, this alone can not
create such an unusually hard power-law. In the case of non-relativistic
shocks, this process happens commonly in oblique shocks when low levels
of turbulence are assumed, with no impact on the power-law index: it
still retains its \teq{\sigma = (r+2)/(r-1)} value (see Jokipii 1982;
Lee 1984; Armstrong et al., 1985; Decker and Vlahos 1986; Webb, Axford
\& Teresawa 1983; Pesses et al., 1982; Decker, 1988; Vandas, 2001). This
is because the energy gained through shock drift acceleration (SDA) is
exactly canceled by decreases in the efficiency of diffusive shock
acceleration (DSA) in oblique shocks, since the angular distribution is
effectively isotropic in the NIF and fluid frames. Furthermore, this
isotropy guarantees that select particles are not trapped in the shock
layer for long durations, because they have the same probability of
being reflected at each shock encounter.  However, in the case of
oblique subluminal shocks whose HT frame speed is approaching the speed
of light, we find different circumstances.

As discussed in section \ref{sec:parallel}, in relativistic, parallel
shocks in an SAS scenario, the field-aligned component of the
distribution function is depleted at the shock.  While the first shock
encounter is that of an isotropic plasma boosted by the upstream flow
speed with a strongly field-aligned population of particles, the
distribution of particles returning after being reflected the first time
is missing these field aligned particles because the population has, in
general, had insufficient time for any of them to become field aligned
in the downstream direction before re-encountering the shock. This is a
result of the fact that relativistic shocks preclude particle speeds
from far exceeding the relevant flow speeds and becoming effectively
isotropic in all reference frames. Subluminal, oblique relativistic
shocks exhibit this same behavior. Upstreaming particles must initially
have pitch angle \teq{\mu_{\hbox{\sixrm HT}} \approx -1} in order to
travel upstream faster than the field line they are on is being
convected downstream. Once they diffuse in pitch angle, they are quickly
swept back into the shock before they achieve isotropy in the HT frame.
Thus, as in the case of parallel shocks, the field aligned,
\teq{\mu_{\hbox{\sixrm HT}} \approx 1}, component of the angular
distribution function is depleted at the shock.  Yet only such
field-aligned particles are capable of penetrating the magnetic
bottleneck at the shock. The net result of this dearth of field-aligned
particles is a significant reduction in the fraction of particles
transmitted through the shock at each  encounter. This enhances the
probability of reflection hardening the power-law. As the HT frame speed
approaches c, the transmission region can become completely depopulated
leading to near 100\% reflection and the observed \teq{\sigma=1}
power-law.

The value of this index value is dictated by energy arguments.  For
long-lived trapping of select particles in the shock layer, energy in
the particle population is transmitted from one Lorentz factor bin
\teq{[\gamma,\, \gamma + d\gamma]} to the next one above, with miniscule
loss.  The energy content of this bin is \teq{\gamma N}, and when
deposited in the next bin above, it is increased by SDA by an amount
\teq{\propto dy\propto d\gamma} according to
Eq.~(\ref{eq:SDA_workdone}), which is independent of the value of
\teq{\gamma}, but just on the electromagnetic structure of the shock
layer. It then follows that the energy increment in going from adjacent
bins is \teq{\gamma dN\propto d\gamma}, so that \teq{dN/d\gamma \propto
1/\gamma}, i.e. \teq{\sigma = 1}.  Introducing significant losses
reduces this energy increment, and thereby steepens the spectrum. Along
with the offering in Baring \& Summerlin (2009), this is the first
identification of the important role shock drift acceleration can play
in determining the spectral index in relativistic shocks. In
non-relativistic shocks, SDA does not influence the spectral index.

Note that the highly enhanced action of SDA is restricted to high
\teq{\eta} and SAS regimes. Invoking a LAS scenario completely
eliminates the enhanced probability of reflection as particles are
isotropic after their first upstream scattering and encounter the shock
as such: all subsequent shock encounters have the same probability of
reflection as the first shock encounter. For SAS scattering, increasing
the amount of turbulence present (reducing \teq{\eta}) allows particles
to scatter into the transmission cone and subsequently softens the
power-law as shown in Figs.~\ref{U1_U5_kirk_comp}
and~\ref{U1_071_r302_r371}.  Trajectories for such \teq{\eta\lesssim
10^2} cases (not shown) exhibit a more ``wonky'' gyration and reveal
prompt convection downstream, shutting down the opportunity for repeated
episodes of coherent acceleration.

A few comments are necessary on the feasibility of encountering
parameters that drive hyper-efficient SDA.  The extremely large values
of \teq{\eta} required correspond to very low levels of turbulence that
are not anticipated near shocks: it would require a truly unusual set of
physical parameters to produce power laws significantly harder than
\teq{\sigma=1.3} using this mechanism.  It should also be noted that the
bulk thermal particles that create these strong shocks must be cold
compared to the flow speed in order for the shock to form. Thus, despite
the fact that they receive substantial kinetic heating during their
first shock crossing, they will not meet the \teq{p\gg\Gamma_1 \uoneHT}
criterion and will have a reduced chance for reflection. Particles may
have only a couple gyro-orbits that pass through the shock during an
encounter and the range of phases that permit them to reflect is reduced
dramatically. For \teq{v<\uoneHT}, it is physically impossible for
particles to diffuse upstream along field lines and some amount of
cross-field diffusion is necessary for particles to return to the shock
at all. This creates an injection problem; see Ellison, Baring \& Jones
(1995) for a discussion of this in non-relativistic, oblique shock
contexts.

\begin{figure*}[t]
\twofigureoutpdf{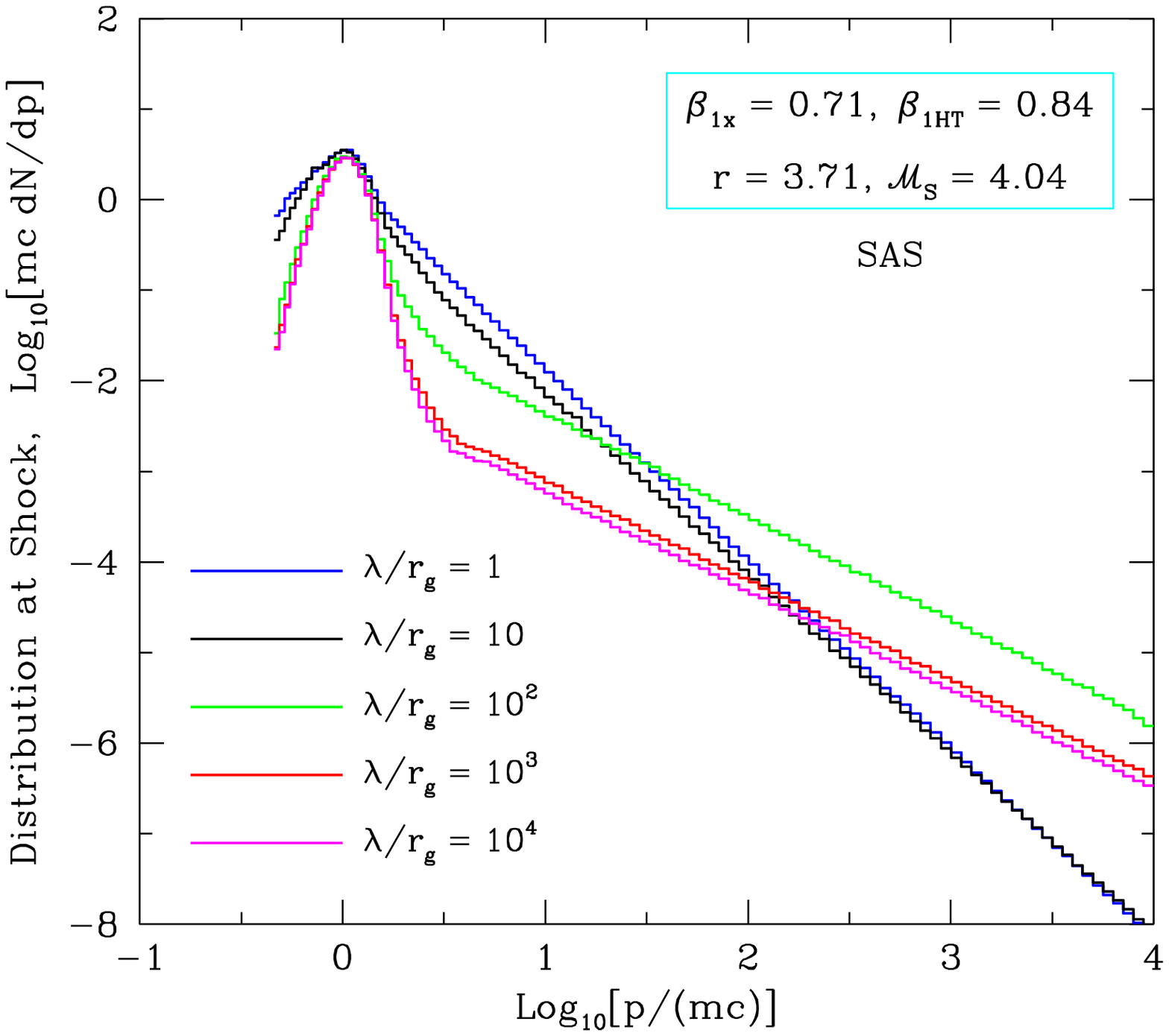}{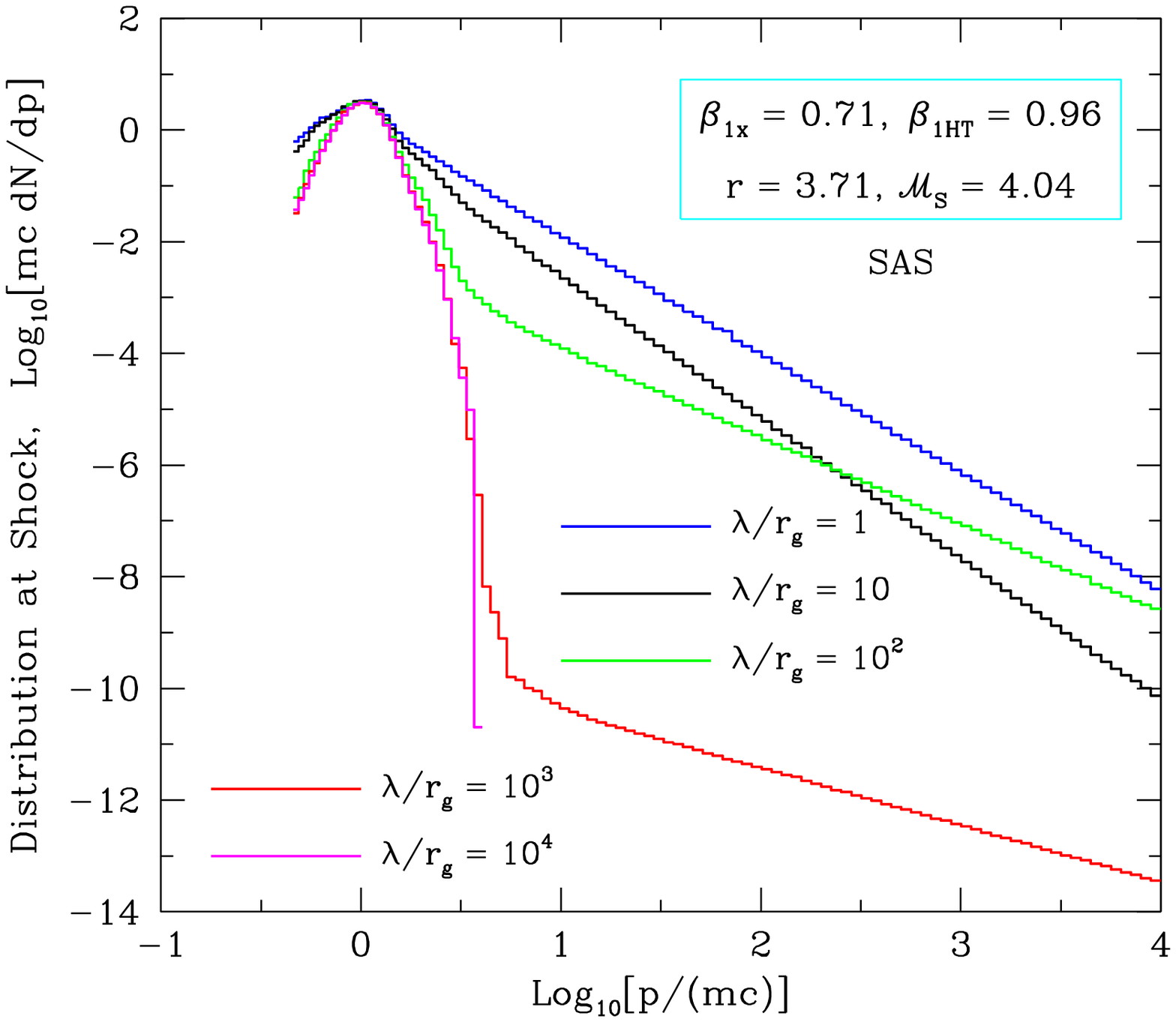}{
Full particle distributions for simulation runs in the small angle
scattering (SAS) limit, for the strong mildly-relativistic shocks of
upstream flow speed \teq{\beta_{1x}\equiv u_{1x}/c =0.71} whose indices
are displayed in Fig.~\ref{U1_071_r302_r371}.  Here the de
Hoffmann-Teller frame upstream flow speed was set at
\teq{\beta_{\hbox{\sevenrm 1HT}}=0.75} (\teq{\ThetaBfone \approx
48.2^{\circ}}), and the five values of the diffusive mean fee path
\teq{\lambda/r_g=1,10,10^2,10^3,10^4} correspond to those in
Fig.~\ref{U1_071_r302_r371} -- the color-coding of the distributions and
the spectral index results in the respective Figures coincides. The
velocity compression ratio was fixed at \teq{r=u_{1x}/u_{2x}=3.71}, and
the upstream temperature corresponded to a sonic Mach number of
\teq{{M}_{\rm S}=4.04} (\teq{T=5.45\times 10^7}K for an
\teq{e^-}-\teq{e^+} shock).
 \label{inj_eff_u1x_71_UHT84_96} }
\end{figure*}

One naturally asks how these energetic particles that so efficiently
participate in SDA get accelerated to high energies in the first place.
Fig.~\ref{inj_eff_u1x_71_UHT84_96} illustrates the injection problem for
\teq{r=3.71} shocks with very warm particles (\teq{\machson=4}). The
distributions in these plots were used to determine the spectral slopes
displayed in Fig.~\ref{U1_071_r302_r371} for each value of
\teq{\betaoneHT}. High turbulence environments are clearly able to
inject the particles efficiently, but the power-law is steeper, namely
\teq{\sigma} is higher. Low turbulence environments have almost no
injection until particles achieve \teq{v>\uoneHT}, at which point a
strong, low \teq{\sigma} power-law develops as particles become trapped
in the shock drift mechanism. This becomes particularly pronounced in
the \teq{\betaoneHT = 0.96} case, an almost luminal shock situation,
where convection of thermal upstream particles through and downstream of
the shock is extremely rapid.  The rapid decline of injection efficiency
with \teq{\eta} is an important factor in interpreting the action of
shock acceleration in astronomical sources, an issue discussed in
Section~\ref{sec:blazar}.

\subsection{Oblique, Superluminal Shocks}
 \label{sec:obq_superluminal}
In superluminal shocks, it is physically impossible for even particles
moving at the speed of light to diffuse upstream along the field lines.
Without any cross-field diffusion, the particles are inexorably swept
downstream after passing through the shock without reflection,
regardless of their energy; see Begelman \& Kirk (1990) for illustration
of such trajectories. Thus, the particles in the low turbulence
environment that relied on reflection for retention in the system are
lost and not accelerated.  Consequently, the power-law becomes very
soft. Particles in high turbulence environments are not truly affected
by the change in the obliquity of the magnetic field because they can
travel across field lines as easily as they can along them. It is
generally found that for \teq{\betaoneHT<1} decreased turbulence
enhances acceleration in the power-law tail, and for \teq{\betaoneHT>1},
the opposite holds. This result adds new perspective to the paper by
Ellison and Double (2004), which presented results showing that the
power-law tail indices in ultra-relativistic (\teq{\Gamma_1\gg 1})
shocks are extremely sensitive to both \teq{\eta} and the obliquity of
the magnetic field, with the power-law index increasing sharply as these
parameters increase. These dependences are also seen in the
mildly-relativistic shocks discussed here, but with somewhat less
sensitivity to \teq{\eta} and \teq{\ThetaBfone}.

While it is clear that an increase in \teq{\eta} will soften the
power-law in oblique, superluminal shocks, the exact values of the
power-law index are simulation-dependent. Therefore, it is prudent to
compare our results with those of Ellison and Double (2004, hereafter
ED04). The code used in ED04 is a Monte Carlo simulation that is
algorithmically very similar to the simulation presented in this paper.
However, the simulations were developed independently and can each serve
as an objective test of the other. For non-relativistic shocks, both
simulations find the standard results of $\sigma=(r+2)/(r-1)$ where $r$
is the compression ratio of the shock and $\sigma$ is defined such that
$dn/dp = p^{-\sigma}$. In the case of ultra-relativistic, parallel
shocks, both simulations also find the theoretical results
$\sigma=-2.23$. In the regime of oblique, relativistic shocks, ED04
focused predominantly on superluminal cases of high Mach numbers, with
high levels of turbulence near the Bohm diffusion limit.

In Fig.~\ref{ell_doub_comp_7} we compare results from our simulation
(histograms) to the results from the top panel of Fig.~7 in ED04 (solid
lines) for a relativistic shock with upstream \teq{\Gamma_1=10} and
compression ratio, \teq{r=3.02}, for different values of the upstream
magnetic field obliquity, \teq{\ThetaBfone}, and
\teq{\eta=\lambda/r_{g}}.  Both sets of results are in the SAS limit,
with the Rankine-Hugoniot solutions for the MHD shock obtained using the
prescription of Double et al. (2004), as outlined in ED04, as opposed to
the J\"uttner-Synge EOS scenario; see Sec.~\ref{sec:EOS} for details.
For \teq{\ThetaBfone=0^{\circ}}, both simulations produce a result very
close to the canonical \teq{\sigma=-2.23} power-law (note that the
y-axis is multiplied by $p^{2.23}$ such that a horizontal line is the
canonical result). However, at $\ThetaBfone=60^{\circ}$,
$u_{1x}/\cos\ThetaBfone \approx 2$ making the shock decidedly
superluminal. Thus, cross-field diffusion is essential in order for
particles to be able to return to the shock and increasing \teq{\eta}
increases the power-law index considerably. This trend is identified by
both simulations in this \teq{\eta\lesssim 6} domain; in our simulation
it continues to somewhat higher \teq{\eta} before statistical
degradation inhibits determination of the spectrum. In general
character, results from the two simulations are clearly similar, and
numerically they are close.  Yet there is a real difference, with the
index determination differing between the two sets of results within the
range of \teq{1-2}\%.  The numerical precision of \teq{\sigma} in our
simulation is of the order of 1\%. Comparisons were made with other
superluminal shock results published in ED04, finding similar levels of
agreement. The origin of the small spectral differences evinced in
Fig.~\ref{ell_doub_comp_7} is not yet clear; we believe that the
thorough comparison of our indices and angular distributions with the
semi-analytic approaches of Kirk \& Heavens (1989) and Kirk et al.
(2000), among other simulation checks, advocates for the robustness of
our results.

The simulation developed by Niemiec and Ostrowski (2004, hereafter NO04;
see also Niemiec \& Ostrowski 2006) provides another opportunity for
comparison. NO04 uses a fundamentally different mechanism for particle
scattering from our Monte Carlo simulation. Instead of
phenomenologically scattering particles and specifying a momentum
dependence for the mean free path, their simulation injects a prescribed
spectrum of turbulent magnetic field structure that is superposed on the
bulk magnetic field. Variations in the magnetic field perturb the
gyro-orbits of the particles, and, are intended to mimic turbulence that
a particle might encounter. Because NO04 uses a turbulent magnetic
field, particle trajectories must be integrated over much shorter time
steps than is possible in the Monte Carlo code presented in this paper.
This necessarily results in longer run times and poorer statistics.
These poor statistics are particularly evident in the angular
distributions produced in Figs. 4, 8, and 11 of their work. We find
qualitatively similar general behavior for results from the two
techniques.

Consider Fig.~2 of NO04, where distribution functions for accelerated
particles in subluminal, mildly-relativistic oblique shocks with
compression ratio, \teq{r=5.11} are exhibited. This value of \teq{r}
exceeds the non-relativistic, strong shock limit of \teq{r=4}, and is
appropriate for the mildly relativistic electron-proton shocks studied
in Heavens \& Drury (1988), from whom they acquired their compression
ratios. Though there is not a simple relationship between the two
turbulence parameters, (\teq{\delta B/B} in their work, and \teq{\eta}
in our simulation) they are correlated with low values of \teq{\delta
B/B} corresponding to high values of \teq{\eta}. In subluminal shocks
for small \teq{\delta B/B}, the power law index for the distribution
function, $f(p)$, which they call $\alpha$, is approximately 3. This is
related to the power-law index for the differential density distribution
by $\sigma=\alpha-2$ or $\sigma\approx 1$ in their low turbulence limit.
This agrees with both our results and those of KH89. Additionally large
amounts of turbulence soften the power-law as we observed in our
simulations runs. Finally, it is interesting to note that at high
energies, where particles are resonant with larger wavelengths than the
stirring scale of their simulation, turbulence disappears for particles
at these energies ($\eta \rightarrow \infty$) and the power-law becomes
\teq{\sigma=1}, in good agreement with our results. In superluminal
oblique shocks such as those shown in Fig. 5 of their paper, although
the statistics are somewhat poor, it is clear that low turbulence is no
longer an asset to the acceleration process. In the shocks shown there,
the power-law is consistently softer in the low turbulence cases, and in
some cases no acceleration occurs without significant turbulence. In the
high turbulence case, power-laws are still produced, but then the
definition of the shock obliquity and the details of particle diffusion
become more important in determining the resulting power-law index,
rendering comparison with our results less insightful.

\figureoutpdf{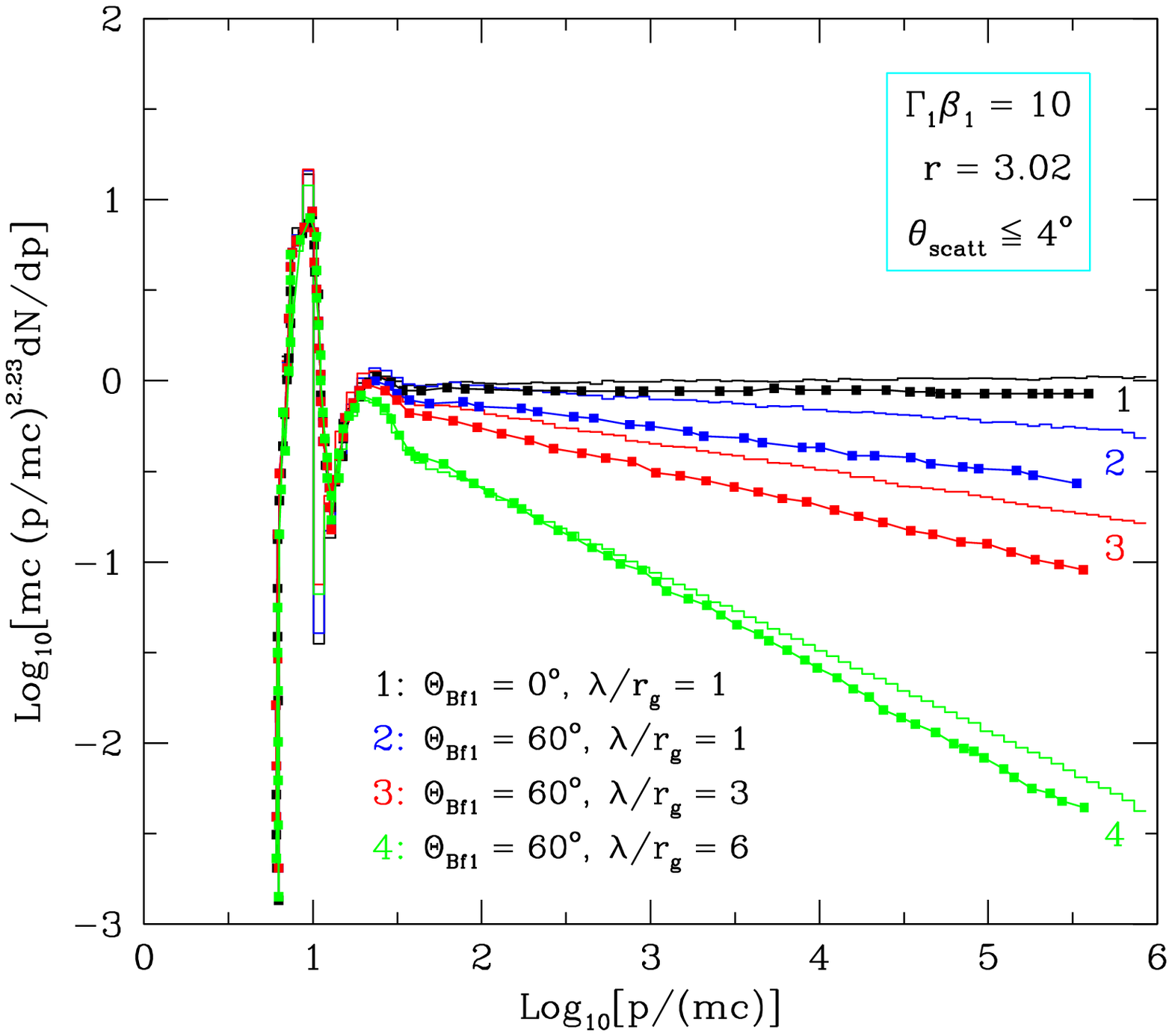}{3.8}{0.0}{-0.1}{
A direct comparison between distribution results from our simulation,
the histograms, and those Ellison and Double (2004), the solid lines,
for a relativistic shock with upstream Lorentz factor \teq{\Gamma_1=10}
and compression ratio, $r=3.02$, for different values of the upstream
magnetic field obliquity, \teq{\ThetaBfone}, and \teq{\eta=\lambda/r_g},
the turbulence parameter. Specifically, the results from the top panel
of their Fig. 7 are compared here with the y-axis multiplied by
\teq{p^{4.23}} to match the presentation in that figure. Both
simulations identify the same trends with only minor differences in the
value of the power-law slope.
 \label{ell_doub_comp_7} 
}

\subsection{Large Angle Scattering Domains}
 \label{sec_LAS}
One aspect of the simulation parameter space that has been neglected
until now is the impact of varying the microphysics of the turbulent
interactions; all previous results have focused on the small angle
scattering limit. In this section, we explore such using our Monte Carlo
simulation to model relativistic parallel shocks, by varying
\teq{\thetascatt}, the angular width of the conical sector into which
the particle's momentum vector is scattered at each encounter with
magnetic turbulence. A value of \teq{\thetascatt =\pi} corresponds to
large angle scattering (LAS), where the particles scatter \teq{\sim 1}
time per mean free path; this is the domain first highlighted by Ellison
et al. (1990a). A small value corresponds to SAS, where the particles
scatter \teq{N} times per mean free path, where \teq{N} is given by
Eq.~(\ref{eq:delthetamax}).

Fig.~\ref{fig:SAS_LAS_plot} depicts accelerated particle distributions
for two different shock speeds, illustrating the multitude of power-law
indices available while varying only the scattering angle,
\teq{\thetascatt}, and fixing the obliquity at
\teq{\ThetaBfone=0^{\circ}}.  Observe that for such parallel shocks, the
distributions are independent of the diffusion parameter \teq{\eta}
since they are measured just downstream of the shock. This depiction
complements results published in Fig.~2 of Stecker, Baring \& Summerlin
(2007), and illustrates two primary results. The first is that LAS
scattering produces a step-like structure in the accelerated
distribution, a characteristic first identified by Ellison et al.
(1990a).  Each step corresponding to particles with increasing numbers
of shock transits. In other words, the first step consists almost
entirely of particles that have crossed the shock 3 times. Particles in
the second step have almost all crossed the shock 5 times, etc.; see
Baring (2004) for an illustration of this correlation. The precise
correlation between shock transit number and particle energy weakens as
the structure damps into a power-law. The prevalence of the step
structure, and how high in energy it extends before relaxing into a
power-law, increases with the Lorentz factor \teq{\Gamma_1} of the
shock.

\figureoutpdf{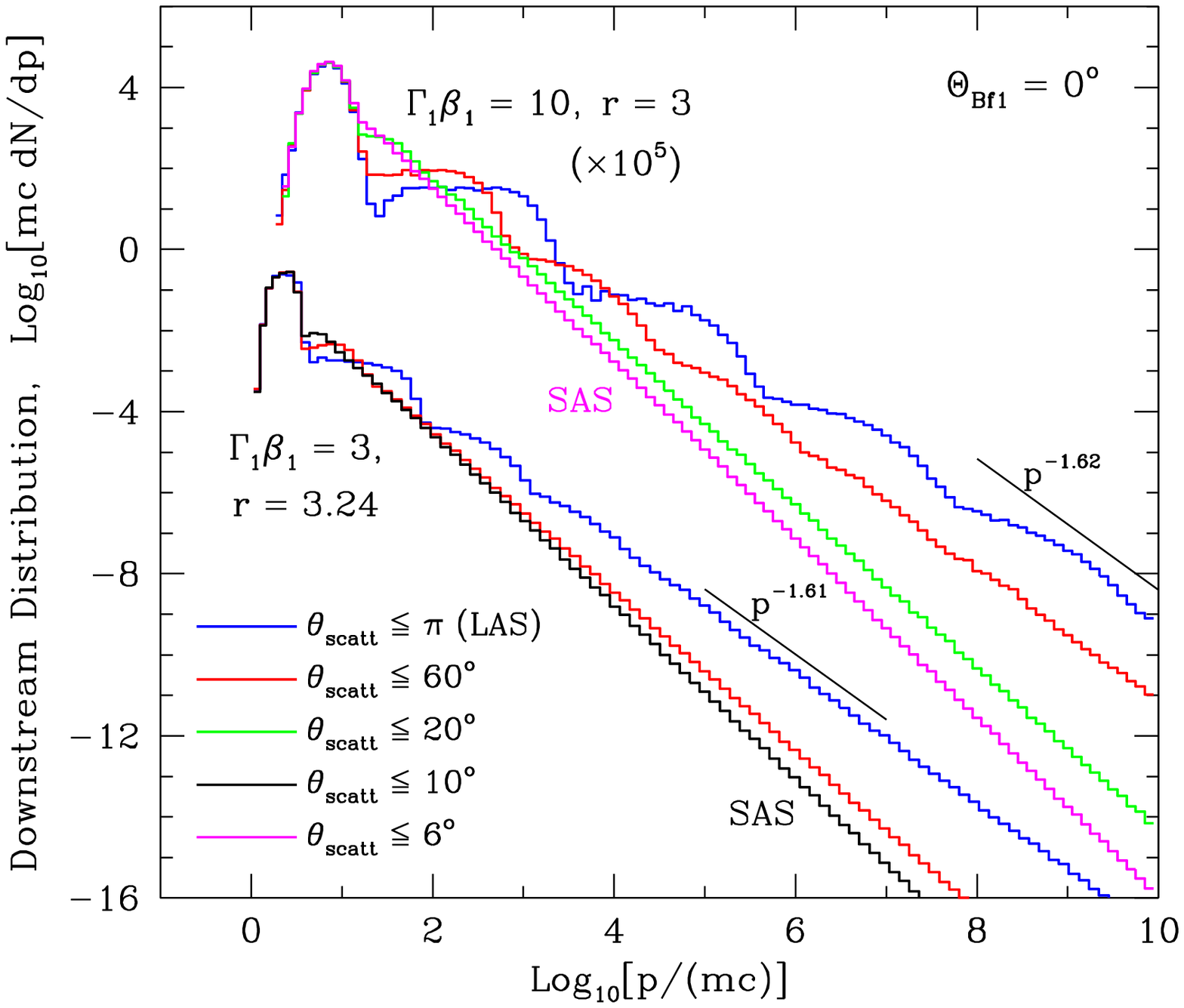}{3.8}{0.0}{-0.1}{
Particle distribution functions \teq{dN/dp} from parallel shocks that
are either mildly-relativistic (\teq{\Gamma_1\beta_1=3}, i.e.
\teq{\beta_1=u_1/c \approx 0.949}) or ultra-relativistic
(\teq{\Gamma_1\beta_1=10}, i.e. \teq{\beta_1=u_1/c \approx 0.995};
multiplied by \teq{10^5} to effect clarity of depiction), of velocity
compression ratios \teq{r=u_1/u_2\approx 3.24} and \teq{r \approx 3},
respectively. For these simulation runs, scattering off hydromagnetic
turbulence was modeled by randomly deflecting particle momenta by an
angle within a cone of half-angle \teq{\thetascatt}, whose axis
coincides with the particle momentum prior to scattering. Values of
\teq{\thetascatt} span the range from large angle scattering (LAS:
\teq{\thetascatt\leq\pi\gg 1/\Gamma_1}) to small angle scattering or
pitch angle diffusion, when \teq{\thetascatt\lesssim 1/\Gamma_1} and the
distributions become independent of the choice of \teq{\thetascatt}. All
distributions asymptotically approach power-laws \teq{dN/dp\propto
p^{-\sigma}} at high energies.  For two LAS cases, these power-laws are
indicated by lightweight lines, with indices of \teq{\sigma =1.61}
(\teq{\Gamma_1\beta_1=3}) and \teq{\sigma =1.62}
(\teq{\Gamma_1\beta_1=10}).
 \label{fig:SAS_LAS_plot} }

\newpage

The second major result is that decreasing the scattering angle removes
this structure but at the same time, softens the resulting power-law. A
complete investigation of why the power-law is harder in the LAS
scenario is deferred to future work, yet the origin of this trend in
\teq{\thetascatt} centers on the distribution function at the shock.
While LAS produces a beamed isotropic distribution similar to the red
histogram of the right panel of Fig. \ref{fig:ang_dist_gambet10} at the
shock, SAS generates a distribution like that shown in the black
histogram of the same panel. Both the probability of return to the shock
from the downstream side, and the energization per shock crossing, are
functions of this angular distribution function (Bell, 1978; Peacock,
1981; Blasi \& Vietri 2005). For non-relativistic shocks, the
distribution function is necessarily isotropic to leading order, which
restricts the power-law index to be only a function of the compression
ratio. Relativistic beaming is the probable cause for breaking this
degeneracy in \teq{\thetascatt} space. Evolution of the angular
distribution with \teq{\thetascatt} can be inferred from Fig.~6 of Blasi
\& Vietri (2005).  Note that introducing magnetic field obliquity can
alter the nature of this trend, as is indicated in Morlino, Blasi \&
Vietri (2009).  Future work will explore the variation of asymptotic
values of the index \teq{\sigma} as functions of \teq{\thetascatt} and
\teq{\ThetaBfone}, and also compare the Monte Carlo simulation values
with those obtained from the semi-analytic, transport equation approach
of Blasi \& Vietri (2005) and Morlino, Blasi \& Vietri (2009).

\section{Observational Connection: The Impact on Blazar Gamma-ray Interpretation}
 \label{sec:blazar}

To briefly outline how these simulation results are relevant to
astrophysical contexts, we discuss blazars, the subset of active
galactic nuclei possessing relativistic jets of material emanating from
the supermassive black holes at their centers; these jets are oriented
virtually towards the observer.  Blazars were discovered as a class of
gamma-ray sources by the EGRET experiment on the Compton Gamma-Ray
Observatory (Hartmann et al. 1992), and subsequently detected by
ground-based \c{C}erenkov telescopes at TeV energies (Punch et al.
1992). The EGRET blazar measurements have been built upon in the last
three years by {\it Fermi} Large-Area Telescope (LAT) detections of
dozens of blazars, offering improved spectroscopy.  The TeV-band signals
typically exhibit steep photon spectra (e.g. see Krennrich et al. 2002;
and Aharonian et al. 2003, for observations of Mrk 421) that include the
absorption due to pair producing interactions \teq{\gamma\gamma\to
e^+e^-} with infra-red and optical light generated by the intergalactic
medium along the line of sight to the observer.  Extremely flat particle
distributions are inferred in some blazars after correcting for this
attenuation (see, for example, Stecker, Baring \& Summerlin 2007), with
indices as low as \teq{\sigma\lesssim 1.5} in high redshift sources.
Coupled with the TeV-band capability, the {\it Fermi}-LAT detections of
blazars enable refined diagnostics by extending the observational window
over a much larger energy range, and most crucially, including below the
\teq{\gamma\gamma\to e^+e^-} attenuation window. Accordingly, {\it
Fermi} observations can probe more directly the underlying radiating
particle population.  The implications of this we explore here.  The
reader can consult Baring (2011) and references therein for the
interpretation of relativistic shock acceleration in gamma-ray burst
contexts.

Pertinent blazar data from the {\it Fermi}-LAT and TeV telescopes can be
found in the GeV-TeV blazar ``compendium'' in Abdo et al. (2009). There
is also the more extensive AGN catalog of {\it Fermi} in Abdo et al.
(2010). For the purposes of discerning indices \teq{\sigma} of particle
populations generating the gamma-ray emission, it is important to
consider photon spectra below any turnovers that may appear in the LAT
band.  This biases the data selection to below 1 GeV, and a nice
tabulation of this for {\it Fermi}-LAT blazars is given in Tables 5 and
6 of Abdo et al. (2009).  Therein, and in the various spectral plots
given in that paper, it is clear that there is a modest spectral
steepening above around 1 GeV in around 50\% of {\it Fermi}-LAT blazars;
this becomes much more pronounced above 100 GeV. From this data
compilation, we use photon indices of \teq{\alpha_{\gamma} = 1.72} for
PKS 2155-304, \teq{\alpha_{\gamma} = 1.78} for Mrk 421, and
\teq{\alpha_{\gamma} = 1.97} for 3C 66A, as a sample blazar selection. 
The uncertainties on these indices are of the order of \teq{\pm 0.1},
which propagate into inferred particle indices \teq{\sigma}. Note that a
sizable fraction of LAT band AGN indices in the catalog of Abdo et al.
(2010) fall below \teq{\alpha_{\gamma}\sim 2}, so that our choice here
is reasonable. Note also that 3C 279 is considerably steeper in the LAT
window, which could be a signature of a low-energy onset of the spectral
turnover, or the operation of the Klein-Nishina regime of inverse
Compton scattering.

Consider first a standard leptonic model interpretation of blazar
emission as emanating from inverse Compton scattering by
shock-accelerated electrons upscattering low energy photons. The
relationship between the particle index \teq{\sigma} and the photon one
is then \teq{\sigma = 2\alpha_{\gamma}-1} (e.g. see Chapter 7 of Rybicki
\& Lightman 1979), if there is insignificant radiational cooling. This
applies to synchrotron self-Compton (SSC) scenarios, where a single
population of electrons emits the synchrotron radiation that it then
upscatters to the gamma-ray band, or to an external supply of seed
photons. For our select blazars, these indices fall in the range
\teq{2.44 < \sigma < 2.94}, and are marked on the left panel of
Fig.~\ref{U1_071_r302_r371}. They have an uncertainty
\teq{\Delta\sigma\sim \pm 0.2}.  From this it is clear that for
shock-layer diffusive scattering in the SAS regime, acceleration at
mildly superluminal oblique shocks provides a good description for all
three of the blazars. This requires strong scattering, \teq{\lambda
/r_g\lesssim 10}. If \teq{\lambda /r_g\lesssim 2}, i.e., near the Bohm
diffusion limit, then Mrk 421 and PKS 2155-304 could be modeled with
subluminal shocks.

To contrast this, consider an alternative picture of ``strong cooling''
by inverse Compton scattering (or synchrotron radiation).   This
corresponds to rapid acceleration of the leptons at shocks, followed by
convection and diffusion away downstream into a larger radiative zone
where the gamma-ray signal is generated over longer timescales.  Then,
as is well known, the time-averaged effective electron distribution that
radiates is a power-law of index \teq{\sigma +1} (e.g., see Blumenthal,
1971, for an analysis).  The steepening reflects a pile-up at lower
Lorentz factors \teq{\gamma_e} induced by the fact that the energy loss
rate for both synchrotron and inverse Compton cooling of electrons
scales as \teq{\gamma_e^2}. The consequence is that now the relationship
between the particle index \teq{\sigma} and the photon one is 
\teq{\sigma = 2\alpha_{\gamma}-2}. These indices are marked on the right
panel of Fig.~\ref{U1_071_r302_r371}, and exhibit a shift of unity down
from those in the uncooled case. Again, they have an uncertainty
\teq{\Delta\sigma\sim \pm 0.2}. Now, subluminal regimes are clearly
suggested, and in the cases of Mrk 421 and PKS 2155-304, move the
inferences into \teq{\eta = \lambda/r_g > 10} territory.  This opens up
the question of whether such weak turbulence can persist in a blazar jet
shock.  However, there is considerable uncertainty in the spectral data,
and increasing the scattering angle \teq{\thetascatt} somewhat above the
SAS limit will reduce the value of \teq{\eta} required to generate a
particular value of \teq{\sigma}.

These inferences should be viewed only as general guidelines, modulo the
uncertainties in \teq{\sigma} spawned by the precision of {\it
Fermi}-LAT spectral index determination. We have selected one particular
shock speed \teq{\beta_{1x}} and restricted the discussion to SAS
regimes.  Clearly, there is a range of shock speeds, field obliquities,
scattering angles \teq{\thetascatt} and turbulence parameters \teq{\eta}
that can satisfy a measured gamma-ray index.  In addition, this
discussion has focused on leptonic models; in hadronic ones mediated by
pion decay, generally \teq{\alpha_{\gamma}} approximately traces the
particle index \teq{\sigma}, mimicking the situation in the right panel
of Fig.~\ref{U1_071_r302_r371}.  Significant Klein-Nishina modifications
to the spectrum in the LAT window can further complicate the
interpretation.  While more extensive study can hone the parameter
space, broader gamma-ray coverage below 100 MeV and multiwavelength
modeling are necessary to make significant strides.  The multiwavelength
aspect is more immediate in terms of its possibilities.  For example, if
the acceleration at the shock is limited by synchrotron cooling in
controlling the maximum electron Lorentz factor, then the turnover in
the synchrotron component is at an energy of \teq{\sim m_ec^2/(\fsc\eta
)} (for \teq{\fsc = e^2/\hbar c} as the fine structure constant), a
well-known result that is discussed in Garson, Baring \& Krawczynski
(2010) with reference to Mrk 421.  Evidently, this energy must match
that seen in hard X-ray/soft gamma-ray observations, providing
additional constraints on \teq{\eta} that often may not be near the Bohm
limit of \teq{\eta = 1}.  This illustration serves to motivate future
multiwavelength models of blazar spectra using complete distribution
functions from acceleration simulations like those presented in this
paper.

\section{Conclusions}
 \label{sec:conclusions}
This paper has presented new results from a robust Monte Carlo
simulation that complement and extend previous semi-analytic and
computational results. It employs the simulation technique devised by
Ellison et al. (1981) that was extended to relativistic shocks by
Ellison et al. (1990a). The simulation produces steady state
distribution functions for planar shocks of infinite extent for large
ranges of shock speeds, energies, and positions, simulating both the
injection and acceleration of particles via first-order diffusive shock
acceleration. By using the unique advantages that a simulation has over
semi-analytic, diffusion-convection equation solution techniques, we are
able to expand upon the work of previous authors by examining various
turbulence regimes and probing individual particle trajectories.  This
affords specific insights that cannot be gleaned from idealized cases
that are analytically tractable.  Our body of results leads to several
key conclusions:

\begin{itemize}
\item The power-law index in relativistic shocks samples a 
considerable range of values, and depends critically on the nature 
and magnitude of turbulence, the shock speed, and the shock field obliquity.
This range extends from extremely hard power-laws with \teq{\sigma \approx 1} 
to extremely steep distributions where simulation statistics preclude discernment 
of acceleration beyond the thermal injection domain.  Notably, 
ultra-relativistic shocks do not necessarily possess the ``canonical'' 
\teq{\sigma \approx 2.23} power-law distribution,
a result evident in the previous works of Kirk \& Heavens (1989),
Ellison \& Double (2004), and Stecker, Baring \& Summerlin (2007).
\item When small angle scattering (SAS) is invoked, the 
value of \teq{u_{1x}/\cos\ThetaBfone} defines
a critical division point in the parameter space.  When it is 
less than \teq{c}, oblique shocks in low levels of turbulence accelerate 
high energy particles extremely efficiently via shock drift acceleration, 
but when this quantity is greater than \teq{c}, turbulence becomes vital to 
injection and acceleration in oblique, superluminal shocks.
In both cases, weak levels of turbulence strongly inhibit
injection from the thermal population.
\item Invoking large angle scattering produces significant 
structure in the high energy particle distributions in relativistic shocks, 
a phenomenon first identified by Ellison, Jones \& Reynolds (1990),
but also generates slightly harder distributions than a similar shock 
in the SAS scenario, where there is only a power-law with little 
discernible structure.
\end{itemize}
These results represent important advances for determining the nature of
turbulent shock environs in blazar and gamma-ray burst jets, and in
other astrophysical objects.   Such interpretations are, admittedly,
complicated by the particular spatial environment and radiation emission
mechanism chosen for generating the observed photon spectra from these
sources.  Yet global insights such as deciding between subluminal or
superluminal shock environments are now possible.

To develop our model to aid future interpretations of astrophysical
shocks, additional details of shock physics will be incorporated into
the simulation. In shocks such as those discussed above with \teq{\sigma
< 2}, a majority of the energy in the system will be found in the
accelerated particles. The Rankine-Hugoniot jump conditions for the
shock must then be modified, since a step function shock profile is no
longer a valid approximation. This yields a non-linear acceleration
phenomenon that is already seen clearly in the Earth's bow shock, models
of supernova remnant shocks, and the heliospheric termination shock
(e.g., Ellison, Jones \& Baring 1999). Additionally, while this paper
worked primarily with large Alfv\'{e}nic Mach number shocks, in
principle, low Alfv\'{e}nic Mach numbers are possible in jet systems as
well. While the Rankine-Hugoniot solutions presented above are fully
capable of determining the appropriate jump conditions, low Alfv\'{e}nic
Mach number shocks may produce significant second-order Fermi
acceleration due to the motion of the scattering centers (Alfv\'{e}n
waves) in the upstream and downstream rest frames. The code currently
assumes scattering centers that are stationary in their respective fluid
frames but can easily be adapted to include non-stationary scattering
centers for the case of low Alfv\'{e}nic Mach number shocks.  In
addition, preliminary work has also been done laying the ground work for
future inclusion of cross-shock potentials in the simulation (Baring \&
Summerlin, 2007). The simulation is currently a single fluid model,
treating electrons/pairs or ions.   For electron-proton shocks, the
disparate diffusion scales of the two species will cause their
distribution functions to react to the presence of the shock on
different length scales. This charge separation at the shock
discontinuity induces an electric field that acts to restore
quasi-neutrality, and can lead to significant energy exchange between
ions and electrons.  The inclusion of these effects and the
determination of their impact on injection and acceleration of protons
and electrons will be the focus of future work.

\acknowledgements
We thank the anonymous referee, Don Ellison and John Kirk for some
comments helpful to the polishing of the manuscript. MGB is grateful for
the generous support of the NASA Astrophysics Theory and Heliospheric
Physics Programs through grants NNX10AC79G and NNG05GD42G, and the
National Science Foundation through grant PHY-0758158. MGB is also
grateful to the Kavli Institute for Theoretical Physics, University of
California, Santa Barbara for hospitality during part of the period when
this research was performed, a visit that was supported in part by the
National Science Foundation under Grant No. PHY05-51164.

\section{References}

\reference
Abdo, A.~A. et al. 2009, \apj,\vol{707}{1310}
\reference
Abdo, A.~A. et al. 2010, \apj,\vol{715}{429}
\reference
Aharonian, F. et al. 2003, \aa,\vol{410}{813}
\reference
Arfken, G. B. \& Weber, H. J. 2001, {\it Mathematical Methods for Physicists, 5th Edition}
(Academic Press, Orlando).
\reference
Armstrong, T. P., Pesses, M. E., \& Decker, R. B. 1985, Collisionless Shocks 
in the Heliosphere: Reviews of Current Research, 
(Washington, DC: American Geophysical Union), p.~271.
\reference
Baring, M. G. 1999, in Proc. XXVI Intl. Cosmic Ray Conf. \vol{4}{5} [{\tt astro-ph/9910128}]
\reference
Baring, M. G. 2002, Publ. Astron. Soc. Aust., \vol{19}{60}
\reference
Baring, M.~G. 2004, Nuclear Physics B, Proc. Supp., \vol{136}{198}
\reference
Baring, M. G. 2011, \asr, \vol{47}{1427}
\reference
Baring, M.~G., Ellison, D.~C., Reynolds, S.~P., Grenier, I.~A., Goret, P. 1999, \apj,\vol{513}{311}
\reference
Baring, M.~G.,  Ogilvie, K.~W., Ellison, D.~C. \& Forsyth, R.~J. 1997, \apj,\vol{476}{889}
\reference
Baring, M.~G. \& Summerlin, E. J., 2007, \apss,\vol{307}{165}
\reference
Baring, M.~G. \& Summerlin, E.~J. 2009, in
     ``Shock Waves in Space and Astrophysical Environments,''
     eds. X. Ao, et al.,  R. Burrows \& G.~P. Zank 
     (AIP Conf. Proc. 1183, New York) pp.~74--83. {\tt [astro-ph/0910.1072]}
\reference
Ballard, K. R. \& Heavens, A. F. 1991, \mnras,\vol{251}{438}
\reference
Bednarz, J. \& Ostrowski M. 1998, \prl,\vol{80}{18}
\reference
Begelman, M.~C. and Kirk, J. G. 1990, \apj,\vol{353}{66}
\reference
Bell, A. R. 1978, \mnras,\vol{182}{147}
\reference
Bell, A. R., Schure, K.~M. \& Reville, B. 2011 \mnras, in press.
\reference
Blandford, R.~D. \&  Eichler, D. 1987, Phys. Rep., \vol{154}{1} 
\reference
Blandford, R.~D. \& McKee, C.~F. 1976, Phys. Fluids, \vol{19(8)}{1130}
\reference
Blasi, P. \& Vietri, M. 2005, \apj,\vol{626}{877}
\reference
Blumenthal, G.~R. 1971, \prd,\vol{3}{2308}
\reference
Decker, R. B. 1988, \ssr,\vol{48}{195}
\reference
Decker, R. B., Pesses, M. E., \& Krimigis, S. M. 1981, \jgr,\vol{86}{8,819}
\reference
Decker, R.~B. \& Vlahos, L. 1986, \apj,\vol{306}{710}
\reference
de Hoffmann, F., \& Teller, E. 1950, \pr,\vol{80}{692}
\reference
Double, G. P., Baring, M. G., Jones, F. C., \& Ellison, D. C. 2004, \apj,\vol{600}{485}
\reference
Drury, L. O'C., 1983, Rep. Prog. Phys., \vol{46}{973}
\reference
Ellison, D. C., Baring, M. G. \& Jones, F. C. 1995, \apj,\vol{453}{873}
\reference
Ellison, D. C., Baring, M. G. \& Jones, F. C. 1996, \apj,\vol{473}{1029}
\reference
Ellison, D.~C. \& Double, G.~P. 2002, \app,\vol{18}{213} 
\reference
Ellison, D. C. \& Double, G. P. 2004, \app,\vol{22}{323}
\reference
Ellison, D. C. \& Eichler, D. 1984, \apj,\vol{286}{691}
\reference
Ellison, D. C., Jones, F. C. \& Baring, M. G. 1999, \apj,\vol{512}{403}
\reference
Ellison, D.~C., Jones, F.~C. \& Eichler, D. 1981, J. Geophys., \vol{50}{110}
\reference
Ellison, D. C., Jones, F. C. \& Reynolds, S. P. 1990a, \apj,\vol{360}{702}
\reference
Ellison, D.~C., M\"obius, E. \& Paschmann, G. 1990b \apj,\vol{352}{376}
\reference
Forman, M. A., Jokipii, J. R. \& Owens, A. J. 1974, \apj,\vol{192}{535}
\reference
Gallant, Y.~A.,  Hoshino, M.,  Langdon, A.~B.,  Arons, J., \& Max, C.~E. 
	1992, \apj,\vol{391}{73}
\reference
Garcia, A.~L., 2000, {\it Numerical Methods for Physics, 2nd Edition}
(Prentice Hall, Upper Saddle River, New Jersey).
\reference
Garson, A.~B., Baring, M.~G. \& Krawczynski, H. 2010, \apj,\vol{722}{358}
\reference
Gerbig, D. \& Schlickeiser, R. 2011, \apj,\vol{733}{32}
\reference
Giacalone, J., Burgess, D., \& Schwartz, S. J. 1992, in Study of the Solar-Terrestrial
System, (Noordwijk: ESA Special Publication), 65.
\reference
Gosling, J.~T., Thomsen, M.~F., Bame, S.~J., \& Russell, C.~T. 1989, \jgr,\vol{94}{3555}
\reference
Hartman, R.~C. et al. 1992, \apj,\vol{385}{L1}
\reference
Heavens, A. F.  \& Drury, L. O'C. 1988, \mnras,\vol{235}{997}
\reference
Hededal,\ C.~B., Haugbolle,\ T., Frederiksen,\ J.~T. \& Nordlund,\ A 2004, \apj,\vol{617}{L107}
\reference
Jokipii, J. R. 1982, \apj,\vol{255}{716}
\reference
Jones, F.~C. 1978, \apj,\vol{222}{1097}
\reference
Jones, F.~C. \& Ellison, D.~C. 1991, \ssr,\vol{58}{259}
\reference
Kirk, J. G., Guthmann, A. W., Gallant, Y. A. \& Achterberg, A. 2000, \apj,\vol{542}{235}
\reference
Kirk, J. G. \& Heavens, A. F. 1989, \mnras,\vol{239}{995}
\reference
Kirk, J.~G. \& Schneider, P. 1987, \apj,\vol{315}{425}
\reference
Krennrich, et al. 2002, \apj,\vol{575}{L9}
\reference
Landau, L.~D. \& Lifshitz, E.~M. 1975, {\it The Classical Theory of Fields}, 4th Edition
(Butterworth-Heinemann, Oxford).
\reference
Lee, M.~A. 1983, \jgr, \vol{88}{6109}
\reference
Lee, M.~A. 1984, \asr,\vol{4}{295}
\reference
Liang, E.~P. \& Nishimura, K. 2004, \prl,\vol{92}{5005}
\reference
Mason, G. M., Gloeckler, G., and Hovestadt, D. 1983, \apj,\vol{267}{844}
\reference
Medvedev, M. V., Fiore, M., Fonseca, R. A., Silva, L. O., \& Mori W. B. 2005, \apj,\vol{618}{L75}
\reference
M\'esz\'aros, P. 2001, Science, \vol{291}{79}
\reference
M\"obius, E., Scholer, M., Sckopke, N., Paschmann, G., \& Luehr, H. 1987, \grl,\vol{14}{681}
\reference
Morlino, G., Blasi, P. \& Vietri, M. 2007, \apj,\vol{658}{1069}
\reference
Niemiec, J., \& Ostrowski, M. 2004, \apj,\vol{610}{851}
\reference
Niemiec, J., \& Ostrowski, M. 2006, \apj,\vol{641}{984}
\reference
Nishikawa, K.-I., Hardee, P., Richardson, G., Preece, R., Sol, H., and Fishman, G. J. 2005, \apj,\vol{622}{927}
\reference
Ostrowski, M. 1991, \mnras,\vol{249}{551}
\reference
Peacock, J. A. 1981, \mnras,\vol{196}{135} 
\reference
Pesses, M.~E. \& Decker, R.~B. 1986, \jgr,\vol{91}{4143}
\reference
Pesses, M. E., Decker, R. B., and Armstrong, T. P. 1982, \ssr,\vol{32}{185}
\reference
Piran, T.  1999, Phys. Rep., \vol{314}{575}
\reference
Pryadko, J. M. and Petrosian, V. 1997 \apj,\vol{482}{774}
\reference
Punch, M. et al. 1992, Nature, \vol{358}{477}
\reference
Rybicki, G.~B. and Lightman, A.~P. 1979, \it Radiative Processes in
Astrophysics \rm (Wiley, New York).
\reference
Sarris, E. T., \& Van Allen, J. A. 1974, \jgr,\vol{79}{4,157}
\reference
Silva, L. O., Fonseca, R. A., Tonge, J. W., Dawson, J. M., Mori, W. B., and Medvedev, M. V. 2003, \apj,\vol{596}{L121}
\reference
Scholer, M., Hovestadt, D., Ipavich, F. M., \& Gloeckler, G. 1980, \jgr,\vol{85}{4,602}
\reference
Sironi, L. \& Spitkovsky, A. 2011, \apj,\vol{726}{75}
\reference
Smolsky, M.~V. \& Usov, V.~V. 1996, \apj,\vol{461}{858}
\reference
Spitkovsky, A. 2008, \apj,\vol{682}{L5}
\reference
Stecker, F.~W. Baring, M.~G. \& Summerlin, E. J. 2007, \apj,\vol{667}{L29}
\reference
Stoer, J., and Bulirsch, R. 1980, {\it Introduction to Numerical Analysis} (New York: Springer-Verlag), $\oint$ 7.2.14.
\reference
Synge, J. L. 1957, {\it The Relativistic Gas} (North Holland, Amsterdam).
\reference
Summerlin, E. J. \& Baring, M. G. 2006, \asr,\vol{37(8)}{1426}
\reference
Tan, L. C., Mason, G. M., Gloeckler, G., \& Ipavich, F. M. 1988, \jgr,\vol{93}{7,225}
\reference
Terasawa, T. 1979, Planet. Space Sci.,\vol{27}{193}
\reference
Vandas, M., 2001, \jgr,\vol{106}{1859}
\reference
Webb, G.~M., Axford, W.~I. \& Terasawa, T. 1983, \apj,\vol{270}{537}

\vspace{1.25in}
\appendix
\section{Appendix A: The Angular Distribution of Particles at the 
Return Plane}

Downstream of the shock, once the particle distribution has realized isotropy 
in the local fluid frame at some position \teq{x}, it maintains isotropy as an 
asymptotic state at all positions further downstream.  Tracking diffusion 
downstream of \teq{x} to assess the momentum components 
of particles that return to position \teq{x} is CPU-intensive.  A much faster 
method is to compute the probability of return to \teq{x} and these 
momentum components statistically, subject to the condition of isotropy in the 
local fluid frame, and constancy of the returning particle's fluid frame momentum 
\teq{p_f}.  This was the expedient approach of Ellison et al. (1990a).  
In this Appendix we develop the formalism for angular distributions of 
returning particles for {\it arbitrary} \teq{p_f}, not just ultra-relativistic particles,
as has been the restriction of previous expositions.  At \teq{x}, the flow speed 
will be \teq{\beta c}, i.e. of Lorentz factor \teq{\Gamma} (subscripted 2 in the main text).
The particle will have a dimensionless momentum \teq{p_f=\gamma_f\beta_f} 
(\teq{p_s=\gamma_s\beta_s}) and angle cosine with respect to the shock normal
of \teq{\mu_f} (\teq{\mu_s}) in the fluid (NIF) frame.  The non-covariant momentum 
distribution function downstream of the {\it probability of return plane} located 
at \teq{x} (and all positions further downstream) assumes the form 
\begin{equation}
   f(\vect{p}_{f}) \; =\; \dover{N\delta(p_{f}-p_{0})}{4\pi p_{f}^{2}}
   \quad ,\quad 
   N\; =\; \int  f(\vect{p}_{f}) d^{3}\vect{p}_{f}
   \;\equiv\; \int  f_{s}(\vect{p}_{s}) d^{3}\vect{p}_{s} \quad .
 \label{eq:phase_space}
\end{equation}
in the local fluid frame.  Observe that \teq{N}, the total number of particles, is a
Lorentz invariant under boosts along the shock normal, and \teq{f} is to be 
distinguished from the phase space density \teq{f(\vect{r},\, \vect{p})}.   Equation 10.2 of 
Landau and Lifshitz (1989) then gives the transformation of the distribution functions 
as \teq{\gamma_s f_{s}(\vect{p}_{s})=\gamma_f f_{f}(\vect{p}_{f})}; in covariant 
formulations, the particle Lorentz factor is absorbed into the definition of \teq{f}.
This is employed in the second integral in Eq.~(\ref{eq:phase_space}), along with 
the fluid frame distribution function.  To integrate over the delta function, it is 
necessary to change variables \teq{dp_s\to dp_f}, so that
\begin{equation}
   N\; =\; 2\pi \int \dover{p_s^2\, dp_s}{\gamma_s}\, \gamma_f\, f_{f}(\vect{p}_{f})\, d\mu_{s} 
   \;\equiv\; \dover{N}{2} \int \dover{\gamma_s\beta_s^2}{\gamma_f\beta_f^2}\,
   \left\vert \dover{\partial p_s}{\partial p_f} \right\vert\,  d\mu_s\quad .
 \label{eq:N_eval}
\end{equation}
Hereafter, the identity \teq{p_f=p_0} will be assumed.
The partial Jacobian \teq{\vert \partial p_s/\partial p_f\vert} can be determined by 
first writing the Lorentz boost relations for the particle momentum:
\begin{eqnarray}
   p_s \mu_s &=& \Gamma (\gamma_f \beta + p_f\mu_f) \nonumber\\[-5.5pt]
 \label{eq:Lorentz_boost}\\[-5.5pt]
  p_{s}\sqrt{1-\mu_{s}^{2}} &=& p_{f}\sqrt{1-\mu_{f}^{2}}\nonumber
\end{eqnarray}
After moderate algebraic manipulation, one can eliminate \teq{\mu_f} and solve for 
\teq{p_s/p_f}  as a function of \teq{p_f} and \teq{\mu_s}:
\begin{equation}
   \dover{p_s}{p_f} \; =\; \dover{\gamma_f}{\Gamma^2}\, \dover{S+\chi\mu_s}{1-\beta^2\mu_s^2}
   \quad \hbox{for} \quad
   S \; =\; \dover{\Gamma}{\gamma_f} \sqrt{1-\chi^2(1-\mu_s^2)}
   \quad ,\quad
   \chi\; =\; \dover{\Gamma\beta}{\gamma_f\beta_f}\quad .
 \label{eq:ps_over_pf}
\end{equation}
Additionally, holding the integration variable, \teq{\mu_s}, constant, one can derive
\begin{equation}
   \left\vert \dover{\partial p_s}{\partial p_f} \right\vert \; =\;
   \dover{\Gamma\gamma_s}{S\gamma_f^2}
 \label{eq:Jacobian_part}
\end{equation}
These allow us to rewrite Eq.~(\ref{eq:N_eval}) in the form
\begin{equation}
   N \; =\; \int_{-1}^1 \dover{dN}{d\mu_s}\, d\mu_s
   \qquad ,\qquad
   \dover{dN}{d\mu_s} \; =\; \dover{N\gamma_f}{2\Gamma^3}
      \,\dover{(S+\chi \mu_s)^2}{S(1-\beta^2\mu_s^2 )^2}
   \; =\; \dover{N\gamma_f}{2\Gamma^3}
      \, \dover{d\Sigma}{d\mu_s}
   \quad \hbox{for}\quad
   \Sigma\; =\; \dover{\mu_s (S + \chi\mu_s)}{1-\beta^2\mu_s^2 } \quad .
 \label{eq:dNdmus_form}
\end{equation}
The angular distribution \teq{dN/d\mu_s} describes the beaming appearing
in the NIF shock frame of an isotropic distribution in the fluid frame
of fixed, specified momentum \teq{p_f}. It is applicable to arbitrary
\teq{\gamma_f}, not just ultra-relativistic cases \teq{\gamma_f\gg 1},
the usual restricted consideration (e.g. see Peacock 1981). The identity
for \teq{dN/d\mu_s}, casting it in terms of a perfect derivative
\teq{d\Sigma/d\mu_s}, can be established using a moderate amount of
algebra; it nicely facilitates the derivation of the integral identity
for \teq{N} in Eq.~(\ref{eq:dNdmus_form}).

To formulate probabilities of transmission and return at position
\teq{x}, the angular distribution \teq{dN/d\mu_s} must be weighted by
the flux of particles incident upon the plane at \teq{x} that is
parallel to the shock plane.  In the NIF, this weighting factor is
proportional to the density of particles, which is proportional to
\teq{\gamma_s}, and also to the velocity component \teq{v_s\vert
\mu_s\vert} along the \teq{x}-direction.  In this way, we form {\it flux
angular distributions} using Eq.~(\ref{eq:ps_over_pf}) as follows:
\begin{equation}
   \dover{d{\cal F}}{d\mu_s}\; =\; C\, \vert \mu_s\vert \, 
      \dover{(S+\mu_s\chi)^3}{S\, (1-\beta^2\mu_s^2)^3}
   \;\equiv\; C\, \vert \Sigma\vert \, \dover{d\Sigma}{d\mu_s}\quad ,
 \label{eq:flux_ang_dist}
\end{equation}
for \teq{C} being a constant of normalization.  This formula generalizes that 
employed in Peacock (1981), which 
is restricted to \teq{\gamma_f\gg 1} cases.  It is easily seen that the integrands in 
Eq.~(26) of Peacock's paper are proportional to the \teq{\gamma_f\to\infty} limit
of Eq.~(\ref{eq:flux_ang_dist}), and can essentially be derived using 
light aberration considerations.  The total probability \teq{P_r} of return to 
\teq{x} of particles of fixed \teq{p_f} incident from the upstream side is then the ratio 
of two integrals over the flux distribution:
\begin{equation}
   P_r\; =\; \dover{{\cal F}_{d\to u}}{{\cal F}_{u\to d}}\; =\; \left( \dover{\beta_f-\beta}{\beta_f+\beta} \right)^2
   \quad ,\quad
   {\cal F}_{d\to u}\; =\; \int_{-1}^0  \dover{d{\cal F}}{d\mu_s}\, d\mu_s
   \quad ,\quad
   {\cal F}_{u\to d}\; =\; \int_0^1  \dover{d{\cal F}}{d\mu_s}\, d\mu_s\quad .
 \label{eq:Pr_ret}
\end{equation}
This simple expression for \teq{P_r} is valid for any flow speed \teq{\beta} and 
any particle speed \teq{\beta_f\geq \beta}; it was first derived for non-relativistic shocks 
(i.e. \teq{\beta\ll 1}) by Bell (1978) and for relativistic shocks 
with high-speed (\teq{\beta_f\approx 1}) particles by Peacock (1981).
Again, recognizing the appearance of a perfect derivative \teq{d(\Sigma^2)/d\mu_s} in
the flux distributions expedites the integrations in Eq.~(\ref{eq:Pr_ret}).  The direction 
of the magnetic field, and the level of cross-field diffusion are irrelevant to the derivation 
of the flux angular distribution and probability of return.  Technically these formulae, 
as derived, must be applied in {\it a downstream normal incidence frame}; in practice,
for many Rankine-Hugoniot solutions such as for high Alfv\'enic Mach numbers
(e.g., see Figs.~\ref{fig:RH_ThetaB=5deg} and~\ref{fig:RH_ThetaB=85deg}), 
the flow deflection at the shock is small, and the 
downstream and upstream NIF frames are almost coincident.

The simulation computes the statistical probability of return to the plane at \teq{x}
according to \teq{P_r}.  Those particles that are deemed to escape are eliminated 
from the simulation.  For those that return, their returning value of \teq{\mu_s\leq 0} is
selected randomly from the distribution in Eq.~(\ref{eq:flux_ang_dist}) when it 
is normalized to unity on \teq{-1\leq \mu_s\leq 0}.  Then the constant of proportionality
is \teq{C=2\gamma_f^2\beta_f^2/\Gamma^6/(\beta_f-\beta)^2}.  
Upon return, the particle is placed at \teq{x} with the same \teq{(y,z)}
coordinates it originally crossed with; the system is uniform in the dimensions 
transverse to the shock normal, so this step introduces no bias.  
After \teq{\mu_s} is determined, the phase 
of the momentum vector about the shock normal is selected randomly from 
a uniform distribution on \teq{[0, 2\pi ]}.  The returning particle momentum vector is then 
totally specified, and is routinely cast in rotated coordinates to identify variables 
connected to gyration about oblique magnetic fields. 

\end{document}